# Capacitación docente en Competencias Digitales Inclusivas en la ESO. El alumnado con Trastornos del Espectro Autista

José María Fernández-Batanero
Pedro Román-Graván
(*eds*.)

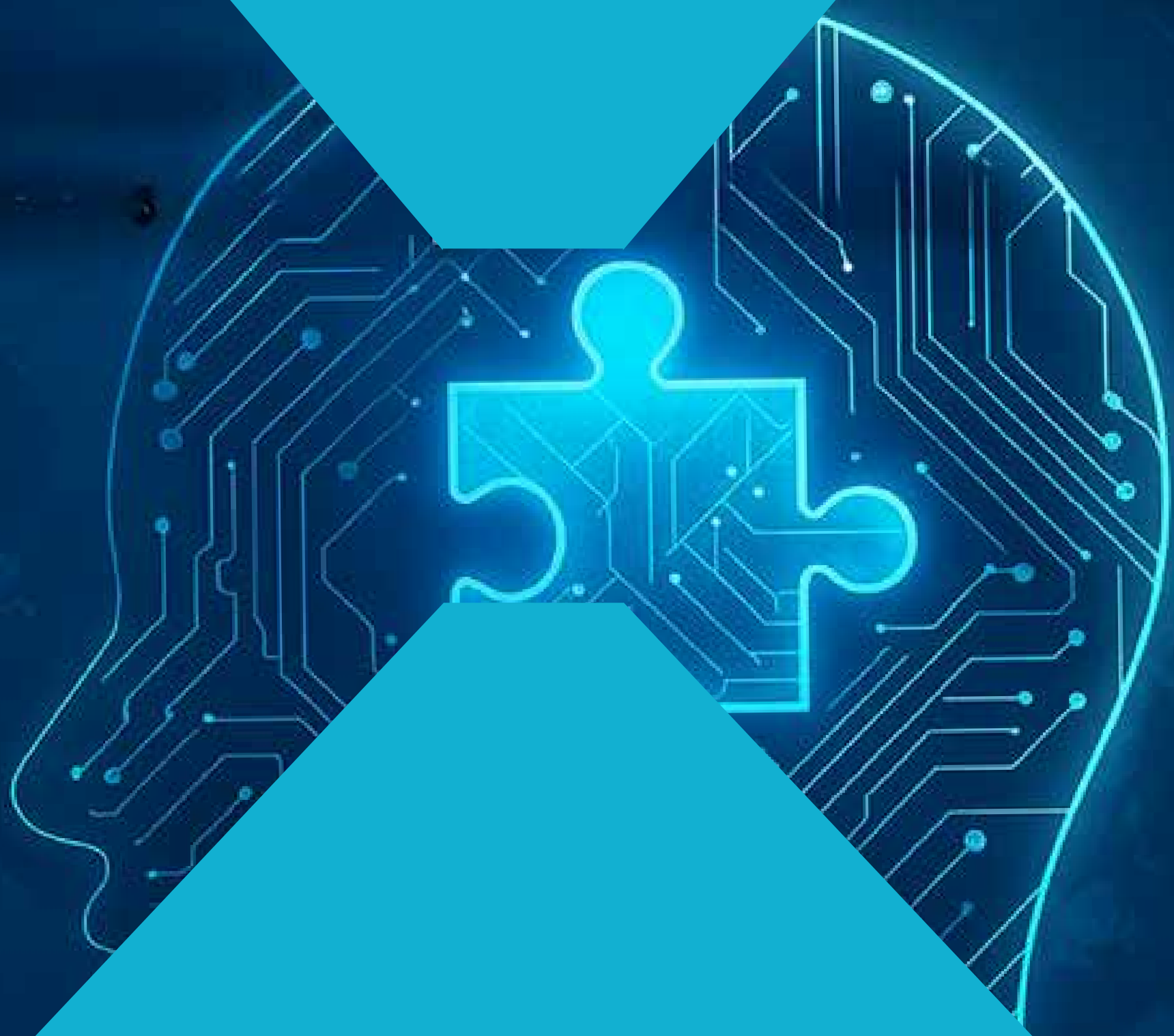

**José María Fernández-Batanero**

Catedrático de Universidad de Didáctica y Organización Educativa de la Universidad de Sevilla. Su actividad investigadora se centra en tres líneas de trabajo: formación del profesorado, atención a la diversidad y tecnología educativa como apoyo a las personas con necesidades educativas especiales. Forma parte del Grupo de Investigación "GID" (HUM-390) y del Grupo de Tecnología Educativa de la Universidad de Sevilla. Director de la Red Latinoamericana de TIC aplicadas a personas con discapacidad (ReLaTICyD) y Miembro del Consejo Consultivo del Centro de Estudos em Educaçao e Inovaçao (CI&DEI, Portugal). Miembro nato de la Cátedra Institucional de Educación en Tecnologías Emergentes, Gamificación e Inteligencia Artificial (EduEmer) de la Universidad Pablo de Olavide (Sevilla). Evaluador de la Agencia Nacional de Evaluación y Prospectiva (ANEP) y asesor de la Comisión Nacional de Investigación Científica y Tecnológica – CONICYT, del Gobierno de Chile. Investigador en el prestigioso ranking World's Top 2% Scientists List 2024 Stanford University.

Ha impartido docencia en los tres niveles del sistema educativo, siendo en la actualidad funcionario docente en Primaria y Universidad (Primaria excedencia voluntaria). Está en posesión de 6 diplomas a la Excelencia Docente Universitaria y de la Insignia de Oro de la Ciudad de Sevilla en reconocimiento a la Excelencia Docente.

**Pedro Román-Graván**

Profesor de EGB, Licenciado en Ciencias de la Educación (Pedagogía), Experto Universitario en Evaluación Educativa y Doctor en Ciencias de la Educación. Es Profesor Titular de Universidad en el Departamento de Didáctica y Organización Educativa de la Universidad de Sevilla. Su línea de investigación principal se relaciona con los procesos educativos mediados por la Tecnología Educativa y de formación en contextos de diversidad, evaluación, robótica educativa, realidad aumentada, impresión 3D y drones.

Forma parte del Grupo de Investigación "GID" (HUM-390) y del Grupo de Tecnología Educativa de la Universidad de Sevilla. Evaluador de la Agencia Nacional de Evaluación y Prospectiva (ANEP). Ha impartido docencia en los tres niveles del sistema educativo: Educación Primaria, Educación Secundaria Obligatoria y Universidad, siendo en la actualidad funcionario docente en la Universidad de Sevilla. Posee cuatro sexenios de investigación reconocidos por la Comisión Nacional Evaluadora de la Actividad Investigadora (CNEAI) ANECA: tres de Investigación y uno de Transferencia. Ocupa la posición 12/51 entre los investigadores de Educación más citados en el Área de Didáctica y Organización Escolar de su Universidad. Desde 2012 hasta 2023 ha sido miembro de Comisión Asesora Aula de la Experiencia de la Universidad de Sevilla, donde imparte la asignatura de Nuevas Tecnologías.

# Capacitación docente en Competencias Digitales Inclusivas en la ESO

## El alumnado con Trastornos del Espectro Autista

# Capacitación docente en Competencias Digitales Inclusivas en la ESO

## El alumnado con Trastornos del Espectro Autista

José María Fernández-Batanero
Pedro Román-Graván
*(eds.)*

Esta obra ha sido sometida a evaluación externa por pares ciegos, aprobada por el Comité Científico y ratificado por el Consejo Editorial de acuerdo con el Reglamento de la Editorial de la Universidad de Cantabria.

Esta publicación es parte del proyecto PID2022-138346OB-I00, financiado por MICIU/AEI/10.13039/501100011033 y por FEDER, UE.

Diseño de colección | digitalización: Gema M. Rodrigo | Manuel Ángel Ortiz Velasco [emeaov]

# SUMARIO

# PRÓLOGO

*José María Fernández-Batanero*
*Pedro Román-Graván*
Universidad de Sevilla

En la sociedad contemporánea, marcada por una evolución tecnológica acelerada, la educación se enfrenta al reto —y a la oportunidad— de incorporar nuevas herramientas digitales que transformen el aprendizaje, haciéndolo más inclusivo, flexible y significativo. Este libro se inscribe en esa línea de compromiso con la innovación educativa y la equidad, poniendo el foco en un colectivo que requiere una atención especializada y sensible: el alumnado con Trastorno del Espectro Autista (TEA).

Lejos de abordar la tecnología desde una perspectiva meramente instrumental, esta obra propone un enfoque profundamente humano, donde las tecnologías emergentes —como la realidad aumentada y virtual, los entornos inmersivos, los sistemas de comunicación aumentativa, las aplicaciones móviles o la inteligencia artificial— se convierten en aliadas para favorecer la autonomía, la autorregulación emocional, el desarrollo de habilidades sociales y la inclusión real del alumnado con TEA en los entornos educativos.

Cada capítulo ofrece una aportación rigurosa, fruto de la experiencia, la investigación y la reflexión crítica de profesionales del ámbito de la educación, la psicología, la tecnología y la atención a la diversidad. Desde estudios de caso hasta propuestas metodológicas, pasando por herramientas prácticas y marcos teóricos actualizados, los autores y autoras comparten conocimientos que nacen del trabajo colaborativo y del contacto directo con la realidad de las aulas.

Este volumen se enmarca en el contexto del proyecto de I+D titulado Proyecto Capacitación docente en competencias digitales inclusivas como apoyo al alumnado con Trastornos del Espectro Autista (CODITEA), financiado por el Ministerio de Ciencia, Innovación y Universidades / Agencia Estatal de Investigación (MICIU/AEI) y por el Fondo Europeo de Desarrollo Regional (FEDER, UE), con la referencia PID2022-138346OB-I00. El proyecto persi-gue, entre otros objetivos, sensibilizar y formar al profesorado, alumnado y familias sobre el uso consciente, ético y eficaz de las tecnologías como facilitadoras de procesos inclusivos con estudiantes TEA.

Con esta obra, esperamos ofrecer un recurso útil, actualizado y comprometido que sirva de guía e inspiración para docentes, investigadores, familias y responsables de políticas educativas. Porque apostar por la inclusión no es solo una cuestión de justicia social, sino también una muestra del potencial transformador de la educación cuando se apoya en la colaboración, la empatía y el conocimiento.

Los editores

# TECNOLOGÍA Y TEA EN LA EDUCACIÓN SECUNDARIA

*Pedro-José Arrifano-Tadeu*
Centro de Estudos em Educação e Inovação (Ci&DEI),
Escola Superior de Educação, Comunicação e Desporto,
Instituto Politécnico da Guarda (Portugal)

*José-María Fernández-Batanero*
Facultad de Ciencias de la Educación. Universidad de Sevilla (España)

## Introducción

Los Trastornos del Espectro del Autismo (TEA) son condiciones del neurodesarrollo que afectan significativamente a una proporción creciente de la población. Según los Centros para el Control y la Prevención de Enfermedades (CDC, 2023), la prevalencia del TEA ha aumentado considerablemente, estimándose en 1 de cada 36 niños de 8 años, lo que equivale aproximadamente al 2,8% de la población infantil. Este incremento puede atribuirse a una mayor concienciación, mejoras en los métodos de diagnóstico y una ampliación de los criterios diagnósticos, lo que ha permitido identificar casos que anteriormente podrían haber pasado desapercibidos.

Las personas con TEA enfrentan desafíos significativos en su integración laboral. Aunque las cifras exactas varían, se estima que una proporción considerable de adultos con TEA está desempleada o subempleada. Por ejemplo, un estudio indica que solo el 19,3% de la población con diversidad funcional en Estados Unidos participa en la fuerza laboral, lo que sugiere que las personas con TEA podrían estar enfrentando tasas de desempleo aún más altas (García & López, 2020). Esta situación resalta la necesidad de desarrollar políticas inclusivas y programas de apoyo que faciliten la integración efectiva de las personas con TEA en el mercado laboral y en la sociedad en general.

Además, se ha observado un aumento en los diagnósticos de TEA entre niñas y mujeres en los últimos años. Tradicionalmente, el autismo se ha diagnosticado con mayor frecuencia en niños, pero estudios recientes sugieren que las niñas y mujeres con TEA han sido subdiagnosticadas debido a diferencias en la presentación de los síntomas y a sesgos en los criterios diagnósticos (La Opinión, 2024). Este reconocimiento tardío implica que muchas mujeres con TEA no han recibido el apoyo necesario durante su desarrollo, lo que puede afectar su calidad de vida y oportunidades laborales.

Por otro lado, hay que decir que las personas con TEA constituyen el colectivo de la discapacidad con una mayor tasa de desempleo, de entre el 76% y el 90%, según datos de Autismo Europa (2021).

En el contexto educativo español los datos más recientes del Ministerio de Educación y Formación Profesional (2022) apuntan a que la presencia de este alumnado en la educación no universitaria se ha incrementado en un 160% desde el curso 2011/2012, hasta el punto de que uno de cada cuatro alumnos con necesidades específicas de apoyo educativo tiene autismo (Confederación Autismo España, 2021). Las cifras de los últimos cursos académicos (2020/2021) muestran una distancia significativa entre el número de estudiantes con TEA que cursan educación secundaria y el número que accede posteriormente a Bachillerato y Formación Profesional (Ministerio de Educación y Formación Profesional, 2021). Concretamente se produce un descenso de 26,99% del alumnado con TEA que cursa la ESO al 3,03% en bachillerato y el 4,23% en formación profesional. Esta bajada de alumnos con TEA que no continúan en la formación postobligatoria (Bachillerato y FP), constituye una barrera para avanzar en el sistema educativo y para su inserción en la sociedad.

En definitiva, estamos ante un alumnado que se encuentra en desventaja educativa, situándose en muchos casos en las fronteras de la marginación y la estigmatización, presentando niveles generalizado de estrés y ansiedad (White *et al.*, 2014), pudiendo tener un impacto profundo en la capacidad de funcionamiento del individuo y puede conducir a comportamientos problemáticos (Stephenson *et al.*, 2016).

Por otra parte, las Tecnologías de la Información y Comunicación (TIC) constituyen un apoyo esencial al aprendizaje, a la hora de llevar a cabo intervenciones educativas personalizadas. Son muchos los estudios que destacan que las TIC ayudan a estructurar y organizar el entorno de interacción del alumno con TEA, al configurarse como un medio muy predictible que ofrece contingencias comprensibles para este alumnado, ayudando a aumentar la autonomía de los participantes e igualando la participación de los sujetos a las condiciones del resto de estudiantes (Hu & Han, 2019).

Ahora bien, una variable educativa fundamental para proporcionar una educación de calidad y satisfacer las necesidades educativas de los estudiantes con TEA de forma adecuada se relacionan con la experiencia y formación del profesorado para la atención de las necesidades educativas (Hu & Han, 2019; Palomino Bastias & Marcelo García, 2021; Briones-Ponce *et al.*, 2021). Al mismo tiempo, se hace hincapié en la importancia que tiene que el profesional en educación cuente con una adecuada formación sobre estrategias de aprendizaje vinculadas al uso de tecnologías inclusivas con esta tipología de alumnado, como ha sido puesto de manifiesto en el 13th Autism-Europe International Congress, celebrado del 7 al 10 de octubre de 2022 en Cracovia (Polonia).

## Tecnología y TEA

La incorporación de las tecnologías en la educación especial, particularmente en el contexto del TEA, constituye un avance significativo en la mejora de las experiencias educativas de los alumnos (Suhaila & Nordin, 2022). Estas herramientas tecnológicas no solo facilitan el acceso a recursos educativos especializados, sino que también actúan como potentes facilitadores para abordar los desafíos específicos asociados a las necesidades educativas de los estudiantes. La flexibilidad de las TIC permite personalizar el proceso de enseñanza, adaptándolo a las necesidades individuales de cada alumno, ya sea en el ámbito de la comunicación, la interacción social o los estilos de aprendizaje.

En el ámbito de la investigación educativa, se ha dedicado un creciente interés al uso de las TIC para apoyar a estudiantes con autismo. El TEA es una condición del desarrollo neurológico que afecta áreas clave como la comunicación, la interacción social y el comportamiento, y que se manifiesta de manera altamente variable en cada individuo (Alcalá & Ochoa Madrigal, 2022). Debido a los retos particulares que estos estudiantes enfrentan en el entorno escolar, resulta crucial investigar y desarrollar estrategias tecnológicas que contribuyan a su desarrollo académico y social. La aplicación efectiva de las TIC en este contexto tiene el potencial de transformar no solo la manera en que los alumnos aprenden, sino también cómo interactúan con su entorno y participan en la sociedad.

En este contexto, las TIC ofrecen una serie de ventajas y oportunidades significativas para el alumnado con autismo (Hu & Han, 2019; Valencia *et al.*, 2019):

- Personalización y adaptación: Las TIC permiten adaptar el contenido y el ritmo de aprendizaje de manera individualizada, lo que es especialmente beneficioso para estudiantes con TEA, que pueden tener necesidades y estilos de aprendizaje diversos.
- Comunicación y socialización: Las aplicaciones y plataformas en línea pueden facilitar la comunicación y la interacción social de los estudiantes con TEA, proporcionando entornos controlados y estructurados para practicar habilidades sociales y de comunicación.
- Apoyo a la atención y la organización: Las herramientas tecnológicas pueden ayudar a los estudiantes con TEA a mantenerse enfocados, seguir instrucciones y organizar sus tareas y actividades diarias.
- Estimulación sensorial: Algunas aplicaciones y dispositivos TIC pueden ser utilizados para proporcionar estímulos sensoriales controlados, lo que puede ser beneficioso para estudiantes con TEA que tienen sensibilidades sensoriales específicas.
- Apoyo a la adquisición de habilidades funcionales: Las TIC pueden emplearse para enseñar habilidades prácticas de la vida cotidiana, como el autocuidado, la autonomía en el hogar y la toma de decisiones.
- Motivación: la motivación es uno de los beneficios notables del uso de las TIC con este alumnado. Las aplicaciones y herramientas digitales pueden diseñarse de manera atractiva y lúdica, lo que puede aumentar el interés y la participación de los estudiantes con TEA en las actividades educativas.

Las herramientas digitales desempeñan un papel esencial en la promoción del aprendizaje de estudiantes con necesidades educativas especiales, al introducir nuevas metodologías y estrategias didácticas que facilitan la comunicación y la interacción, independientemente de las diferencias individuales. Diversos autores han subrayado que el acceso a las TIC proporciona un nivel significativo de igualdad de oportunidades para todas las personas, independientemente de sus condiciones. Aunque el uso de tecnologías para apoyar a estudiantes con discapacidades ha sido objeto de investigación durante varias décadas, solo en la última ha cobrado un papel fundamental en el apoyo a este grupo (Fernández Batanero, Montenegro Rueda y Fernández Cerero, 2021).

No obstante, la efectividad de las intervenciones basadas en TIC para estudiantes con TEA puede variar considerablemente según las características y necesidades individuales de cada estudiante. Uno de los principales desafíos en la implementación exitosa de estas tecnologías en el ámbito educativo radica en la formación digital del profesorado. Para que las TIC se utilicen de manera efectiva y significativa, los educadores deben estar capacitados y actualizados en el uso de estas herramientas, además de comprender cómo adaptarlas a las necesidades específicas de los estudiantes con TEA. Esta formación no solo requiere habilidades técnicas, sino también una comprensión profunda de cómo estas tecnologías pueden apoyar el aprendizaje y el desarrollo de habilidades (Romero & Harari, 2017). Estudios recientes destacan la falta de formación digital del profesorado y el desconocimiento general sobre las posibilidades que ofrecen estas herramientas (Gallardo-Montes & Capperucci, 2021).

En esta línea, el estudio realizado por Fernández Batanero, Cabero, Rodríguez Palacios y Román (2022) reveló que la capacitación tecnológica del profesorado en competencias digitales para apoyar a estudiantes con diversidad funcional es insuficiente. Este estudio mostró que dicha capacitación no depende de variables como el género o la edad, sino que está más relacionada con la experiencia en la enseñanza y la afiliación institucional, factores que a menudo están interconectados. Aunque se reconoce la necesidad de mejorar la formación docente en educación superior, también se destaca el valor de las TIC como herramientas de accesibilidad que benefician no solo al alumnado con TEA, sino a toda la comunidad educativa.

Por otro lado, la integración efectiva de las intervenciones tecnológicas en el entorno educativo no debe considerarse un reemplazo de las estrategias pedagógicas tradicionales, sino un complemento valioso. Las TIC pueden enriquecer el proceso de enseñanza-aprendizaje al ofrecer un conjunto versátil de herramientas personalizables que se adaptan a las necesidades específicas de cada estudiante. Para lograr este objetivo, es fundamental que los docentes sean capaces de diseñar y gestionar entornos educativos que integren estas tecnologías de manera armoniosa con otras estrategias y métodos pedagógicos.

## Estado actual de la investigación sobre el uso de las TIC en el alumnado con Trastorno del Espectro Autista

La producción científica sobre el uso de las TIC en el alumnado con autismo ha experimentado un aumento significativo en los últimos años, lo que abre nuevas posibilidades en la educación y el desarrollo de habilidades para este grupo de estudiantes. Sin embargo, también se han identificado algunas tendencias notables en los datos recopilados. El aumento significativo en el año 2022 sugiere un creciente interés en la comunidad científica por este tema, lo cual indica que, a pesar de los desafíos, la investigación sobre el uso de las TIC en el alumnado con TEA sigue siendo relevante y está ganando importancia en la comunidad académica.

La figura 1 muestra un mapa que refleja la producción científica por país de origen sobre el tema desde 2014 hasta 2023. En azul se muestran los países que han realizado investigación, siendo los tonos más oscuros los que han producido más producción científica durante el intervalo de tiempo establecido. En este sentido, Estados Unidos se presenta como el país con mayor producción. Siguiendo esta línea, el Reino Unido se posiciona como el país con el segundo mayor número de publicaciones, seguido de Canadá, Italia, India y Malasia. Sin embargo, otros países también han realizado investigaciones, pero en menor medida, como Brasil, España, Francia, China y Nueva Zelanda (Fernández Batanero *et al.*, 2024).

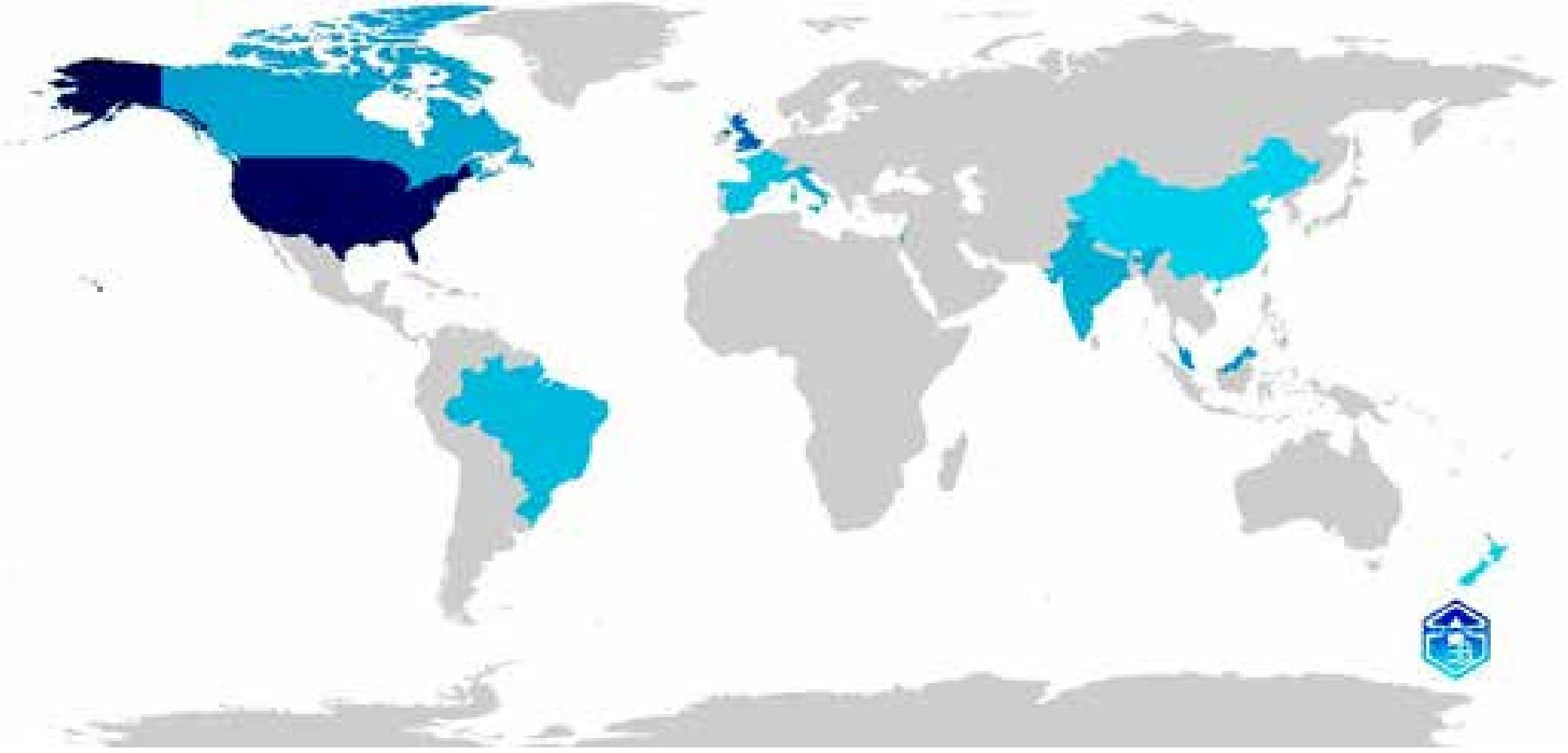

Figura 1. Producción científica por países (Fernández Batanero *et al.*, 2024).

Si ponemos el foco de atención en el impacto de las TIC en el alumnado con Trastorno del Espectro Autista observamos un impacto positivo del uso de herramientas en el alumnado con autismo. Los estudios realizados al respecto han confirmado que conlleva beneficios significativos en varios aspectos. Las TIC permiten la individualización del aprendizaje, adaptando contenido y actividades según las necesidades específicas de cada estudiante, lo que facilita un enfoque de enseñanza personalizado (Bellini *et al.*, 2014).

Las aplicaciones y plataformas en línea desempeñan un papel fundamental en la mejora de la comunicación y la interacción social de los estudiantes con autismo. Estos recursos tecnológicos ofrecen entornos virtuales con-trolados y estructurados que permiten a los estudiantes con TEA practicar y desarrollar sus habilidades sociales y de comunicación de manera efectiva, proporcionando un ambiente seguro donde los estudiantes pueden trabajar en la comprensión de las señales sociales, el lenguaje corporal, la expresión facial y otras habilidades relacionadas con la comunicación y la interacción. En esta línea, como resultado, se obtiene que el alumnado progrese a su propio ritmo y de manera personalizada.

Siguiendo esta línea, la motivación juega un papel fundamental a la hora de aplicar las herramientas digitales. Las TIC ofrecen un entorno propicio para diseñar aplicaciones y herramientas digitales que son atractivas y lúdicas, lo que puede tener un impacto significativo en la participación y el interés de los estudiantes con TEA en las actividades educativas. La clave para comprender este beneficio radica en la capacidad de las TIC para adaptarse a las necesidades individuales de los estudiantes con autismo. Estas herramientas pueden ser personalizadas para abordar los intereses específicos de cada estudiante, lo que les permite conectarse con el contenido de una manera más significativa. Esto puede ser especialmente importante para los estudiantes con autismo, ya que puede ayudar a reducir la ansiedad y fomentar una actitud más positiva hacia el aprendizaje. Asimismo, la evidencia científica refleja demuestra que el uso de la tecnología, puede ser una técnica de instrucción efectiva para mejorar las habilidades de los estudiantes (Pérez-Fuster *et al.*, 2019).

## Futuro de la Tecnología aplicada al alumnado con TEA

El avance tecnológico está transformando la forma en que abordamos la educación inclusiva, especialmente en relación con el alumnado con autismo. La combinación de inteligencia artificial (IA), aprendizaje automático y tecnologías emergentes está permitiendo el desarrollo de herramientas más efectivas para apoyar a esta población. Estas herramientas no solo mejoran el acceso a la educación, sino que también promueven una mayor autonomía, inclusión y personalización del aprendizaje.

La combinación de IA y aprendizaje automático permite que los sistemas educativos analicen grandes volúmenes de datos para comprender mejor las necesidades individuales de los estudiantes con TEA. Estos sistemas pueden identificar patrones de comportamiento, preferencias de aprendizaje y respuestas emocionales que informan la creación de estrategias pedagógicas personalizadas. Este nivel de adaptabilidad es fundamental para los estudian-tes con TEA, ya que sus necesidades pueden variar ampliamente según factores como el nivel de sensibilidad sensorial, las habilidades comunicativas y las metas educativas.

En el ámbito de la comunicación, las herramientas basadas en IA están revolucionando cómo los estudiantes con dificultades para expresarse verbalmente pueden interactuar con el mundo que los rodea. Aplicaciones avanzadas de comunicación aumentativa y alternativa (CAA) no solo proporcionan palabras o frases predefinidas, sino que también aprenden del uso del estudiante, ajustándose a su estilo comunicativo. En el futuro, se espera que estos sistemas sean capaces de interpretar emociones o intenciones no verbales a través de sensores de movimiento o análisis faciales, ampliando así las capacidades comunicativas de las personas con TEA.

Por otro lado, las tecnologías emergentes como la realidad virtual (VR) y la realidad aumentada (AR) están creando nuevas oportunidades para el desarrollo de habilidades sociales y emocionales. Estas herramientas ofrecen entornos simulados que permiten a los estudiantes practicar interacciones sociales, como mantener una conversación o interpretar el lenguaje corporal, sin el riesgo de sentirse abrumados. Al combinarse con IA, estas experiencias se vuelven aún más efectivas, ya que pueden adaptarse a las respuestas individuales del usuario, ajustando la dificultad o la naturaleza de las interacciones en tiempo real.

Además, la IA también está desempeñando un papel crucial en la identificación temprana del TEA y en el seguimiento del progreso de los estudiantes. Herramientas de análisis basadas en datos permiten detectar indicadores tempranos del trastorno, lo que facilita intervenciones más tempranas y efectivas. Una vez diagnosticado, la IA puede monitorear continuamente el aprendizaje y el desarrollo del estudiante, generando informes detallados que ayudan a los educadores y terapeutas a ajustar las estrategias de apoyo.

Podemos decir que las aulas inteligentes impulsadas por IA representan el futuro de la educación inclusiva. Estas aulas integran tecnologías que pueden ajustar elementos como la iluminación, el sonido y el ritmo de la clase para adaptarse a las necesidades sensoriales de cada estudiante. Por ejemplo, un sistema podría detectar cuando un estudiante comienza a sentirse sobrecargado sensorialmente y ajustar el entorno para reducir el estrés. Esto crea un espacio de aprendizaje más cómodo y efectivo para todos los alumnos.

Así pues, la combinación de inteligencia artificial, aprendizaje automático y tecnologías emergentes está transformando la educación para el alumnado con TEA. Estas herramientas no solo están eliminando barreras, sino que también están redefiniendo lo que es posible en términos de inclusión y personalización educativa. Con una implementación ética y accesible, el futuro promete una educación más equitativa y adaptada a las necesidades de todos los estudiantes.

## Conclusiones

La reflexión central recae en cómo las TIC no solo facilitan el acceso a recursos educativos personalizados, sino que actúan como habilitadores para superar barreras comunicativas, sociales y organizativas propias de los estudiantes con TEA. Al adoptar un enfoque adaptativo y centrado en las necesidades individuales, estas herramientas se convierten en potentes aliados para estructurar un entorno de aprendizaje accesible y efectivo.

Al mismo tiempo se vislumbra un problema recurrente en la implementación de TIC: la falta de formación digital del profesorado. Esto es crucial porque, aunque las herramientas tecnológicas ofrecen un potencial transformador, su eficacia depende directamente de la competencia de los docentes para utilizarlas e integrarlas de manera estratégica en sus prácticas pedagógicas. Es evidente la necesidad de programas de formación que no solo capaciten técnicamente, sino que también sensibilicen sobre la importancia de las TIC en la mejora de la calidad de vida educativa para estudiantes con TEA.

Otro punto relevante es la capacidad de las TIC para proporcionar estímulos sensoriales y entornos controlados que favorecen el aprendizaje y la interacción social. Herramientas como la realidad virtual y la realidad aumentada abren nuevas posibilidades para practicar habilidades sociales en un contexto seguro y adaptable, representando un avance significativo hacia una educación inclusiva.

Podemos afirmar que el binomio TIC y TEA nos ofrece una perspectiva optimista respecto al futuro. Al combinarse con tecnologías emergentes como la inteligencia artificial, las TIC prometen transformar no solo la forma de aprender, sino también de enseñar, detectando patrones únicos en el comportamiento y las necesidades de cada estudiante. Sin embargo, se enfatiza que estas tecnologías deben ser vistas como complementos a las estrategias tradicionales, no como reemplazos, para garantizar una educación verdaderamente inclusiva.

## Financiación

## Referencias bibliográficas

# LA REALIDAD VIRTUAL Y AUMENTADA EN CONTEXTOS DE ESTUDIANTES CON TEA

*Víctor Abella-García*
*Vanesa Ausín-Villaverde*
*Sonia Rodríguez-Cano*
*Vanesa Delgado-Benito*
Facultad de Educación. Universidad de Burgos (España)

## Introducción

La tecnología es una herramienta muy valiosa que permite ayudar y apoyar a los estudiantes con Trastorno del Espectro Autista (TEA) en su desarrollo educativo y social. Las intervenciones apoyadas con tecnologías están mostrando un gran potencial para la mejora de la calidad de vida de estas personas, contribuyendo de forma general a mejorar sus habilidades académicas, sociales, comunicativas, etc.

Se ha demostrado que las personas con TEA tienen cierta predilección por los medios electrónicos y fuertes habilidades en cuanto al procesamiento visual, con lo que las intervenciones apoyadas en tecnologías pueden ser muy motivadoras para ellos, aumentando así el mantenimiento de la intervención y las posibilidades de generalización (Cuesta & Abella, 2012). Los principales beneficios que aporta la tecnología a los estudiantes con TEA se centran sobre todo en el desarrollo de sus habilidades comunicativas y sociales, ya que en muchos casos las intervenciones con apoyo tecnológico proporcionan entornos seguros y confortables. La tecnología también se aplica en la adaptación de la conducta y su independencia con la intención de mejorar o facilitar la inclusión de estas personas en la sociedad. Sobre esta línea, no se debe olvidar el impacto sobre las habilidades académicas, como por ejemplo la lectura.

Existen diversos tipos de intervenciones apoyadas por la tecnología, como las intervenciones apoyadas por ordenador, el uso de dispositivos móviles y aplicaciones y tecnologías de generalización más reciente como la Realidad Virtual (RV) y la Realidad Aumentada (RA), las cuales serán objeto de análisis en este capítulo.

A pesar de los avances tecnológicos y el aumento de evidencias científicas que muestran la eficacia en los tratamientos es necesario seguir investigando en la calidad y eficacia de las intervenciones apoyadas en tecnologías con la intención de mantener y generalizar las habilidades aprendidas por parte de las personas con TEA. Es importante tener en cuenta que la tecnología es un vehículo para la intervención en personas con TEA, pero deben ser un apoyo que se base en aquellas condiciones que definen un programa de intervención eficaz y centrado en las necesidades específicas de cada individuo (Cuesta y Abella, 2012).

## Tecnologías emergentes: realidad aumentada y realidad virtual

La Realidad Virtual (RV) así como la Realidad Aumentada (RA) son Tecnologías Emergentes que proporcionan un enfoque innovador para la atención de alumnado con diversas necesidades. Entre las ventajas que pueden proporcionar estas tecnologías se encuentran que ofrecen entornos seguros y controlados, generan mayor motivación, permiten la interactividad, proporcionan feedback inmediato y contribuyen a la mejora de habilidades relacionadas con el procesamiento visual y la memoria de trabajo.

La RA, se remonta al inicio de los años noventa y puede conceptualizarse como una tecnología que complementa la percepción e interacción con el mundo real y permite al usuario estar en un ambiente real interactuando con información sobreimpresa generada digitalmente. De acuerdo con Cabero *et al.* (2022), implica utilizar una colección de herramientas técnicas que complementan los datos físicos con datos virtuales, por lo que sugiere incorporar un componente virtual sintético a lo real.

Por otro lado, la RV es una tecnología emergente, convirtiéndose en los últimos quince años en un poderoso elemento educativo y en una herramienta de evaluación e intervención en el ámbito escolar (Aznar Díaz *et al.*, 2018). Las simulaciones generadas por computadora permiten a los usuarios interactuar con un mundo visual tridimensional sintético u otro entorno sensorial (Concepción & Muñoz, 2024).

Ambas tecnologías se caracterizan por ser inmersivas, generar motivación, permitir la interacción, la transducción (acceso del usuario a estímulos virtuales) así como la inclusión de modelos virtuales 2D y 3D. Todos estos elementos posibilitan su aplicación en el ámbito educativo puesto que ofrecen entornos más lúdicos, seguros y controlados.

La versatilidad que ofrece estas tecnologías emergentes hace que sean adaptables a diferentes contextos, siendo uno de ellos el trabajo con personas con Trastorno del Espectro Autista. En este sentido ayudan a mejorar los procesos conductuales, cognitivos y situacionales y pueden considerarse efectivas para el desarrollo y aprendizaje de áreas específicas como la comunicación, el lenguaje, las emociones o las habilidades sociales (Gallardo *et al.*, 2019).

Investigaciones recientes indican que las tecnologías emergentes impactan positivamente en la mejora de las habilidades sociales de los niños con TEA, ayudan a la integración y estimulación sensorial de los aprendizajes; reducen los niveles de ansiedad generando una sensación de seguridad y aportan experiencias sobre situaciones de diferente naturaleza en base a escenarios reales, donde se propicia el reconocimiento de emociones relacionadas a contextos concretos, favoreciendo la comprensión y uso del lenguaje no verbal (Zhao *et al.*, 2018).

El uso de tecnologías emergentes para la mejora de habilidades sociales y comunicativas en alumnado con TEA representa un área prometedora y de interés, sin embargo, estos recursos sólo son usados por los docentes que poseen el dominio de las competencias digitales necesarias para su implementación en el aula. Por ello, es preciso capacitación y disposición de los docentes para integrarlas en su práctica diaria y facilitar un entorno de aprendizaje más accesible y efectivo para los estudiantes con autismo (Cored *et al.*, 2021).

## Realidad aumentada en estudiantes con TEA

La Realidad Aumentada (RA) ha emergido como una herramienta prometedora en el mundo de la educación por la versatilidad y simplicidad en el manejo, ya que se puede utilizar con dispositivos que están a nuestro alcance (Cabero, 2018; Cabero *et al.*, 2016) y dando la posibilidad que el estudiante se convierta en emisor de mensajes (Ausín Villaverde *et al.*, 2023).

La revisión de la literatura (Guerrero *et al.*, 2024; Gilabert *et al.*, 2019), manifiestan que la RA favorece una enseñanza activa por parte del alumnado, puesto que este puede participar en el proceso de aprendizaje al tomar la decisión de cuándo necesita aumentar la información y combinar lo real con lo virtual (Wedyan *et al.*, 2021).

Diversos estudios (Gallardo *et al.*, 2021, Lee, 2020) han explorado cómo la RA puede mejorar la experiencia educativa y social de los estudiantes con TEA, ofreciendo nuevas oportunidades para el aprendizaje.

En este sentido, esta tecnología puede mejorar la comunicación y la interacción social. Los trabajos de Keshav *et al.*, (2018) y Kinsella *et al.*, (2017) presentan resultados eficaces respecto a la incorporación de la RA para trabajar estas habilidades. Un ejemplo de ello es la aplicación MOSOCO que permite un mayor contacto visual, aumento en las interacciones sociales e integración del alumnado TEA con alumnado neurotípico (Escobedo *et al.*, 2012).

Otro ejemplo es Augmented Reality Concept Map (ARCM) que es un sistema de formación que combina la RA con un mapa conceptual de relaciones sociales entre personas, mediante el cual se pueden ver y participar en situaciones de la vida diaria, con el propósito de mostrar qué están sintiendo y cómo han de comportarse en una situación dada, es decir, qué gestos han de emplear, qué expresiones, qué tipo de saludo, etc. (Lee *et al.*, 2018).

Otra área trabajada es como apoyo en el aprendizaje visual y abstracto. El uso de esta tecnología puede transformar conceptos abstractos en experiencias visuales y tangibles, lo que facilita la comprensión y el aprendizaje (Liu *et al.*, 2017; Wan *et al.*, 2022).

Happy Minion Game (Bhatt *et al.*, 2017) es una aplicación que permite mejorar la coordinación mano-ojo, lo que posibilita el desarrollo de la confianza y la autoestima fruto del juego de esta coordinación, mejorando también el contacto visual y la discriminación de emociones.

La naturaleza interactiva y lúdica de la RA puede aumentar la motivación y el compromiso de las personas con TEA hacia la actividad realizada. Un excelente ejemplo es la aplicación móvil MOBIS que posibilita superponer contenido digital, como imágenes, texto o videos a diferentes objetos físicos para discriminar entre unos y otros. Favorecen el aumento de la motivación, mejora en la atención sostenida y permite el desarrollo de habilidades de comportamiento como la tolerancia con los demás (Escobedo *et al.*, 2014).

Como puede comprobarse, los recursos tecnológicos y más concretamente el uso de la RA está cada vez más presente en el ámbito educativo y esto contribuye a mejorar algunos déficits que presentan las personas con TEA, tales como el lenguaje y la comunicación, la interacción social, la identificación o el reconocimiento de emociones, todos ellos necesarios para un buen desarrollo de habilidades sociales en las personas con TEA.

## Realidad virtual en estudiantes con TEA

La Realidad Virtual (RV) se ha mostrado como una herramienta prometedora para realizar intervenciones con alumnos con Trastornos del Espectro Autista (Krishna *et al.*, 2023; Liu, 2023).

Aunque la utilización de la RV pueda parecer muy reciente, ya en los años 90 del siglo pasado algunos terapeutas mostraron interés por la utilización de las tecnologías inmersivas para el tratamiento del autismo (Shane & Albert, 2008).

Las terapias convencionales han mostrado que, en muchas ocasiones, debido a la dificultad en habilidades de generalización que presentan las personas con TEA, les resulta muy complicado transferir las habilidades trabajadas en el tratamiento a situaciones del mundo real. En este sentido, la literatura científica ha mostrado que las personas con TEA se desempeñan bien en entornos virtuales, con lo que aumenta la probabilidad de que transfieran esta respuesta a situaciones del mundo real (Bekele *et al.*, 2014).

Como se ha indicado, la RV permite simular situaciones sociales del mundo real y repetir todas las veces que se desee dicha situación en un entorno no amenazador. De esta manera se ha comprobado que la RV es eficaz en la mejora de algunas habilidades sociales, como el reconocimiento de emociones, la reciprocidad o la atribución social, contribuyendo así a reducir el riesgo de aislamiento social puesto que ayuda a los estudiantes con TEA responder de forma apropiada en situaciones sociales (Didehbani, 2016; Krishna *et al.*, 2023; Liu, 2023). También se ha utilizado para la mejora de las habilidades emocionales y cognitivas (Herrero y Lorenzo, 2019; Didehbani, 2016), incluyendo entre estas últimas las funciones ejecutivas y habilidades de resolución de problemas.

Desde el punto de vista de las emociones la RV se ha utilizado para trabajar una exposición controlada a estímulos sociales y emocionales. A modo de ejemplo, VirTEA es una aplicación de Realidad Virtual móvil que recrea distintos escenarios de la vida de una persona con TEA con el fin de facilitar la manera de afrontar situaciones fuera de su rutina habitual como, por ejemplo, ir al médico o esperar el autobús. Los beneficios de esta herramienta son tres principalmente: la anticipación, la disminución del estrés y la mejorar de la calidad de vida.

Finalmente, la RV también se ha mostrado eficaz como herramienta educativa (Zhang *et al.*, 2022; Silva *et al.*, 2022) debido a que utiliza una de las fortalezas de los alumnos con TEA, como son las habilidades visoespaciales.

## Desafíos y barreras en el uso de RV y RA en personas con TEA

Aunque la RV y la RA ofrecen grandes oportunidades, su implementación enfrenta varios obstáculos. Uno de los principales desafíos es el acceso y la sostenibilidad económica. Como señalan Khowaja *et al.* (2022), el costo de los dispositivos tecnológicos y la falta de financiación específica para programas centrados en TEA limitan su adopción. Además, los desarrolladores suelen crear soluciones genéricas que no siempre se adaptan a las necesidades sensoriales y cognitivas de las personas con autismo.

Aunque en este campo hay ya iniciativas europeas que buscan desarrollar alternativas sostenibles, otro desafío crítico es la personalización. Las personas con TEA presentan un espectro amplio de características, y las soluciones tecnológicas deben ser lo suficientemente flexibles como para abordar esta diversidad. Sin embargo, como destacan Chen *et al.* (2020), muchos programas carecen de opciones para ajustarse a estas necesidades individuales.

También es fundamental la formación de los profesionales que trabajan con las personas con TEA, pues la falta de formación en el uso de estas tecnologías puede limitar su impacto y generar frustración entre los usuarios y sus familias, además de no conseguir los resultados esperados.

### Interacción de la RV y la RA con el diseño centrado en el usuario

El Diseño Centrado en el Usuario (DCU) desempeña un papel esencial en el éxito de las tecnologías dirigidas a personas con TEA. Este enfoque coloca al individuo en el centro del proceso de desarrollo, asegurando que la tecnología responda a sus necesidades específicas y fomente experiencias positivas, además de proveer un diseño inclusivo, accesible y adaptativo (Alba Pastor, 2022).

En el caso de las personas con TEA esta metodología debe considerar características como la sensibilidad sensorial y las necesidades de aprendizaje individualizado. Por ejemplo, en entornos virtuales, pudiéndose ajustar factores como el nivel de estímulos visuales y auditivos para evitar la sobrecarga sensorial (Maskey *et al.*, 2019). Además, las tecnologías inmersivas permiten practicar habilidades sociales y laborales en escenarios simulados, lo que mejora la transferencia de estas habilidades al mundo real. Otros autores subrayan que la retroalimentación constante de los usuarios, sus familias y los profesionales es crucial para mejorar el diseño y garantizar que la tecnología sea efectiva y aceptada (Makransky y Petersen, 2021).

### Propuestas de políticas educativas para promover el uso de RV y RA en TEA

España en 2015 aprobó la Estrategia Española en Trastornos del Espectro del Autismo como un marco de referencia para diseñar y coordinar políticas y acciones relacionadas con el autismo. Para operativizar esta estrategia, se presentó el Plan de Acción para el periodo 2023-2027.

Respecto al ámbito de la tecnología este Plan de acción plantea medidas específicas que deben ser desarrolladas como el aumento de la financiación de proyectos específicos para TEA, la formación específica para profesionales, la colaboración interdisciplinar, el fomento de la equidad tecnológica y la evaluación de los programas desarrollados.

## Conclusiones

La realidad virtual y aumentada representan una oportunidad transformadora para el apoyo a personas con TEA, al ofrecer entornos personalizables y seguros para el desarrollo de sus habilidades comunicativas, sociales y educativas. Su implementación requiere superar barreras técnicas, económicas y pedagógicas mediante políticas educativas inclusivas y colaborativas. Diseñar soluciones centradas en el usuario, con la participación activa de la comunidad del autismo, es clave para garantizar que estas tecnologías alcancen su máximo potencial y mejoren la calidad de vida de las personas con TEA y sus familias.

## Financiación

## Referencias bibliográficas

# HERRAMIENTAS Y RECURSOS DIGITALES PARA EL DESARROLLO DE HABILIDADES SOCIOEMOCIONALES DEL ALUMNADO CON TEA

*María-Dolores Díaz-Noguera*

*Carlos Hervás-Gómez*

Facultad de Ciencias de la Educación. Universidad de Sevilla (España)

## Introducción

El desarrollo de las habilidades socioemocionales es fundamental para el bienestar y la integración social de los estudiantes con trastorno del espectro autista (TEA). En este contexto, las herramientas y recursos digitales han demostrado ser una solución innovadora y eficaz para abordar las necesidades específicas de este grupo. La capacidad de estas tecnologías para ofrecer intervenciones estructuradas, personalizables y atractivas representa una ventaja significativa frente a los métodos tradicionales. Por ejemplo, la realidad virtual (RV) permite a los estudiantes practicar interacciones sociales en entornos controlados y seguros, fomentando la confianza y reduciendo la ansiedad ante situaciones del mundo real. Asimismo, plataformas gamificadas, como el prototipo "Emotion Adventure", integran habilidades esenciales como el reconocimiento de emociones y la teoría de la mente en actividades interactivas y lúdicas, maximizando el compromiso y la motivación de los usuarios.

Además, las herramientas digitales respaldan enfoques de aprendizaje individualizado, permitiendo adaptar los contenidos y desafíos al ritmo de desarrollo de cada estudiante. Estas características no solo potencian la eficacia de las intervenciones, sino que también facilitan la generalización de las habilidades aprendidas a contextos sociales dinámicos y diversos.

Este capítulo explora cómo la realidad virtual, los juegos interactivos y las evaluaciones basadas en la web están transformando el desarrollo de las competencias socioemocionales en estudiantes con TEA. A través de un análisis detallado de estas herramientas, se destacan sus ventajas, como la adaptabilidad, la capacidad para fomentar la colaboración entre pares y su papel en la personalización de las intervenciones. De este modo, se evidencia cómo estas tecnologías están cerrando brechas en los enfoques terapéuticos tradicionales, ofreciendo oportunidades transformadoras para mejorar el aprendizaje y la integración social de los estudiantes con TEA.

El desarrollo de habilidades socioemocionales es una parte esencial del aprendizaje y el crecimiento personal, particularmente para estudiantes con Trastorno del Espectro Autista (TEA). Estas habilidades, que incluyen la capacidad de reconocer, comprender y gestionar las emociones, así como de interactuar eficazmente con los demás, son fundamentales para fomentar la autonomía y la inclusión en diversos contextos sociales y educativos. En este sentido, las herramientas y recursos digitales han emergido como un apoyo significativo, ofreciendo nuevas oportunidades para abordar los desafíos particulares que enfrentan los estudiantes con TEA.

Las tecnologías digitales han demostrado ser especialmente útiles en el desarrollo de competencias socioemocionales debido a su capacidad para personalizar las experiencias de aprendizaje y crear entornos seguros para la exploración emocional. Las aplicaciones móviles, las plataformas de realidad virtual y aumentada, los juegos educativos y las comunidades online representan recursos valiosos que permiten a los estudiantes con TEA practicar habilidades de socialización y comunicación de manera interactiva y adaptada a sus necesidades individuales. Estas herramientas no solo facilitan el acceso a experiencias enriquecedoras, sino que también promueven un aprendizaje más profundo al ofrecer retroalimentación inmediata y reforzar comportamientos positivos.

## Recursos y Herramientas

El desarrollo de habilidades socioemocionales en alumnado con Trastorno del Espectro Autista (TEA) puede beneficiarse enormemente del uso de herramientas y recursos digitales diseñados para apoyar la comunicación, la autorregulación emocional y la interacción social.

Selección de herramientas y recursos que pueden ser útiles:

- **Simulo VR experience**
  Simulo es un próximo juego sandbox de física 2D desarrollado por Carroted. Está siendo programado en Rust y utiliza el motor de física

Box2D v3, junto con el motor de juegos Bevy. El objetivo de Simulo es proporcionar una plataforma donde los usuarios puedan construir y compartir creaciones, desde simples vehículos hasta maquinaria compleja, utilizando una variedad de herramientas integradas en el juego. Además, ofrece scripting en Lua para una personalización más profunda. Esto significa en términos simples, que te pones unas gafas especiales que te transportan a otro lugar completamente diferente. Podrías estar explorando las profundidades del océano, caminando por una ciudad futurista, o incluso participando en una batalla espacial. Simulo VR Experience te ofrece estas experiencias inmersivas y realistas. Las aplicaciones de Simulo VR Experience son muy variadas, en educación las utilizamos para enseñar conceptos complejos de manera visual e interactiva que es lo que se pretende con estudiantes de estas características.

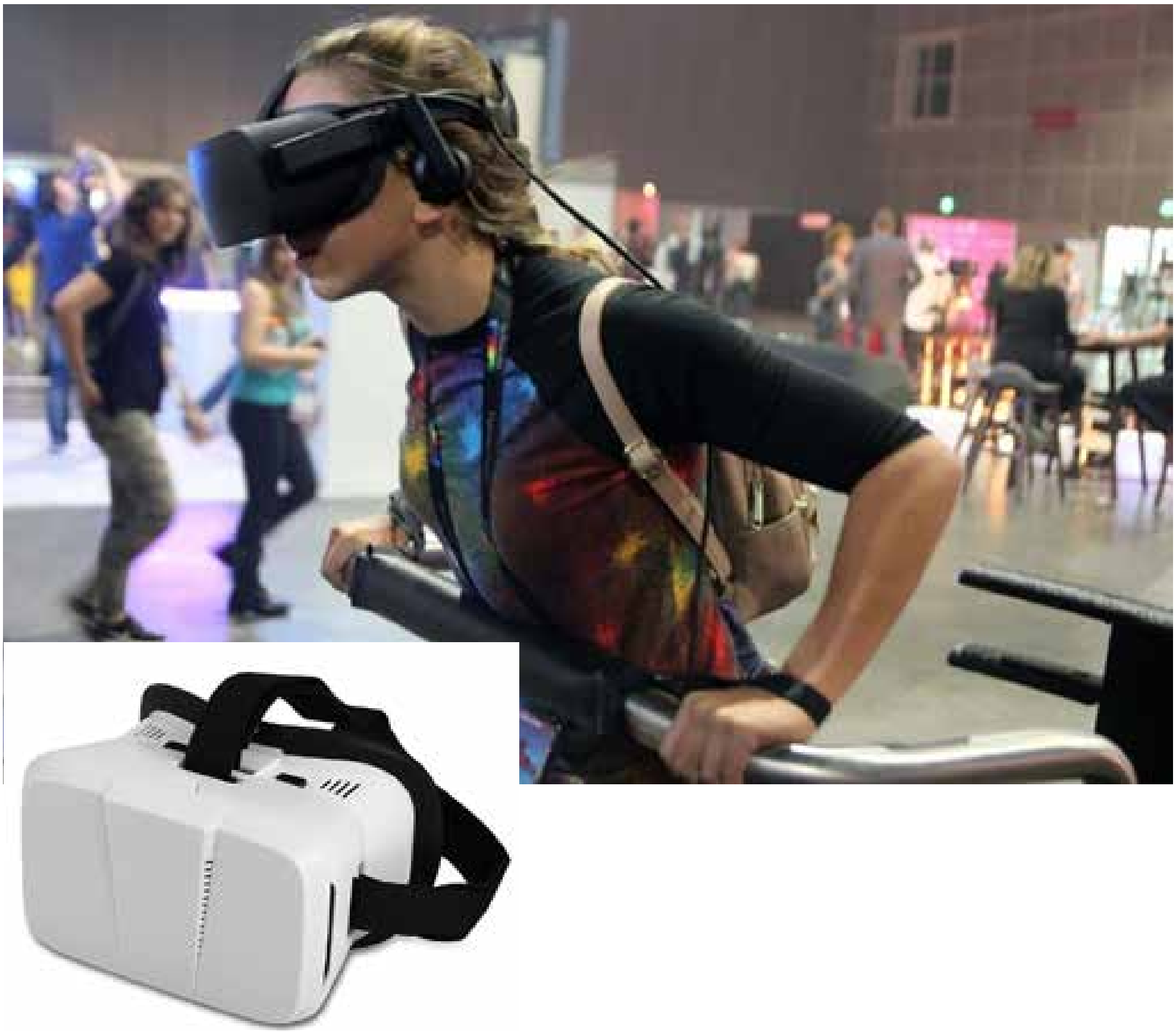

Herramientas de Comunicación Alternativa y Aumentativa (CAA):

– **Boardmaker**
  Crea tableros de comunicación con imágenes, texto y voz para facilitar la expresión y comprensión.

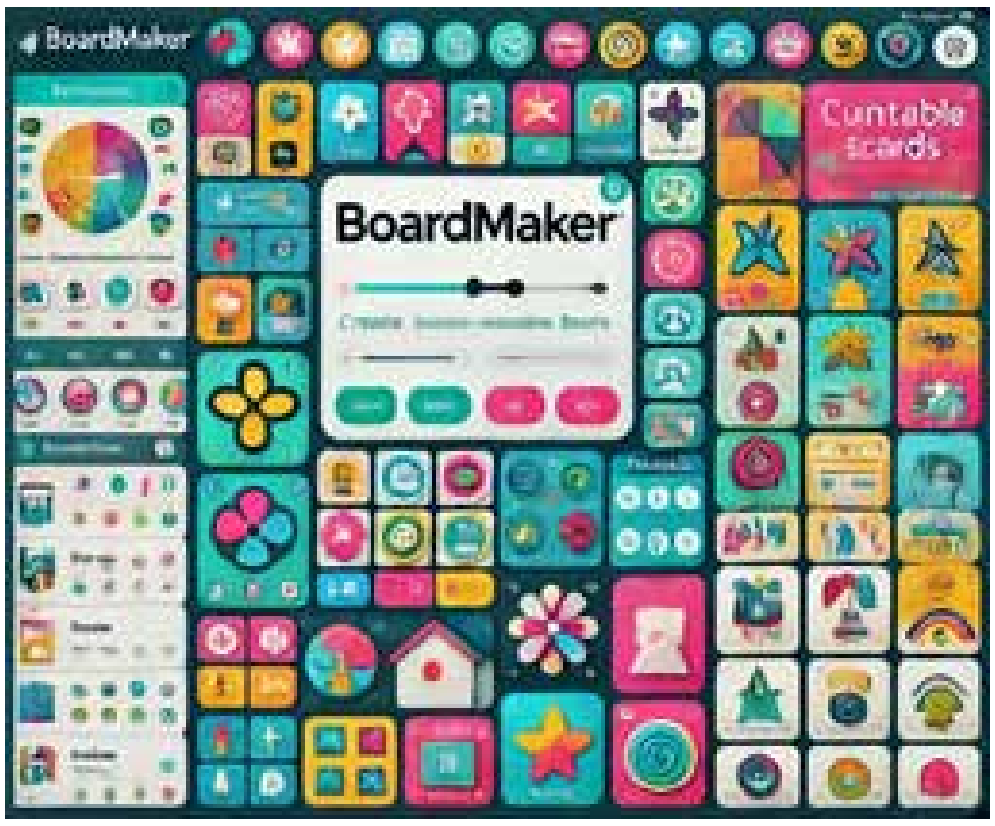

Boardmaker es una plataforma ampliamente utilizada en el ámbito educativo, terapéutico y de asistencia para la comunicación aumentativa y alternativa (CAA). Fue desarrollada por Tobii Dynavox, una empresa especializada en tecnología de apoyo. La aplicación y sus materiales asociados están diseñados para facilitar la comunicación, la enseñanza y la accesibilidad, especialmente para personas con discapacidades de aprendizaje, comunicación o motoras. En nuestro caso, hacemos hincapié en la comunicación. Sus principales características son: permite crear tableros personalizados con símbolos visuales que facilitan la comunicación para personas con TEA y ofrece herramientas para diseñar materiales como horarios visuales, gráficos de conducta, libros de historias sociales, entre otros. Incluye la biblioteca de símbolos PCS (Picture Communication Symbols), con miles de imágenes que representan palabras, conceptos y acciones. Los símbolos son adaptables en cuanto a idioma, tamaño y estilo, lo que permite personalización para diferentes necesidades y contextos culturales. Apoya la creación de lecciones y actividades interactivas para estudiantes con diversas habilidades, por tanto, es compatible con planes de estudio inclusivos, integrando el aprendizaje con adaptaciones específicas. Además de materiales impresos, Boardmaker 7 (la versión más reciente) permite desarrollar actividades interactivas que los usuarios pueden completar en dispositivos como tabletas y computadoras. Y, no olvidamos lo más importante la Ac-cesibilidad, integrados con dispositivos de acceso alternativo, como interruptores, pantallas táctiles y tecnología de seguimiento ocular, para usuarios con movilidad limitada. Dado que tenemos la plataforma online, Boardmaker Online permite a los educadores y terapeutas diseñar, compartir y asignar actividades interactivas de forma remota, lo cual es útil en contextos de enseñanza a distancia o híbrida.

## Aplicaciones para la comunicación y la interacción social

### *Pictogramas y Comunicación Alternativa/Aumentativa (CAA)*

- **Proloquo2Go**
  Es una aplicación de habla generativa que permite a los usuarios construir frases completas y personalizar voces. Por tanto, se crean tableros de comunicación personalizados para ayudar a los alumnos no verbales o con dificultades de comunicación.

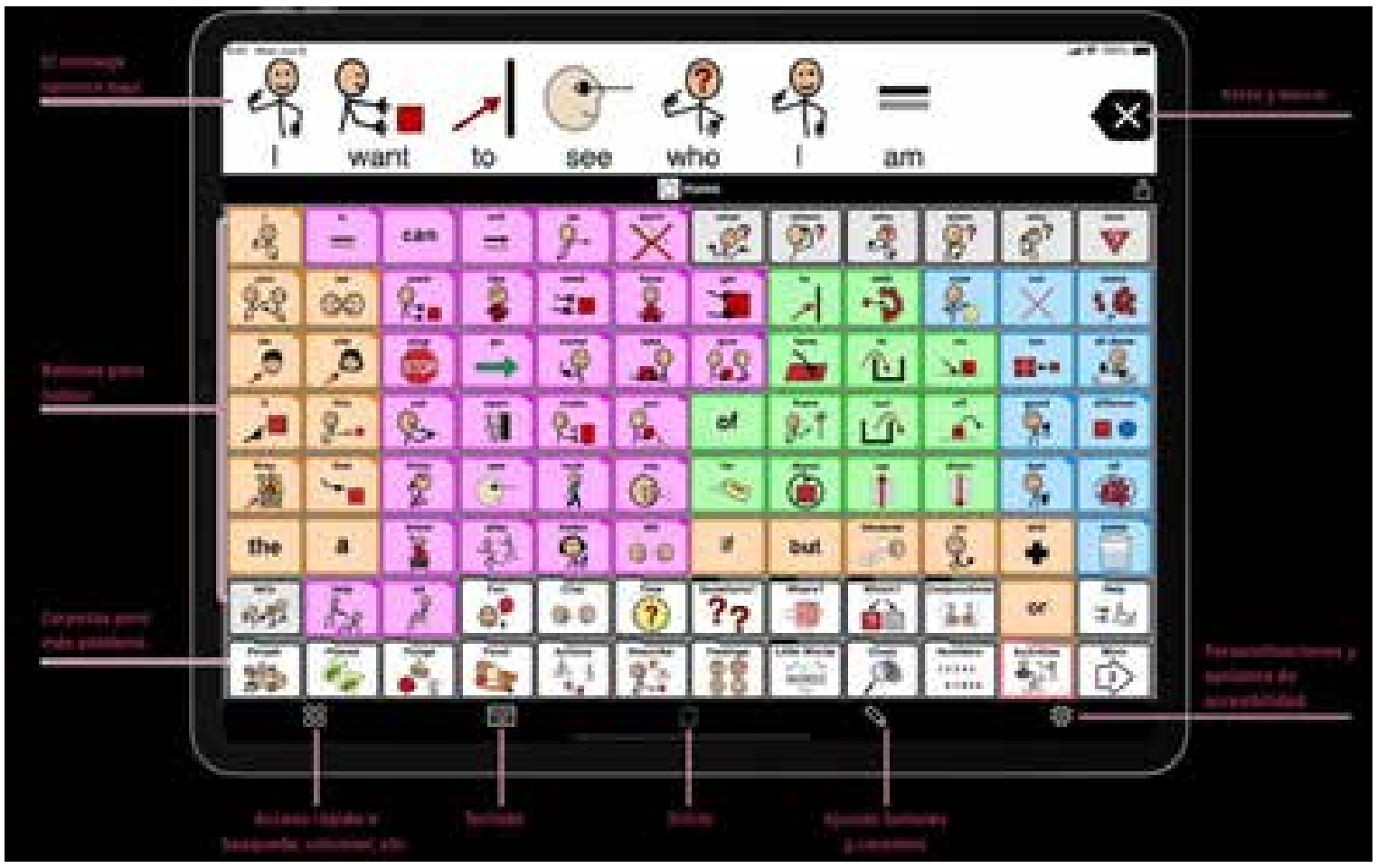

- **Avaz**
  Una herramienta de CAA fácil de usar con pictogramas, voces personalizables y una interfaz amigable.
  Está diseñada para facilitar la comunicación aumentativa y alternativa (CAA), destinada principalmente a personas con discapacidades del habla o dificultades de comunicación. Es ampliamente utilizada por personas con Trastornos del Espectro Autista (TEA), parálisis cerebral, síndrome de Down, apraxia del habla y otras condiciones que afectan la capacidad de comunicarse de manera verbal. Qué nos proporciona, tableros de comunicación basados en pictogramas y texto, utilizando símbolos visuales y texto para ayudar a los usuarios a formar frases y expresar sus pensamientos. Los pictogramas son intuitivos y fáciles de reconocer, lo que facilita la navegación. Otra característica muy significativa es la generación de voz, con un solo toque, las frases seleccionadas en el tablero se convierten en salida de voz. Ofrece diferentes voces y acentos para adaptarse a las necesidades del usuario. Lo más importante, es adaptable y se personaliza, por tanto, las necesidades específicas de los usuarios son cubiertas, se agregan palabras, categorías y fotos personales.

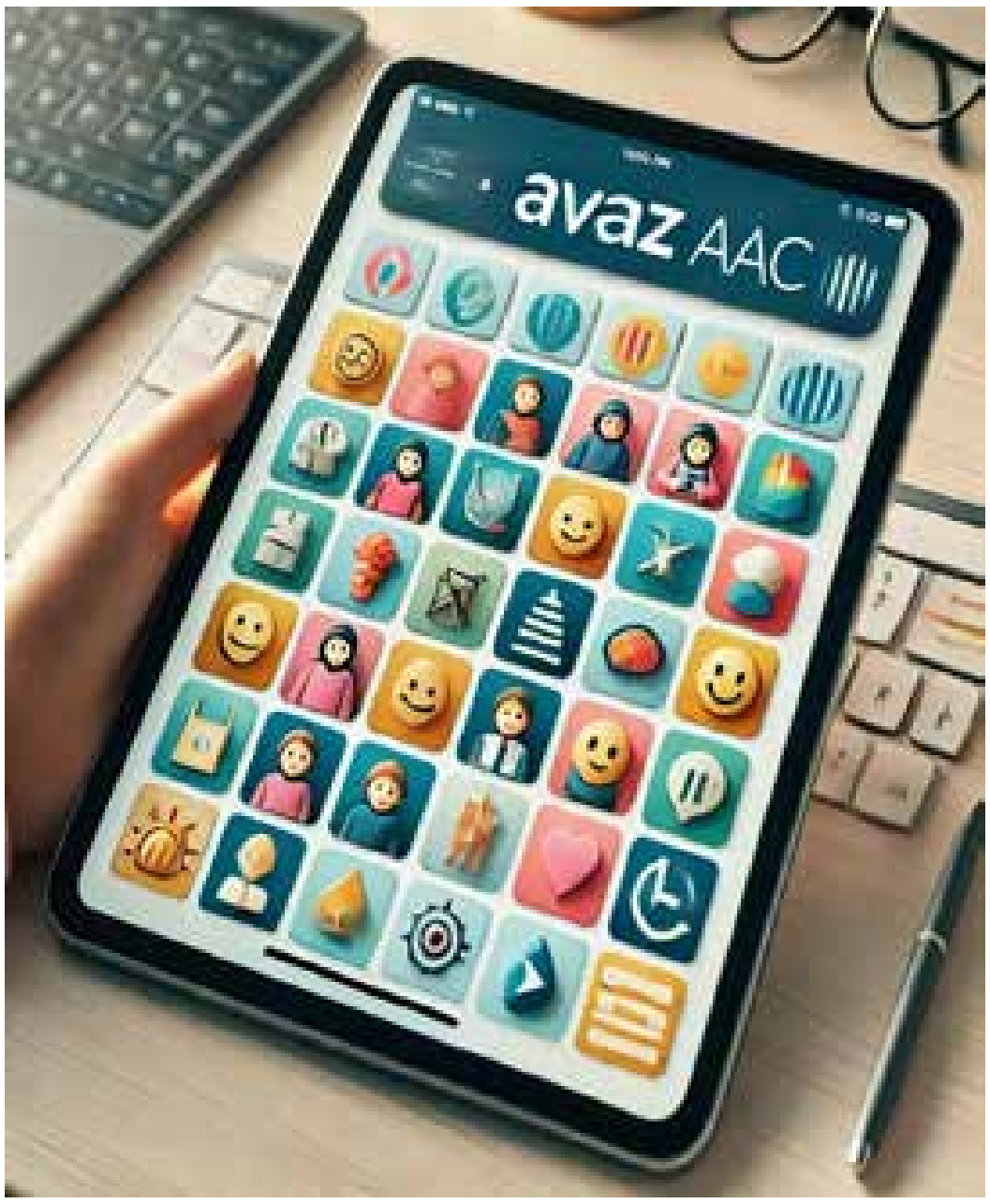

Es compatible en múltiples idiomas, lo que la hace útil para comunidades diversas. Cuando los estudiantes, progresan más allá de los pictogramas, Avaz ofrece un teclado predictivo que les ayuda a escribir y comunicarse de manera independiente. Desarrolla herramientas para padres y terapeutas, donde se Incluyen funcionalidades para realizar un seguimiento del progreso, como informes de uso y análisis de datos. Además de Guías complementarias que ayudan a padres y educadores a apoyar mejor el aprendizaje del usuario. Es compatible con dispositivos de acceso alternativo, como interruptores y tecnología de seguimiento ocular, para usuarios con movilidad limitada. Es muy útil como entrenador para facilitar la alfabetización, porque fomenta la transición gradual de la comunicación basada en pictogramas al uso de palabras escritas. Por tanto, tiene ventajas interesantes como: Interfaz amigable y adecuada para diferentes edades, se puede usar tanto en tabletas como en teléfonos inteligentes y es flexible para adaptarse a necesidades diversas, desde principiantes hasta usuarios avanzados.

La aplicación permite a los usuarios expresar necesidades básicas, emociones y pensamientos, por tanto, comunicación diaria, así como apoya la inclusión en el aula, facilitando la comunicación con maestros y compañeros (entorno educativo) y es útil en sesiones de terapia del habla y del lenguaje (entorno terapéutico). ¿Cómo funciona?

Un ejemplo típico de uso es cuando un usuario selecciona símbolos en la pantalla para construir una frase, como "Quiero comer una manzana". Avaz traduce esta selección en voz, permitiendo que el usuario se comunique eficazmente con los demás. Su disponibilidad, es también un tema importante está disponible en sistemas operativos iOS y Android.

– **LetMeTalk**

  Gratuita, ofrece miles de pictogramas y es ideal para mejorar la comunicación. Es otra aplicación de Comunicación Aumentativa y Alternativa (CAA) gratuita, diseñada para ayudar a personas con dificultades del habla o discapacidades de comunicación a expresarse. Es ampliamente utilizada por personas con condiciones como autismo, parálisis cerebral, apraxia del habla, síndrome de Down y otras necesidades de comunicación. Esta aplicación permite a los usuarios seleccionar pictogramas de una biblioteca visual para construir frases, donde cada pictograma está acompañado de una etiqueta de texto, lo que facilita la comprensión y el aprendizaje del lenguaje. Tal y como hemos explicado en otras aplicaciones tiene salida de voz, convierte las frases seleccionadas en salida de audio eso ayuda a su comunicación de una forma eficaz. Tiene una biblioteca de símbolos amplia y personalizable dado que utiliza una base de datos de imágenes del conjunto de pictogramas ARASAAC, reconocidos por su accesibilidad. Se pueden agregar fotos o imágenes personales para personalizar su experiencia. Funciona sin necesidad de conexión a internet, lo que la hace ideal para usar en cualquier lugar y es compatible con varios idiomas, permitiendo su uso en diferentes contextos culturales y lingüísticos. Su diseño intuitivo la hace accesible para niños, adultos y personas con diversas habilidades.

### *Juegos para la Interacción Social*

La socialización y la comunicación son dos áreas fundamentales en las que los estudiantes con TEA suelen enfrentar dificultades. En este contexto, ya hemos explicado las aplicaciones diseñadas específicamente para fomentar estas habilidades desempeñan un papel crucial. Por ejemplo, aplicaciones como "Proloquo2Go" y "Avaz" ofrecen sistemas de comunicación aumentativa y alternativa (CAA) que permiten a los estudiantes expresar sus necesidades, emociones y pensamientos a través de pictogramas, texto y símbolos visuales. Estas herramientas son especialmente útiles para estudiantes no verbales o con habilidades de comunicación limitadas, ya que proporcionan un medio eficaz para interactuar con su entorno.

Además, plataformas como **Otsimo** combinan juegos interactivos con actividades de aprendizaje para enseñar habilidades sociales básicas, como el contacto visual, el turno de palabra y la interpretación de expresiones faciales. Estas aplicaciones no solo ayudan a mejorar la comunicación, sino que también reducen la ansiedad asociada con las interacciones sociales al permitir que los estudiantes practiquen en un entorno controlado y predecible.

Otsimo es una plataforma educativa y de comunicación diseñada específicamente para niños con necesidades educativas especiales o dificultades de comunicación. Está dirigida principalmente a niños con autismo, síndrome de Down, TDAH, discapacidades del habla, y otras condiciones que requieren un enfoque personalizado para el aprendizaje y la interacción. La plataforma incluye dos aplicaciones principales:

- **Otsimo Special Education (Educación Especial)**
  Esta aplicación ofrece juegos educativos y terapéuticos para niños con necesidades especiales. Su diseño se basa en la metodología ABA (Applied Behavior Analysis), una técnica probada en la enseñanza de habilidades a niños con trastornos del desarrollo. Sus principales características: juegos educativos personalizados que cubren áreas como matemáticas, habilidades lingüísticas, identificación de emociones, resolución de problemas y más. Están diseñados para mejorar habilidades cognitivas, motoras y sociales. Destaca por su gran adaptabilidad, dado que los juegos se ajustan automáticamente al nivel del niño según su progreso y necesidades. Además, realizan un Seguimiento del progreso. Los padres y educadores pueden monitorear el aprendizaje y los avances a través de informes detallados. El diseño es claro y atractivo, pensado para facilitar el uso por parte de niños con dificultades de procesamiento sensorial.

- **Otsimo Speech Therapy (Terapia del Habla)**
  Es una herramienta de terapia del habla basada en juegos y tecnología de reconocimiento de voz, dirigida a niños con discapacidades del habla. Entre sus características se encuentra el reconocimiento de voz avanzado que ayuda a los niños a aprender y practicar palabras, sonidos y frases mediante actividades interactivas. Incluye juegos diseñados en colaboración con logopedas y expertos en terapia del habla, por tanto, es terapéutico. Los ejercicios se adaptan al nivel y progreso del niño. No necesita conexión a internet y algunos ejercicios están disponibles offline, facilitando el acceso en cualquier lugar.
  Otsimo se destaca como una herramienta inclusiva que combina tecnología y pedagogía para mejorar la calidad de vida y el aprendizaje de niños con necesidades especiales.
- **The Social Express**
  Enseña habilidades sociales mediante historias interactivas y actividades prácticas es una aplicación educativa y de desarrollo socioemocional diseñada para ayudar a niños, adolescentes y adultos jóvenes con dificultades para comprender y manejar situaciones sociales. Está especialmente dirigida a personas con Trastornos del Espectro Autista (TEA), TDAH, ansiedad social u otras condiciones que afectan las habilidades interpersonales. El propósito de The Social Express es enseñar habilidades sociales y emocionales esenciales mediante historias interactivas, juegos y actividades. Estas herramientas ayudan a los usuarios a desarrollar competencias como: reconocimiento de emociones, resolución de conflictos, toma de decisiones en situaciones sociales y mejora en la comunicación verbal y no verbal. Entre sus características destacan: historias interactivas, donde se presenta escenarios sociales comunes a través de personajes animados.
  Los estudiantes toman decisiones en diferentes contextos sociales, aprendiendo las consecuencias de sus elecciones. Se enfoca en habilidades específicas, como habilidades prácticas, tales como iniciar conversaciones, interpretar lenguaje corporal, responder a emociones y manejar conflictos donde refuerza conceptos clave como la empatía, la autoconciencia y el trabajo en equipo. El diseño está basado en evidencias, dado que la aplicación está desarrollada en colaboración con expertos en psicología y educación especial y está basada en metodologías de aprendizaje socioemocional (SEL, por sus siglas en inglés). Se adapta a la personalidad de los estudiantes y las actividades van dirigidas a su nivel de desarrollo y progreso, va ajustándose a las diferentes edades y niveles de habilidades adquiridas por él. Diseñada con gráficos animados y actividades interactivas que son atractivas para niños y adolescentes y se suma que es muy fácil de usar tanto para estudiantes como para educadores y padres. Es, sin duda una herramienta para realizar un buen seguimiento, dado que se utiliza en educación y promueve la colaboración de la comunidad educativa y los terapeutas.

– **Peppy Pals**
Ayuda a identificar y comprender emociones a través de juegos y personajes amigables. Es una herramienta divertida y efectiva para apoyar el desarrollo emocional y social de los niños desde una edad temprana, fomentando habilidades que serán esenciales a lo largo de sus vidas. Esta plataforma educativa interactiva enfocada en el desarrollo de habilidades socioemocionales (SEL, por sus siglas en inglés) para niños pequeños, típicamente entre 2 y 8 años. Diseñada como una herramienta divertida y accesible, utiliza historias, juegos y actividades para enseñar a los niños a identificar y gestionar emociones, resolver conflictos y construir relaciones positivas. Su principal propósito, tal y como hemos indicado es fomentar la inteligencia emocional mediante experiencias lúdicas que promuevan habilidades clave como: la empatía, la comunicación efectiva, la resolución de problemas sociales y el manejo de emociones como la tristeza, la alegría, la frustración y el enojo. Sus principales signos distintivos son: los niños interactúan con personajes animales amigables que enfrentan diferentes situaciones sociales y a través de decisiones y observaciones, los niños aprenden sobre emociones, amistades y resolución de problemas. Incluye cuentos animados y videos que presentan conflictos cotidianos y enseñan formas saludables de resolverlos y se centra en temas como compartir, pedir disculpas y entender el punto de vista de los demás. Los juegos y actividades no requieren habilidades de lectura, lo que los hace accesibles para niños pequeños o aquellos que aún están desarrollando el lenguaje. Proporciona recursos y guías para que los padres o cuidadores refuercen los aprendizajes en casa y fomenta el diálogo familiar sobre emociones y relaciones interpersonales. Destacamos los beneficios educativos en el desarrollo de habilidades sociales, por que enseña a los niños a comprender y gestionar sus emociones, interactuar con otros y desarrollar relaciones positivas, por tanto, fortalece la empatía y enseña estrategias para manejar el estrés, la frustración y otras emociones intensas provocando su autorregulación.

### *Otros juegos y actividades socioemocionales y de regulación*

- **Storyboardthat**
  Ofrece la oportunidad a los alumnos de crear historietas digitales sobre temas que enfaticen la conciencia sobre los problemas sociales.
  Los juegos digitales también desempeñan un papel importante en el aprendizaje emocional. Títulos como "Social Adventures" utilizan escenarios virtuales para enseñar habilidades relacionadas con la empatía y la resolución de conflictos. Estas herramientas permiten a los estudiantes practicar reacciones emocionales y sociales en un entorno que minimiza el riesgo de error y fomenta la experimentación.
- **Befunky**
  Permite a los alumnos crear collages con imágenes de personas de diferentes culturas y etnias, fomentando la sensibilidad y el respeto a la diversidad cultural.
  El reconocimiento y la regulación emocional son componentes clave de las habilidades socioemocionales. Las tecnologías digitales han facilitado el desarrollo de herramientas que enseñan a los estudiantes con TEA a identificar y gestionar sus emociones. Aplicaciones como "Mood Meter" y "Zones of Regulation" ofrecen estrategias visuales e interactivas para ayudar a los estudiantes a comprender cómo se sienten y cómo pueden responder de manera adecuada a diferentes situaciones Autista (TEA).

Tabla 1. Herramientas socioemocionales

| | *Herramienta* | *Descripción* |
|---|---|---|
| 1 | Calm Counter | Diseñada para enseñar estrategias de calma y control emocional. |
| 2 | Zones of Regulation App | Basada en el popular marco 'Zonas de Regulación', ayuda a los niños. |
| 3 | Breathe, Think, Do with Sesame | Enseña habilidades básicas de resolución de problemas. |
| 4 | Emotions Flashcards | Aplicaciones con tarjetas visuales para enseñar a reconocer expresiones. |
| 5 | Touch and Learn - Emotions | Juegos interactivos para practicar la identificación y comprensión de emociones. |
| 6 | iSEQUENCES | Presenta secuencias visuales para enseñar habilidades socioemocionales. |
| 7 | Padlet | Ofrece la posibilidad de crear espacios colaborativos en el aula. |
| 8 | Interland: Reino Amable | Un juego interactivo en línea que permite a los alumnos realizar gestos de amabilidad y empatía. |

### *Juegos y actividades digitales para la expresión emocional*

El juego es una herramienta poderosa para el aprendizaje, y las actividades digitales han ampliado sus posibilidades al ofrecer experiencias inmersivas y atractivas. Juegos como "**Emotionary**" enseñan a los estudiantes a identificar y etiquetar emociones a través de actividades interactivas. Estos juegos utilizan gráficos llamativos y retroalimentación positiva para mantener el interés del estudiante y reforzar el aprendizaje.

Otra aplicación destacada es "**Feelings in a Flash**", que utiliza tarjetas digitales para estimular conversaciones sobre emociones y promover la comprensión de estas en diferentes contextos. Estas herramientas no solo ayudan a los estudiantes a expresar sus sentimientos, sino que también fomentan la discusión y el entendimiento compartido entre pares, educadores y familias.

### *Recursos visuales y de historias sociales*

Los recursos visuales son especialmente efectivos para los estudiantes con TEA debido a su preferencia por la información estructurada y concreta. Las historias sociales, diseñadas para enseñar habilidades sociales y conductuales a través de narrativas visuales y textuales, son un ejemplo clave. Plataformas como "**Story Maker for Social Skills**" permiten a los educadores y padres crear historias personalizadas que abordan situaciones específicas, como asistir a una fiesta o visitar un nuevo lugar. Estas historias ayudan a preparar a los estudiantes para experiencias reales al proporcionarles un marco claro y predecible de lo que pueden esperar y cómo deben comportarse.

Además, las aplicaciones que utilizan pictogramas y diagramas visuales, como "**Visual Schedule Planner**", ayudan a los estudiantes a organizar sus actividades diarias y a comprender las expectativas sociales. Estos recursos no solo mejoran la comprensión de las normas sociales, sino que también aumentan la autonomía y reducen la ansiedad asociada con cambios o situaciones inesperadas. Boardmaker: Permite crear materiales visuales personalizados, como calendarios, rutinas y secuencias sociales. Y, StoryMaker for Social Stories: Ayuda a crear y personalizar historias sociales que abordan situaciones específicas y habilidades sociales. Además, CanPlan: Apoya el aprendizaje de tareas paso a paso con instrucciones visuales y recordatorios.

## Plataformas y recursos de formación para docentes

– **ARASAAC**
  Portal con una gran variedad de recursos gratuitos de CAA, como pictogramas, videos y plantillas.

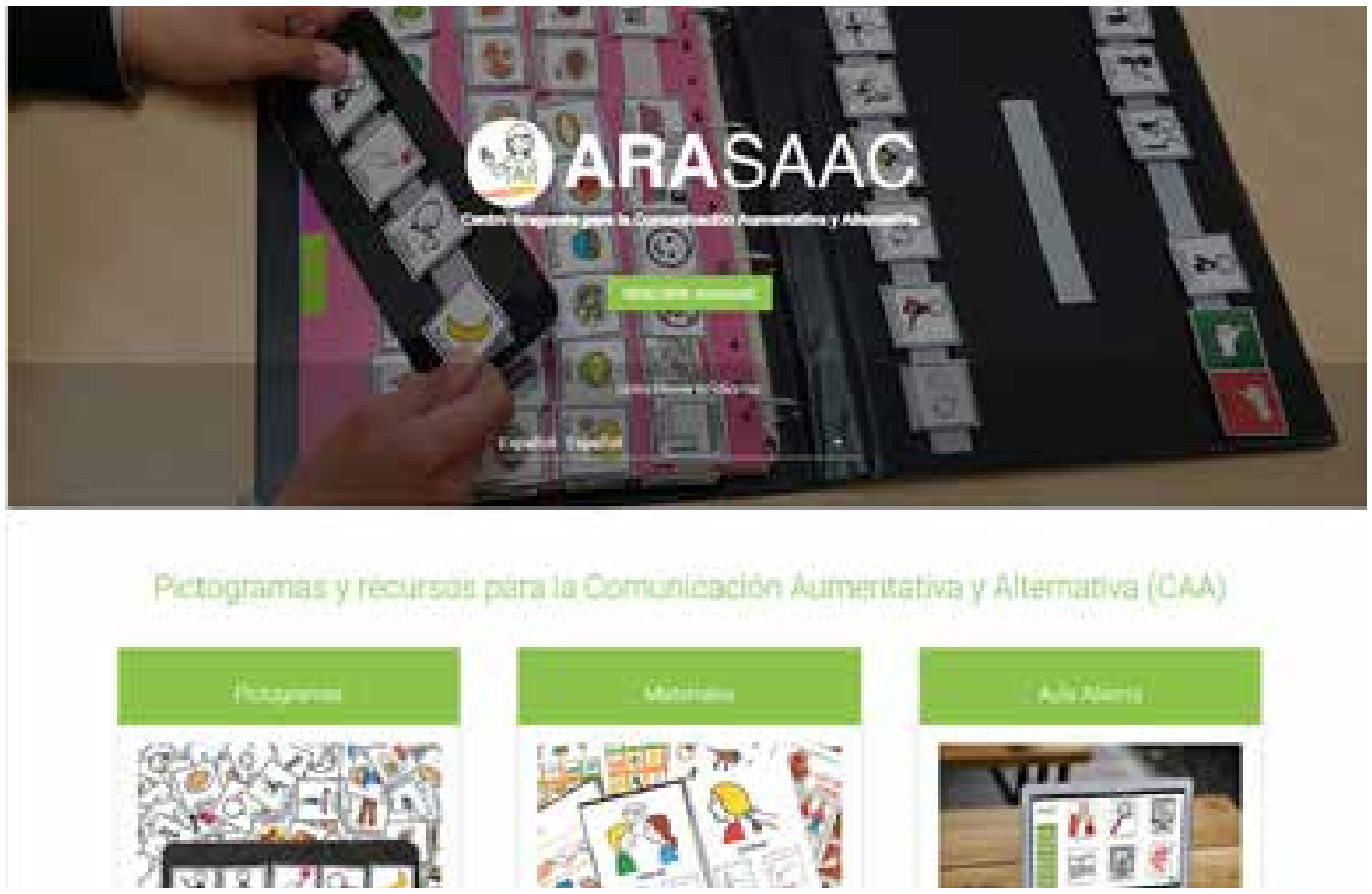

– **Google Classroom**
  Organiza el trabajo en el aula, facilita la comunicación y permite compartir recursos de manera visual.
– **Kahoot!**
  Crea juegos interactivos para repasar conceptos y fomentar la participación en grupo.

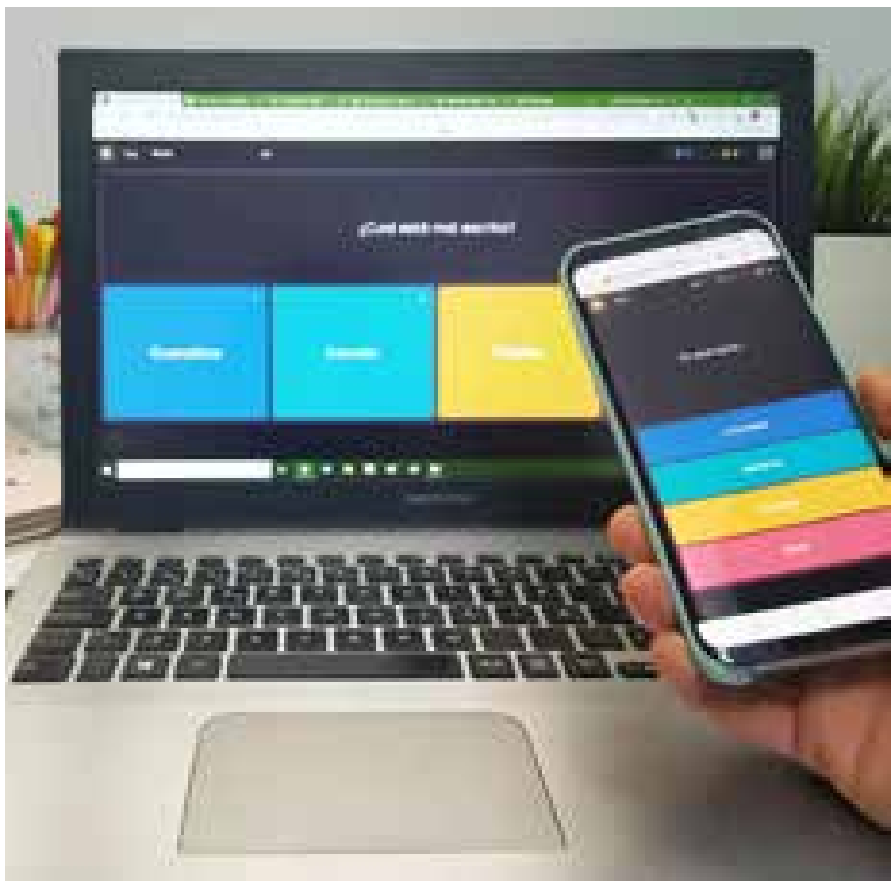

- **Quizizz**
  ¡Similar a Kahoot!, ofrece una amplia variedad de plantillas y permite personalizar los cuestionarios.

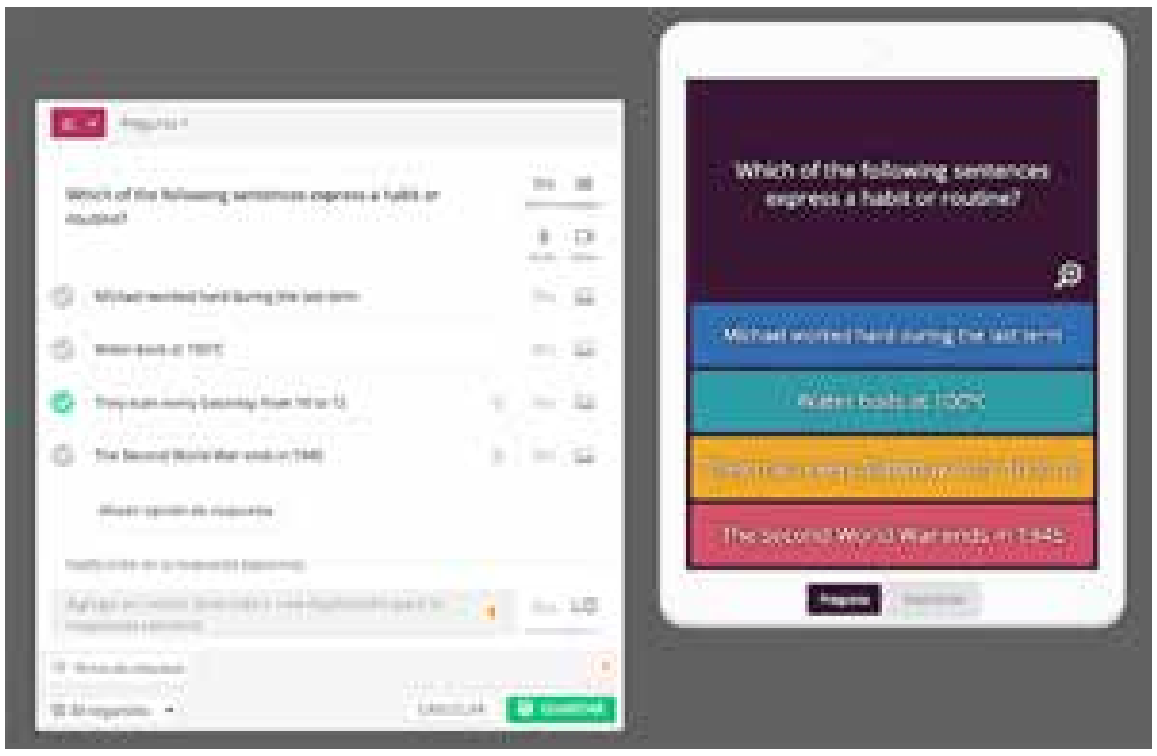

- **Autism Navigator**
  Recursos en línea para entender y enseñar a estudiantes con TEA, con un enfoque en habilidades socioemocionales.
- **Do2Learn**
  Actividades y recursos descargables para desarrollar habilidades de la vida diaria y socioemocionales.
- **Learning A-Z**
  Ofrece herramientas de enseñanza personalizadas para atender las necesidades específicas de cada alumno.
- **Social Stories Creator**
  Permite crear historias sociales personalizadas para enseñar habilidades como la interacción social, la resolución de conflictos o la rutina diaria.

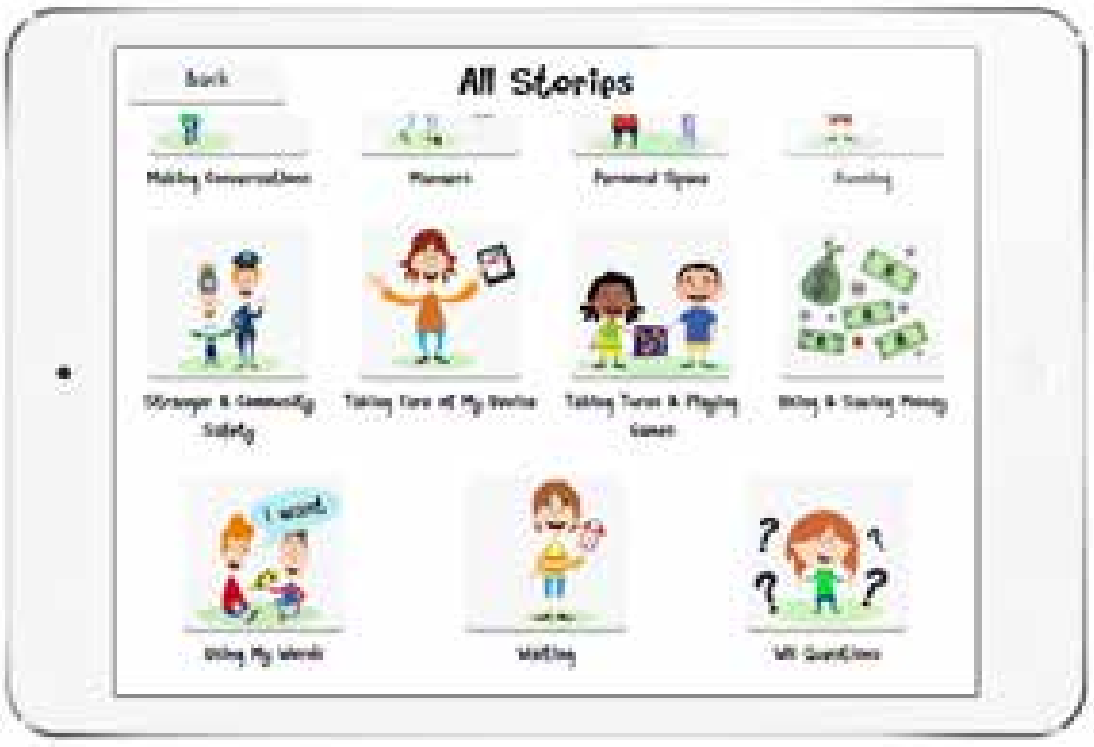

– **Goally**
Ofrece un sistema de recompensas y rutinas visualizadas para fomentar habilidades como la independencia y la organización.

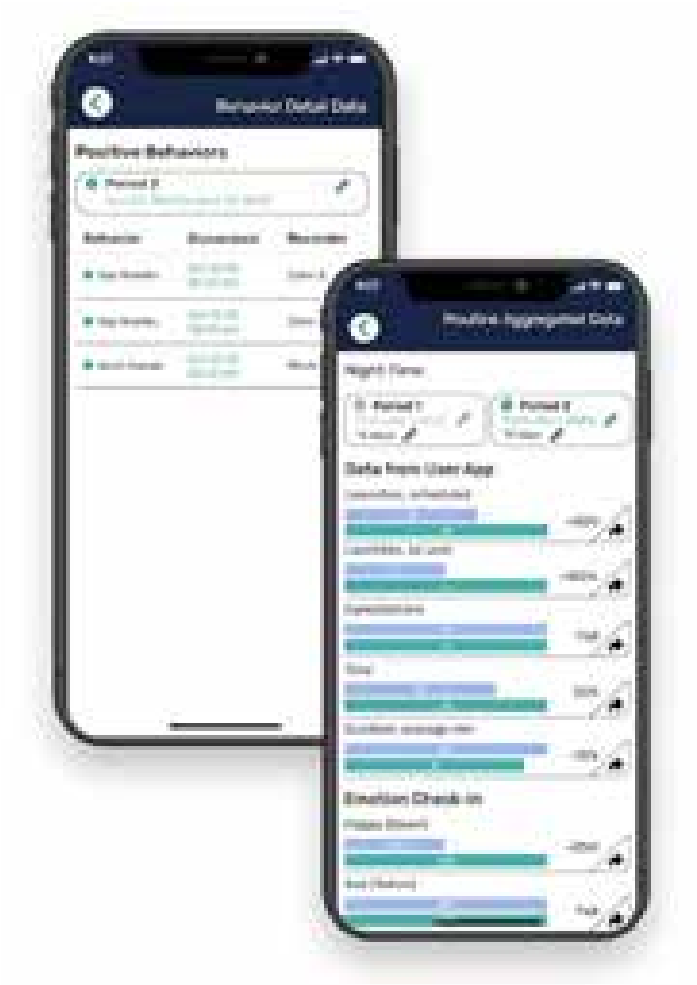

– **Simulo**
Es una plataforma que utiliza la realidad virtual para simular situaciones sociales, ayudando a los estudiantes a practicar habilidades de comunicación e interacción.

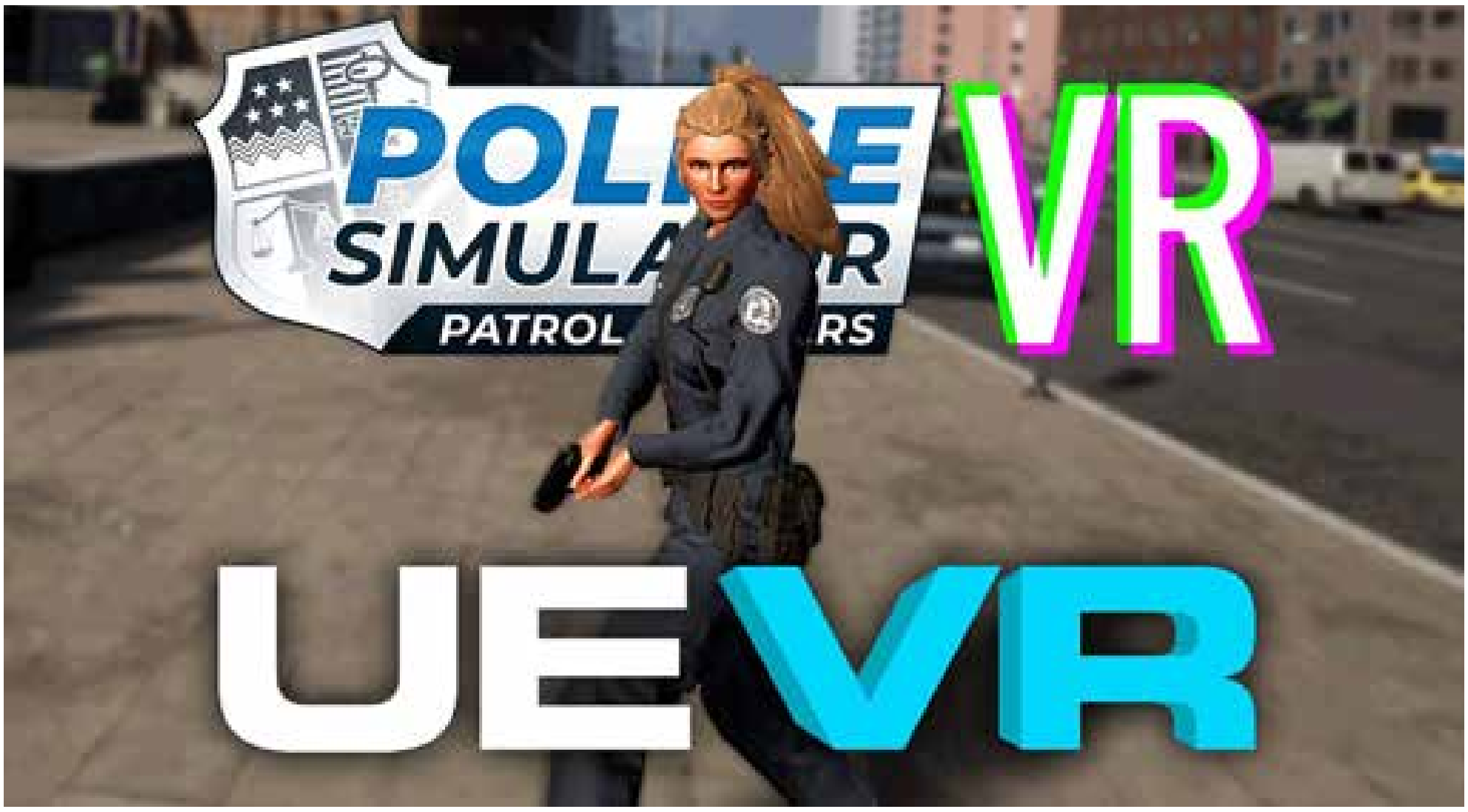

## Software y herramientas de realidad virtual/aumentada

La realidad virtual (RV) y la Realidad Aumentada (RA) ofrecen oportunidades únicas para el aprendizaje socioemocional al proporcionar entornos inmersivos y controlados donde los estudiantes pueden practicar habilidades sociales y emocionales. Herramientas como "Floreo" utilizan RV para recrear situaciones sociales comunes, como una conversación en el aula o una visita al supermercado, permitiendo a los estudiantes practicar y desarrollar su confianza en estas interacciones.

Por otro lado, la RA también se ha utilizado para enseñar conceptos abstractos y emocionales de manera más concreta. Aplicaciones como "Quiver" permiten a los estudiantes interactuar con objetos tridimensionales que representan emociones o conceptos sociales, facilitando la comprensión y el aprendizaje activo.

Evo Assist: Simula interacciones sociales en entornos virtuales para practicar habilidades en tiempo real.

## Blogs y comunidades online de apoyo

Blogs y webs especializadas: Muchos profesionales comparten sus experiencias y recursos en blogs y páginas web. Comunidades online: Plataformas como foros y grupos de Facebook permiten a padres y educadores compartir experiencias y encontrar apoyo. Autism Speaks: Contiene recomendaciones de apps y recursos útiles para padres y educadores.TEACCH Autism Program: Proporciona materiales visuales y estrategias estructuradas para apoyar a estudiantes con TEA. Estas comunidades online desempeñan un papel crucial en el apoyo a los estudiantes con TEA y sus familias. Plataformas como "Autism Speaks Community" y "MyAutismTeam" ofrecen un espacio seguro donde los padres, educadores y terapeutas pueden compartir experiencias, recursos y estrategias para fomentar el desarrollo socioemocional.

La IA también está revolucionando la manera en que se abordan las necesidades de los estudiantes con TEA. Herramientas como "Replika" y "Woebot" actúan como compañeros virtuales que ofrecen apoyo emocional y enseñan habilidades sociales a través de interacciones personalizadas. Estas herramientas no solo proporcionan un feedback constante, sino que también se adaptan a las necesidades individuales del estudiante, haciendo que el aprendizaje sea más efectivo y significativo.

No podemos olvidar, que al fin y al cabo lo más importante es la colaboración entre el equipo educativo, la familia y el estudiante es esencial para el éxito en el desarrollo de habilidades socioemocionales. Esta colaboración permite una comprensión más profunda de las necesidades y fortalezas del estudiante, y asegura que todos los involucrados trabajen hacia objetivos comunes. La comunicación abierta y regular entre maestros, padres y el propio estudiante facilita la identificación de áreas de mejora y la implementación de estrategias efectivas. Además, la colaboración fomenta un ambiente de apoyo y confianza, lo que es crucial para el bienestar emocional del estudiante.

Evaluar el progreso del estudiante es fundamental para ajustar las estrategias de enseñanza según sea necesario. La evaluación continua permite a los educadores y a la familia identificar qué métodos están funcionando y cuáles necesitan ser modificados. Esto puede incluir la observación directa, el uso de herramientas de evaluación específicas y la retroalimentación del propio estudiante. Al ajustar las estrategias basadas en estas evaluaciones, se puede asegurar que el estudiante reciba el apoyo adecuado para su desarrollo socioemocional.

Adaptar las herramientas y recursos a las necesidades individuales de cada estudiante es crucial para su éxito. Cada estudiante con TEA tiene un conjunto único de habilidades y desafíos, por lo que es importante personalizar las intervenciones para que sean efectivas. Esto puede incluir el uso de tecnologías asistidas, programas de aprendizaje personalizados y actividades diseñadas específicamente para abordar las áreas de necesidad del estudiante.

Utilizar las mismas herramientas y recursos de manera consistente facilita la comprensión y el aprendizaje del estudiante. La consistencia en el uso de herramientas digitales y estrategias de enseñanza ayuda a crear un entorno predecible y estructurado, lo cual es especialmente beneficioso para los estudiantes con TEA. Además, la repetición y la práctica regular permiten al estudiante familiarizarse con las herramientas y desarrollar confianza en su uso.

## Conclusiones

Podemos concluir que las herramientas y los recursos digitales son fundamentales para fomentar las habilidades socioemocionales en estudiantes con TEA debido a su capacidad para proporcionar intervenciones estructuradas, atractivas y personalizables que abordan necesidades de aprendizaje únicas.

Por ejemplo, los entornos de RV crean espacios seguros para practicar interacciones sociales, lo que mejora la confianza en escenarios del mundo real. Estas herramientas no solo son inmersivas, sino que también permiten experiencias de aprendizaje individualizadas adaptadas al progreso del usuario, algo de lo que a menudo carecen las intervenciones tradicionales como los juegos de roles. De manera similar, los enfoques basados en juegos, como el prototipo "Emotion Adventure", están diseñados para enseñar la teoría de la mente, una habilidad socioemocional fundamental, a través de un juego interactivo que mejora la participación del usuario y facilita el aprendizaje en un formato agradable (Kim *et al.*, 2023).

Las herramientas digitales también respaldan la instrucción diferenciada, lo que permite a los estudiantes con TEA acceder al contenido de maneras que resuenen con sus estilos de aprendizaje individuales y mejoren la colaboración entre pares (World Journal of Biology Pharmacy and Health Sciences, 2023). Además, las evaluaciones basadas en la web como SELweb demuestran cómo la tecnología puede medir y abordar déficits socioemocionales específicos, ofreciendo información sobre intervenciones personalizadas (Russo-Ponsaran *et al.*, 2019). Estos recursos cierran brechas en los métodos tradicionales, proporcionando vías consistentes, flexibles y efectivas para desarrollar competencias sociales críticas. Cómo la realidad virtual y las plataformas de juegos mejoran el aprendizaje social en estudiantes con TEA.

La RV y las plataformas de juegos mejoran significativamente el aprendizaje social en estudiantes con TEA al proporcionar entornos controlados, inmersivos y atractivos adaptados a sus necesidades únicas. La realidad virtual permite a las personas practicar interacciones sociales del mundo real en un espacio seguro, lo que reduce la ansiedad y permite la exposición repetida a escenarios como saludos, actividades grupales y resolución de conflictos. Estas plataformas a menudo incorporan retroalimentación dinámica, que ayuda a los estudiantes a refinar sus respuestas y generar confianza con el tiempo. Por ejemplo, las intervenciones de realidad virtual han demostrado una mayor eficacia para mejorar las habilidades sociales en comparación con los métodos tradicionales, ya que sumergen a los usuarios en entornos realistas, pero sin riesgos (Sideraki y Drigas, 2023; Bejarano, 2015) En su investigación sobre el diseño de un recurso educativo digital para entrenar el desarrollo de habilidades socioemocionales en niños y jóvenes con TEA, estos autores exploran cómo las herramientas digitales pueden ser utilizadas para intervenciones basadas en teorías psicológicas.

Las plataformas de juegos también fomentan el crecimiento socioemocional al incorporar habilidades críticas como el reconocimiento de emociones, la empatía y la teoría de la mente dentro de marcos interactivos y entretenidos. Por ejemplo, el prototipo de juego "Emotion Adventure" integra estos elementos en el juego, enseñando de manera efectiva conceptos esenciales para la cognición social mientras mantiene altos niveles de participación (Kim *et al.*, 2023). A diferencia de las intervenciones estáticas, las herramientas de realidad virtual y juegos se adaptan al progreso del alumno, ofreciendo desafíos personalizados que se adaptan a los ritmos de desarrollo individuales. Además, el formato gamificado fomenta la motivación y la persistencia, especialmente entre los usuarios más jóvenes, que son más propensos a interactuar con los medios digitales que con las actividades terapéuticas tradicionales.

Estas herramientas también promueven las habilidades de colaboración y comunicación a través de tareas multijugador o cooperativas, fomentando la interacción entre pares en un entorno digital controlado. Al combinar aplicaciones prácticas, entretenimiento y adaptabilidad, las plataformas de realidad virtual y juegos presentan oportunidades transformadoras para desarrollar competencias socioemocionales en estudiantes con TEA, cerrando la brecha entre la terapia y la aplicación en el mundo real.

Las intervenciones tradicionales de habilidades socioemocionales para personas con TEA suelen enfrentar varias limitaciones debido a su falta de personalización, desafíos de generalización y compromiso limitado. Muchos métodos tradicionales, como la terapia de grupo o los juegos de roles, siguen enfoques estandarizados que pueden no abordar los estilos de aprendizaje únicos y las sensibilidades sensoriales de las personas con TEA. Esta falta de personalización puede reducir su eficacia, ya que las personas en el espectro a menudo requieren estrategias personalizadas para satisfacer sus necesidades de desarrollo específicas (Russo-Ponsaran *et al.*, 2019).

Otra limitación importante es la dificultad de generalizar las habilidades aprendidas a entornos del mundo real. Las intervenciones tradicionales pueden enseñar comportamientos sociales apropiados en entornos controlados, pero estas habilidades a menudo no se transfieren a contextos sociales impredecibles y dinámicos. Esta brecha obstaculiza el impacto a largo plazo de tales intervenciones y puede hacer que las personas tengan dificultades para adaptar sus habilidades fuera de las sesiones de terapia (Sideraki y Drigas, 2023).

La participación también es un desafío, ya que los métodos tradicionales pueden resultar repetitivos, monótonos o irrelevantes para las personas con TEA, en las más jóvenes. Estos métodos pueden no aprovechar las fortalezas de las personas, como su afinidad por el aprendizaje visual o interactivo, y a menudo carecen de la adaptabilidad para ajustarse al progreso del alumno. Además, el estigma asociado con la asistencia a las sesiones de terapia también puede disuadir la participación, especialmente en entornos grupales donde las personas pueden sentirse juzgadas o incomprendidas.

Estas limitaciones resaltan la necesidad de enfoques innovadores, como herramientas digitales e intervenciones basadas en realidad virtual, que ofrezcan soluciones personalizadas, atractivas y transferibles para desarrollar de manera efectiva las habilidades socioemocionales en las personas con TEA.

Las evaluaciones basadas en la web se pueden integrar de manera eficaz en planes de terapia más amplios para personas con TEA, ya que brindan un enfoque escalable, eficiente y personalizado para medir las habilidades socioemocionales. Herramientas como SELweb, diseñadas para evaluar dominios clave como el reconocimiento de emociones, la teoría de la mente y la resolución de problemas sociales, ofrecen a los médicos información detallada sobre las fortalezas y los desafíos únicos que enfrentan las personas con TEA (Russo-Ponsaran *et al.*, 2019). Estas evaluaciones complementan los métodos tradicionales al brindar métricas objetivas basadas en datos que se pueden usar para adaptar los planes de terapia.

El formato digital de las evaluaciones basadas en la web garantiza la accesibilidad y la flexibilidad, lo que permite administrarlas en entornos clínicos, educativos o domésticos. Esta adaptabilidad es particularmente útil para el seguimiento continuo del progreso, ya que los terapeutas pueden hacer un seguimiento de las mejoras o identificar áreas que requieren un enfoque adicional a lo largo del tiempo. Además, la naturaleza interactiva y gamificada de muchas herramientas basadas en la web ayuda a involucrar a las personas con TEA, fomentando la participación activa y reduciendo la ansiedad relacionada con los exámenes.

Cuando se integran en los planes de terapia, estas evaluaciones pueden guiar el diseño de intervenciones personalizadas al identificar déficits específicos. Por ejemplo, si un niño presenta dificultades en el reconocimiento de emociones, la terapia puede priorizar ejercicios relacionados utilizando herramientas digitales específicas o escenarios del mundo real. Las evaluaciones basadas en la web también permiten compartir datos entre equipos mul-tidisciplinarios, lo que fomenta la colaboración entre terapeutas, educadores y cuidadores (Rondel & Vera, 2023).

Al unir la evaluación y la intervención, estas herramientas facilitan un enfoque terapéutico más cohesivo y holístico. Su escalabilidad y capacidad para proporcionar datos valiosos y procesables las hacen indispensables para la terapia moderna del TEA, lo que garantiza que las intervenciones sigan siendo adaptativas, precisas y efectivas.

## Financiación

## Referencias bibliográficas

# HERRAMIENTAS Y RECURSOS DIGITALES PARA EL DESARROLLO DE LA GESTIÓN DEL TIEMPO Y DE LAS ACTIVIDADES DEL ALUMNADO CON TEA

*Mª José Navarro-Montaño*
*Elena Hernández-de-la-Torre*
Facultad de Ciencias de la Educación. Universidad de Sevilla (España)

## Introducción

El uso de herramientas y recursos digitales en educación ha experimentado un gran avance en los últimos años, prueba de ello es el número creciente de diseños de instrumentos para el desarrollo de prácticas de aprendizaje, no solo para el alumnado en general, sino también para el alumnado con dificultades para aprender. En este sentido, Cabero *et al.* (2020) señalan que se recomienda mejorar las ofertas de formación docente proporcionando programas de capacitación más amplios que brinden a los maestros la oportunidad de implementar prácticas basadas en la evidencia en aulas inclusivas. Como señala el Centro de Desarrollo de Competencias Digitales de Castilla-La Mancha (2019), especializado en la elaboración de herramientas TIC para alumnado con necesidades específicas de apoyo educativo con Trastorno del Espectro Autista (TEA), es necesario apoyar la socialización y la comunicación verbal de este alumnado, los gestos y la comunicación no verbal, la necesidad de concretar orden y rutinas, disminuir comportamientos repetitivos e intereses específicos, así como aportar un mayor número de estímulos sensoriales.

El apoyo educativo al alumnado con TEA a partir de herramientas y recursos digitales constituye una alternativa y una oportunidad para implementar prácticas inclusivas. El autismo se define como:

> Una condición del neurodesarrollo que afecta a la configuración del sistema nervioso y al funcionamiento cerebral. Se caracteriza por dar lugar a dificultades para la comunicación e interacción social y para la flexibilidad del pensamiento y de la conducta de la persona que lo presenta (Confederación de Autismo de España, 2024).

Entre los rasgos característicos del alumnado con TEA destacan los siguientes: introversión, escasa reacción a los estímulos del medio ambiente, falta de habilidades sociales y buena memoria, por lo que el aprendizaje y la participación se convierten en un desafío constante para el progreso en su educación. En este sentido, es necesario tener en cuenta que este alumnado tiene los mismos derechos, así como sus familias, para mejorar su calidad de vida y conseguir la igualdad de oportunidades en un sistema educativo inclusivo y que atiende a la diversidad (Confederación de Autismo de España, 2024). En relación con el uso de recursos digitales, las TIC en la actualidad se constituyen en una alternativa para ayudar a satisfacer las dificultades que presenta este alumnado con necesidades educativas. "Teniendo en cuenta la gran variedad de necesidades que presenta el alumnado con TEA, se hace necesaria la creación e implementación de diferentes herramientas tecnológicas con el objetivo de lograr una mejora en el proceso de aprendizaje" (Durán, 2021, p. 107)

En nuestro contexto educativo, la LOMLOE (2020) recoge entre sus enfoques, en quinto lugar:

> …la necesidad de tener en cuenta el cambio digital que se está produciendo en nuestras sociedades y que forzosamente afecta a la activad educativa. El desarrollo de la competencia digital no supone solamente el dominio de los diferentes dispositivos y aplicaciones (…) se incluye la atención al desarrollo de la competencia digital de los y las estudiantes en todas las etapas educativas.

Pino, Barriga y González (2019) afirman que la inclusión educativa, no tiene que abarcar exclusivamente el ámbito propiamente educativo, sino que nace de una demanda social que está estrechamente vinculada a la digitalización que también supone implicar al estudiante en su propio proceso educativo, es decir, el profesorado que sigue métodos activos se basa en un enfoque centrado en el alumnado, partiendo de las características individuales de los estudiantes para diseñar una respuesta educativa basada en sus necesidades (Jiménez, González y Tornel 2020). Por tanto, las TIC constituyen una herramienta valiosa para promover la educación inclusiva de todos los estudiantes y específicamente de estudiantes con TEA, la adaptación del proceso de ense-ñanza aprendizaje a las necesidades educativas de estos estudiantes puede facilitarse mediante el uso de herramientas y recursos digitales adecuados. Cabero (2021) se refiere a una serie de factores que caracterizan las TIC como su ergonomía, adaptabilidad a situaciones diversas, motivación del estudiante, individualización de la enseñanza, feedback, accesibilidad, interactividad, facilitación del aprendizaje, participación activa, adaptación a la variabilidad de ritmos de aprendizaje, entre otros. Para Cáceres *et al.* (2021), las TIC favorecen el uso de metodologías activas por sus características y capacidad de adaptación a diferentes contextos. Contribuyen a mejorar la inclusión educativa de todos los estudiantes y específicamente de los estudiantes con TEA ya que, como afirma Marín (2018), son recursos diseñados con la finalidad de promover la formación de calidad para todos considerando las circunstancias de cada uno.

También Berrocal (2022) recientemente se ha referido a la integración de las TIC en la educación como realidad que no puede ignorarse por lo que es necesario abordar cambios de roles en el modelo educativo que implican tanto a docentes como a estudiantes. En esta línea de investigación, Pinto, Pérez y Darder (2020) se refieren al cambio experimentado en el proceso educativo como consecuencia de la implantación de las TIC que ha originado escenarios de aprendizaje diferentes e impulsado cambios en las metodologías y funciones docentes. Por tanto, considerando que la educación es un medio para alcanzar unos objetivos, podemos considerar que las TIC constituyen un medio al que se puede recurrir, tal como afirman Cifuentes y Crespo (2021), quienes consideran que el uso didáctico de las TIC va más allá del simple “cultivo de destrezas” para utilizar los dispositivos tecnológicos.

## Herramientas y recursos digitales

Entre los recursos TIC para los estudiantes con TEA, el Centro de Desarrollo de Competencias Digitales Castilla-La Mancha ha publicado la *Guía TIC TEA* elaborada por Confederación Autismo España, considerando una serie de materiales entre los que se encuentran el álbum de fotos parlante, los gestores de tiempo y atención, el horario visual con salida de voz, un *Software* para la gestión de símbolos, robótica, el ratón de bola, material multisensorial, dispositivos adaptados, etiquetas con salida de voz, comunicador basado en el sistema de tarjetas, etc. Como podemos comprobar, este material se encuentra al servicio del apoyo de este alumnado para el desarrollo del aprendizaje y la participación. Es necesario destacar que el uso de estrategias digitales dependerá de las necesidades específicas de este alumnado con TEA en cada caso particular.

Respecto a las herramientas para trabajar con alumnado TEA, "se ha comprobado que el uso de la tecnología eleva en las personas con TEA su atención y motivación en la tarea y, por lo tanto, da lugar a una mejor predisposición social. La intervención con herramientas de realidad aumentada ha obtenido muy buenos resultados, siendo por lo tanto muy recomendable". (Castillo y Sánchez-Suricalday, 2023, 39)

El Centro de Desarrollo de Competencias Digitales Castilla-La Mancha señala la Guía TICTEA (Recursos tecnológicos para personas con Trastorno del Espectro del Autismo) elaborada por la Confederación Autismo España en colaboración con el Ministerio de Educación, Cultura y Deporte, así como cursos gratuitos en Recursos educativos en Internet como Formación para Formadores.

Entre los recursos digitales elaborada por *Guía TIC TEA* elaborada por Confederación Autismo España se encuentran los siguientes recursos tecnológicos:

- El álbum de fotos parlante, útil para el aprendizaje del vocabulario, comportamientos en situaciones cotidianas o temas del currículo; "se pueden incluir fotografías, pero también otras imágenes y recursos educativos. El dispositivo permite grabar mensajes en cada una de las páginas para combinar el aprendizaje visual con el mensaje en voz (…) se pueden organizar las tareas diarias". Este álbum responde a la necesidad de este alumnado para estructurar los entornos y organizar tiempo y rutinas.
- Gestores de tiempo y atención, diseñados para ayudar al alumnado a controlar la inseguridad en el desarrollo de las tareas; se trata de "explicar al niño o la niña que las actividades que vamos a realizar con ellos se acotan en el tiempo". El Centro de Desarrollo de competencias Digitales señala que, para ello, existen herramientas TIC como GESTIAC (Gestión del Tiempo y la Actividad) o eLIGE 2.0 ayudan a gestionar el tiempo. Ayuda al alumnado a desarrollar actividades y controlar el tiempo que transcurre entre ellas.
- Horario visual con salida de voz, se emplea para "grabar una secuencia de acciones que permitan al niño o la niña conseguir desarrollar una habilidad determinada". Reduce la ansiedad que provoca en este alumnado la incertidumbre de las actividades que realiza, enseña a secuenciar las tareas que realizará durante la mañana.

– *Software* para la gestión de símbolos, para la gestión de pictogramas. Tiene como objetivo "facilitar los procesos de aprendizaje, se emplea en mayor medida en el desarrollo de la lectoescritura". El Centro de Desarrollo de competencias Digitales señala que el software a destacar dentro de esta área serían Comunicate In Print o Comunicate Symwriter.
– Robótica, ayuda a la resolución de problemas complejos, la coordinación y el trabajo en equipo, el pensamiento crítico y la flexibilidad cognitiva. El Centro de Desarrollo de competencias Digitales recomienda consultar "Los 5 robots estrella de la robótica educativa". Se emplean dispositivos robóticos sencillos con poca programación con lenguajes simples como Scratch; además se emplean los robots sociales que interactúan de forma natural con el alumnado.
– Ratón de bola, se emplean como paso intermedio entre la pantalla táctil y los ratones convencionales cuentan con otras funcionalidades que hacen más sencillas acciones habituales como seleccionar un texto o arrastrar un archivo, mejoran las relaciones entre el alumnado.
– Material multisensorial, refuerzan el aprendizaje a través de los sentidos. Ayuda al alumnado con TEA para integrar los sentidos y jerarquizar la información que recibe a través de ellos, así como a decodificar la información.
– Dispositivos adaptados, se trata de dispositivos presentes en el aula con algunos elementos de la zona de juegos como son los dispositivos informáticos o las pizarras digitales. El Centro de Desarrollo de competencias Digitales recomienda *Play with me* que se utiliza para enseñar la relación causa-efecto y trabajar turnos durante los juegos, asimismo el alumnado pueda asociar cada color a una acción determinada.
– Etiquetas con salida de voz, crea ambientes espaciotemporales estructurados que reducen la ansiedad, convierten el entorno en un lugar más accesible y potencian la participación del alumnado en actividades desarrolladas en clase. Amplían el vocabulario dentro del aula y se utilizan como comunicador para realizar actividades más complejas
– Comunicador basado en el sistema de tarjetas, funciona a través de pictogramas con los que creamos frases completas que serán reproducidas mediante la voz previamente grabada. El Centro de Desarrollo de competencias Digitales recomienda el Logan Proxtalker, recurso TIC creado por un padre de un niño autista para ayudarle en su desarrollo.

Otras herramientas y recursos digitales que pueden contribuir a optimizar la gestión del tiempo y promover la participación de estudiantes con TEA en las actividades de clase son las que a continuación se relacionan:

- VIRTEA: Es una aplicación desarrollada por la compañía murciana "Answare Tech", ha sido diseñada para trabajar con personas con TEA, recrea distintos escenarios rutinarios que pueden crear desasosiego en el/la estudiante, como ir al dentista, esperar el autobús, etc. Algunos beneficios que ofrece esta aplicación a estudiantes con TEA son: desde el móvil se pueden calibrar los tiempos y los escenarios de forma previa a la puesta en funcionamiento con la persona con la que se desea trabajar, se aprende a manejar la anticipación, preparando al niño/a o adulto con TEA a mejorar su actitud ante situaciones que se salgan de su rutina habitual y disminuye los niveles de estrés de la persona que la esté utilizando en esos momentos que pueden provocarle temor o malestar, mejorando su calidad de vida y la de sus cuidadores. https://www.virtea.io
- ARASAAC: Esta página web ofrece recursos gráficos y materiales adaptados con licencia Creative Commons (BY-NC-SA) para facilitar la comunicación y la accesibilidad cognitiva a estudiantes con TEA así como a estudiantes con diversidad funcional en general. Este recurso es de gran ayuda en el aula, pues posibilita que el alumno con dificultades en el lenguaje o en la compresión pueda interactuar adecuadamente con sus iguales y con el docente a través del uso de pictogramas, sistemas aumentativos y alternativos de comunicación, etc. https://arasaac.org
- MINECRAFT EDUCATION EDITION: Es una plataforma que proporciona una selección de actividades donde el alumnado puede crear y desarrollar diferentes mundos dentro de un entorno seguro a partir de los materiales de aprendizaje ofrecidos, el estudiante puede ejecutar diferentes tareas y encontrar nuevas metodologías para su resolución. Especialmente recomendado para al alumnado que presente TEA o TDAH ya que el juego convierte el aprendizaje en una experiencia inmersiva controlada por el docente, quién puede visualizar en todo momento qué se encuentra realizando cada estudiante desde su ordenador, lo que permite realizar un seguimiento, con control del tiempo, de la actividad. https://education.minecraft.net
- PADLET: Es una plataforma que facilita la realización de múltiples actividades quedando registradas, el objetivo es que el alumnado adquiera destrezas para un correcto uso de esta plataforma promoviendo la interacción y el pensamiento crítico. El alumnado con TEA podría tener dificultades en la comprensión de las instrucciones, dificultando su participación en las actividades o juegos y dificultades para organizar y procesar la información. Para dar respuesta a esta necesidad, la plataforma contiene estructuras visuales y concretas facilitando la organización de la información. Además, ofrece a los estudiantes la oportunidad de participar, pudiendo elegir ellos cómo desarrollar y planificar la intervención. https://es.padlet.com

– WORDWALL: Es una herramienta educativa versátil y dinámica que se utiliza en el aula para crear una amplia variedad de actividades interactivas y recursos de aprendizaje. Los docentes pueden diseñar de manera sencilla y personalizada juegos, ejercicios y cuestionarios basados en palabras, imágenes, audio o vídeo, adaptados a los objetivos de aprendizaje y a las necesidades de los estudiantes. Ofrece la posibilidad de utilizar plantillas prediseñadas o de crear contenido desde cero, lo que permite a los educadores desarrollar actividades educativas estimulantes y atractivas para motivar la participación. Para estudiantes con TEA, puede ser una herramienta muy útil debido a su capacidad para ofrecer actividades visuales interactivas, estructuradas y repetitivas que facilitan el aprendizaje y la comprensión de conceptos. Además, se pueden personalizar las actividades para adaptarse a las preferencias y necesidades específicas de cada estudiante con TEA, utilizando imágenes familiares o recompensas visuales para reforzar la participación y el aprendizaje. También, ofrece la posibilidad de crear actividades que fomenten la comunicación, la interacción social y la autonomía del estudiante, puede resultar muy útil en el contexto educativo de un estudiante con TEA. https://wordwall.net/es

## Conclusiones

Los recursos digitales constituyen herramientas eficaces para la atención a las necesidades del alumnado con TEA. Uno de los problemas que surgen en este ámbito de la atención a las necesidades educativas es la formación del profesorado, ya que la mayoría de los docentes afirman que no han recibido suficiente formación para investigar y llevar a cabo estas prácticas. “Los profesionales se enfrentan a un doble desafío: saber qué tipo de aplicaciones están disponibles y en qué evidencia se fundamentan para respaldar su uso con el alumnado con TEA”. (Gómez-León, 2024, p. 223). Asimismo, el autor señala que “es fundamental que el conocimiento técnico y/o científico se integre en un proceso de toma de decisiones más amplio que necesariamente se verá afectado por el conocimiento conceptual, la formación práctica y las experiencias y conocimientos del profesional” (p. 235).

A continuación, destacamos algunas reflexiones finales sobre los beneficios del uso de herramientas y recursos digitales para alumnado con TEA, en relación con la gestión del tiempo y la realización de actividades:

- El uso es de gran importancia para este alumnado cuando tienen dificultades en la adquisición, comprensión y expresión del lenguaje pues favorece la interacción con los iguales y con su entorno.
- La accesibilidad de las herramientas y recursos digitales promueve la participación de todos los estudiantes en el aula y especialmente la de estudiantes con TEA que podrían presentar dificultades para interaccionar con los iguales.
- Estimulan la innovación y la creatividad, además de permitir a los docentes el desarrollo de tutoriales en línea para trabajar habilidades y conocimientos que respondan a las necesidades de los estudiantes con TEA.
- Impulsan la creación de comunidades de aprendizaje dentro del aula, donde todo el alumnado puede interactuar para alcanzar un objetivo. El diseño de las explicaciones es más atractivo; los desafíos ponen a prueba las competencias del alumnado; y los biomas y mundos conforman un espacio virtual donde explorar, descubrir y aprender siempre orientados por el docente.
- Facilitan un aprendizaje inmersivo a través de proyectos cuya orientación, seguimiento y control del tiempo de realización de las actividades es seguido por el/la docente.

Para concluir, la mayoría de los estudios describen un impacto evidente del uso de las tecnologías en el proceso de enseñanza-aprendizaje para el alumnado con TEA, "ya que se adaptan al ritmo de aprendizaje de los alumnos y a un aprendizaje individualizado" (Durán, 2021, p. 120). Como señala Gómez-León (2023), "el desafío será mantener el conocimiento sobre qué tecnología está disponible y qué evidencia existe para respaldar el uso de esa tecnología con el alumnado con TEA" (p. 132).

## Financiación

## Referencias bibliográficas

# HERRAMIENTAS Y RECURSOS DIGITALES PARA LA AUTODETERMINACIÓN Y AUTORREGULACIÓN DEL ALUMNADO CON TEA

*Rocío Piñero-Virué*
*Miguel-María Reyes-Rebollo*
Facultad de Ciencias de la Educación. Universidad de Sevilla (España)
*Gloria-Luisa Morales-Pérez*
Escuela Universitaria de Osuna, Sevilla (España)

## Introducción

Las directrices en las que fundamentamos este capítulo se centran en poder acrecentar una adecuada y completa personalidad en sujetos con Trastorno de Espectro Autista (TEA), tomando como piezas fundamentales la autodeterminación como la capacidad para tomar decisiones, establecer metas y controlar sus acciones de forma independiente y la autorregulación implicando la capacidad de gestionar emociones, comportamientos, y pensamientos para alcanzar objetivos específicos. El educando con este tipo de trastorno requiere de un trabajo continuo y coordinado por parte de los profesionales especialistas y de la familia para que resulte exitoso, por tanto, para Hipólito *et al.* (2024) cuestionar y repensar la participación en los ámbitos educativos es un reto de la sociedad actual (p. 70).

Y en este recorrido toman un papel fundamental las herramientas tecnológicas y los recursos digitales como vías de refuerzo para el tratamiento del sujeto TEA (Fernández-Batanero *et al.*, 2023). Para ello, se hace necesario la formación de profesionales y de familias para que puedan continuar el trabajo tanto en el aula como en casa; trabajo que requiere de una alta complicidad entre alumno, profesionales y familias para que se vea potenciado.

En este contexto, las herramientas tecnológicas y los recursos digitales emergen como aliados clave para generar tanto la autodeterminación como la autorregulación en el alumnado con TEA. Estas herramientas facilitan la planificación, la gestión emocional y el aprendizaje adaptativo mediante aplicaciones y programas diseñados específicamente para responder a las necesidades de este colectivo, permiten una personalización del aprendizaje, adaptándose a las necesidades individuales de cada educando. Además, permiten establecer puentes efectivos entre el trabajo en el aula y en el hogar, promoviendo una colaboración estrecha entre profesionales y familias para incrementar el desarrollo integral del estudiante. Desde aplicaciones interactivas que facilitan la comunicación hasta plataformas que fortalecen funciones ejecutivas, estas tecnologías contribuyen significativamente a la mejora de la calidad de vida y a la inclusión social de las personas con TEA y refuerzan funciones ejecutivas clave, como la resolución de problemas, la toma de decisiones y la planificación a largo plazo, o impulsar específicamente en la mejora de habilidades cognitivas, fundamentales para el aprendizaje autónomo y adaptativo de los estudiantes con TEA. Por tanto, estas herramientas y estrategias proporcionan una mejora en la persona, puesto que no sólo en su calidad de vida, sino que también favorecen su inclusión social y su empoderamiento dentro de la comunidad educativa y en su vida diaria.

## Desarrollando una plena autonomía en el alumnado TEA: autodeterminación y autorregulación como ejes para la construcción de la personalidad

La autodeterminación en sujetos con TEA se refiere a su capacidad para tomar decisiones, establecer metas y controlar sus acciones de manera independiente. Este concepto es fundamental para fomentar su participación en la vida cotidiana y promover su calidad de vida. De acuerdo con Andrés-Gárriz *et al.* (2023), comunicación, interacción social y flexibilidad, es posible cultivar su autodeterminación mediante estrategias adaptadas a sus necesidades.

La implicación de las familias y educadores es esencial, brindando oportunidades para que los niños participen en decisiones, tanto en casa como en la escuela (Zambrano-Mendoza & Lescay-Blanco, 2022).

Además, promover la autodeterminación promueve la autoestima y la autonomía, habilidades esenciales para su crecimiento personal. Un entorno que valore sus preferencias y respeten sus elecciones contribuye a que se sientan comprendidos y valorados. Este enfoque no sólo fortalece su independencia, sino también su capacidad para enfrentar los desafíos de manera efectiva.

La autorregulación es un componente esencial en el desarrollo del aprendizaje y la adaptación social en sujetos con TEA. Este proceso, que implica la capacidad de gestionar emociones, comportamientos, y pensamientos para alcanzar objetivos específicos, presenta desafíos particulares en los niños con TEA debido a las características propias del trastorno, como dificultades en la comunicación, en la flexibilidad cognitiva y en la interacción social (Lledó *et al.*, 2023). Sin embargo, al promover estrategias efectivas de autorregulación, se pueden crear entornos más inclusivos y potenciar el aprendizaje autónomo y significativo. La autorregulación se compone de tres dimensiones principales: regulación emocional, conductual y cognitiva. Estas dimensiones son fundamentales para el aprendizaje, ya que permiten a los niños responder de manera adecuada a las demandas del entorno, concentrarse en las tareas y superar desafíos. En la persona con TEA, la regulación emocional es una de las áreas más afectadas. Suelen presentar reacciones desproporcionadas a estímulos sensoriales, frustración frente a cambios en la rutina, y dificultad para comprender y expresar emociones. Estas dificultades pueden derivar en problemas conductuales que interfieren con el aprendizaje y las relaciones sociales (Fernández & Mederos, 2024).

En el contexto educativo, esto incluye habilidades como esperar turnos, seguir instrucciones, y mantener la atención. La persona con TEA frecuentemente presenta déficits en esta área debido a su tendencia a la hiperactividad, comportamientos repetitivos, o intereses restringidos. También, la autorregulación cognitiva incluye habilidades como la planificación, la resolución de problemas y la autoevaluación. Estas habilidades son esenciales para la construcción del aprendizaje autónomo y adaptativo, pero suelen estar afectadas en los niños con TEA debido a sus dificultades en la flexibilidad cognitiva (Gutiérrez *et al.*, 2024). El progreso de la autorregulación en el educando con TEA requiere un enfoque estructurado y adaptado a sus necesidades individuales. Algunas estrategias efectivas incluyen: estrategias visuales, entrenamiento en habilidades sociales y emocionales, ambientes estructurados y predecibles, técnicas de mindfulness y relajación, refuerzo positivo y estrategias de motivación y modelado y andamiaje. Cuando los niños con TEA desarrollan habilidades de autorregulación, experimentan mejoras no sólo en su capacidad para aprender, sino también en su bienestar general y en sus relaciones interpersonales (Rogel *et al.*, 2024). Logran mayor autonomía, lo que les permite participar activamente en su proceso de aprendizaje y alcanzar metas a largo plazo. Además, la autorregulación reduce la incidencia de comportamientos desafiantes, facilitando su integración en entornos escolares y sociales.

## Las herramientas y recursos digitales como vías de apoyo en el ámbito de la autodeterminación y autorregulación del alumnado con TEA

El uso de herramientas y recursos digitales constituye un medio fundamental para desarrollar tanto la autodeterminación como la autorregulación del alumnado con TEA. Estas herramientas no sólo facilitan la comunicación y el aprendizaje, sino que suponen un apoyo a la mejora de habilidades clave para la independencia y la adaptación social. Según Gutiérrez *et al.* (2024), la tecnología educativa permite abordar las necesidades individuales de este colectivo mediante soluciones personalizadas y flexibles, promoviendo así la inclusión y el empoderamiento del alumnado con TEA.

La autodeterminación implica que los individuos con TEA puedan tomar decisiones sobre su vida de forma autónoma y actuar según sus preferencias. Las herramientas digitales han demostrado ser especialmente útiles para este propósito. Aplicaciones de planificación visual, como Choiceworks, permiten organizar sus actividades diarias de forma autónoma, proporcionando estructuras claras y personalizables que se adaptan a sus necesidades específicas (Lledó *et al.*, 2023).

Además, plataformas como Boardmaker y SymbolStix han revolucionado la comunicación aumentativa y alternativa (CAA), permitiendo a los estudiantes con dificultades en el lenguaje expresar sus deseos y necesidades de manera efectiva. Estos sistemas apoyan no sólo la comunicación, sino también el desarrollo de habilidades para la toma de decisiones, fortaleciendo su sentido de control y autonomía.

Un aspecto clave en el uso de estas herramientas es su capacidad para adaptarse a las preferencias y ritmos de cada estudiante. Tecnologías como los dispositivos portátiles o tabletas equipadas con aplicaciones educativas permiten personalizar el contenido, desde horarios visuales hasta juegos interactivos, promoviendo así una mayor participación y motivación en el proceso de aprendizaje (Zambrano-Mendoza & Lescay-Blanco, 2022).

En el ámbito de la autorregulación, las herramientas digitales desempeñan un papel crucial al proporcionar estrategias estructuradas para la gestión de emociones, comportamientos y pensamientos. Aplicaciones como Zones of Regulation ayudan a los estudiantes a identificar y gestionar sus estados emocionales mediante actividades interactivas y retroalimentación inmediata (Fernández & Mederos, 2024). Estas plataformas facilitan la comprensión de las emociones propias y ajenas, promoviendo así habilidades sociales y de comunicación esenciales para su integración en entornos educativos y sociales.

Otro ejemplo es Calm Counter, una herramienta diseñada para ayudar a los estudiantes a manejar episodios de frustración o ansiedad que utiliza técnicas de relajación y mindfulness adaptadas a las necesidades de las personas con TEA, permitiendo una regulación efectiva en situaciones desafiantes. Según Rogel *et al.* (2024), el uso de este tipo de herramientas mejora significativamente la capacidad del alumnado para enfrentarse a demandas cotidianas, reduciendo la incidencia de comportamientos desafiantes.

Las herramientas digitales también pueden apoyar el desarrollo de habilidades cognitivas relacionadas con la autorregulación, como la planificación y la resolución de problemas. Ejemplos como Lumosity y CogniFit ofrecen ejercicios diseñados para fortalecer funciones ejecutivas, lo que resulta especialmente beneficioso para los estudiantes con dificultades en la flexibilidad cognitiva.

El impacto de las herramientas digitales se extiende al hogar, facilitando la continuidad del aprendizaje y el impulso de habilidades. Escobar-Villacrés *et al.* (2023) resalta la importancia de involucrar a las familias en el uso de estas herramientas, formando a los cuidadores, con una aplicación consistente y efectiva de las estrategias en todos los ámbitos. Las tecnologías también impulsan la inclusión mediante la eliminación de barreras comunicativas y el aumento de la participación. Por ejemplo, las pizarras interactivas y las plataformas de aprendizaje gamificadas, como ClassDojo, motivan a los estudiantes a alcanzar sus metas a través de recompensas visuales y retroalimentación positiva. En el ámbito familiar, aplicaciones como See.Touch.Learn. ofrecen recursos adaptados que permiten a las familias trabajar en la mejora de sus habilidades sociales y emocionales, facilitando la identificación de patrones de comportamiento y la implementación de estrategias preventivas, reduciendo así el estrés asociado al manejo de situaciones desafiantes (Zurita-Díaz & Calleja-Reina, 2024).

A pesar de sus ventajas, el uso de herramientas digitales también enfrenta retos significativos, como la falta de accesibilidad y formación adecuada. Muchos educadores y familias carecen de conocimientos sobre cómo integrar estas tecnologías de manera efectiva en sus prácticas diarias. Según Sánchez-Rivas *et al.* (2020), es fundamental diseñar programas de formación que aborden estas brechas y aseguren un uso adecuado de las herramientas digitales. De cara al futuro, la investigación deberá centrarse en el desarrollo de soluciones más inclusivas y accesibles que respondan a las diversas necesidades del alumnado con TEA. Tecnologías emergentes como la IA y la realidad aumentada podrían transformar la manera en que se abordan la autodeterminación y la autorregulación, proporcionando experiencias más inmersivas y personalizadas a las preferencias y contextos del alumnado con TEA (Fernández-Batanero *et al.*, 2024).

### Trabajando desde el aula a casa y, viceversa, para fomentar una adecuada personalidad del alumnado con TEA

La personalidad es un concepto complejo que ha sido estudiado por profesionales durante décadas. Podemos afirmar que cada persona es única, tiene unos rasgos personales que lo hacen diferentes al resto, influyéndole la genética y el ambiente; aunque también existen los trastornos de la personalidad. Por tanto, podemos justificar la necesidad de desarrollar una adecuada personalidad para que el sujeto pueda construir su aprendizaje, estamos argumentando según la RAE (2025) sobre la autodeterminación como la capacidad de una persona para decidir por sí misma algo y, la autorregulación como el hecho de saber determinar las reglas o normas a que debe ajustarse alguien o algo. Y aunque no se considere un rasgo central, se viene detectando que las personas con TEA suelen presentar dificultades significativas en su capacidad de regulación emocional (Zurita-Díaz & Calleja-Reina, 2024).

La autodeterminación es importante porque mejora la calidad de vida, potencia la independencia, aumenta la autoestima, y promueve la inclusión social: facilita la participación en la comunidad y en las relaciones sociales. Ante ello, las personas con TEA se encuentran con una serie de barreras puesto que tienen dificultades en la comunicación, dificultades para adaptarse a cambios o imprevistos y limitación en la capacidad de tomar decisiones espontáneas; y dificultades en las habilidades sociales. En este sentido, se hacen necesarias una serie de estrategias para poder fomentar la autodeterminación, como, por ejemplo: empezar desde la infancia para incentivar la autonomía, enseñar una comunicación efectiva, favorecer la interacción social y la comprensión de las normas sociales, la necesidad de ofrecer un apoyo individualizado, involucrarlos en la toma de decisiones sobre su vida diaria, establecer metas a largo plazo; para lo que se necesitan unos recursos y programas centrados en habilidades para la vida. Y para ello, las familias han de participar (Aguiar *et al.*, 2020) y también requieren recibir una formación a la vez que se le forma a sus hijos, puesto que las familias han de trabajar de manera coordinada, para lo que se aprende en la escuela o en las sesiones con los diferentes profesionales, pueda continuarse en casa, ya que al proporcionar ese sustento, las familias podrán ayudarles en el avance de las habilidades necesarias para tomar decisiones sobre sus propias vidas y alcanzar una mayor independencia y calidad de vida. A ello, cabe resaltar la máxima de Escobar-Villacrés *et al.* (2023) que: debemos recordar que cada niño con TEA es único, y el enfoque debe adaptarse a sus necesidades individuales (p. 85).

Por su parte, para Álvarez-Aguado *et al.* (2024), la autodeterminación es un constructo clave para garantizar el crecimiento de habilidades que permitan a personas con TEA tomar el control de sus vidas. Ello implica estar capacitado de autonomía, libertad, responsabilidad e incluso empoderamiento. Para educandos TEA, se hace necesario recibir una formación que les permita adquirir una serie de hábitos para alcanzar esta autodeterminación. Y para ello, se puede trabajar técnicas que emerjan la autonomía desde la infancia donde se les enseñe desde niños a tomar decisiones adecuadas a su edad y edad mental; desarrollar habilidades para la toma de decisiones, considerando las consecuencias y asumiendo riesgos calculados; cultivar la confianza en uno mismo haciéndoles creer en sus propias capacidades y habilidades; rodearse de personas que los apoyen y respeten sus decisiones; y enseñarles a defender sus derechos. La autodeterminación es un derecho fundamental que nos permite vivir una vida plena y significativa. Al ejercer nuestra capacidad de elegir y tomar decisiones, podemos construir un futuro que sea coherente con nuestros valores y aspiraciones, se posee mayor calidad de vida, independencia, aumento de la autoestima y plena inclusión social.

Para alumnos con TEA es un verdadero desafío el alcanzar esta meta, pues las limitan las dificultades que expusimos anteriormente, y las estrategias podrían servir para reducir las mismas. Por ello, se hace fundamental, el papel de la familia y su adecuada coordinación con los educadores para poder ir trabajando la autodeterminación; estableciendo rutinas claras y visuales, enseñar habilidades para la vida cotidiana, fomentar la toma de decisiones, incrementar habilidades sociales, y promover la independencia en el autocuidado. Y de nuevo, resaltar que cada individuo es único y se habrá de adaptar la enseñanza a sus características, por lo que la coordinación con la familia es fundamental para mantener una comunicación abierta y regular con ella para coordinar las estrategias y reforzar los aprendizajes en el hogar. Siempre teniendo en cuenta, que la autonomía en personas con TEA es un objetivo elemental para mejorar su calidad de vida y su inclusión social, y en este sentido, se han de abordar estrategias específicas en el área social, el área académica y el área de la vida diaria.

## Conclusiones

La personalidad es un constructo complejo que influye en todos los aspectos de nuestra vida. Comprender nuestra propia personalidad y la de los demás puede ayudarnos a relacionarnos mejor con nosotros mismos y con los demás. En este caso, nos centramos en uno de los trastornos de personalidad como es el TEA, considerado como un trastorno del neurodesarrollo que afecta a la comunicación, a la interacción social y al comportamiento (Sauer *et al.*, 2021). Y en este sentido, desde este capítulo, nos hemos centrado en dos factores concretos, la autorregulación y la autodeterminación como ejes fundamentales a emprender en sujetos TEA para que puedan adquirir estos individuos una adecuada personalidad y puedan sentirse incluidos en la sociedad. Por ello, instamos a una implicación y coordinación tanto por parte del colegio como de las familias ya que el hecho de trabajar la autonomía en personas con TEA requiere un enfoque integral que involucre a la escuela, la familia y otros profesionales. Al proporcionar un entorno estructurado y apoyo personalizado, podemos ayudar a las personas con TEA a mejorar en habilidades necesarias para vivir de manera más independiente y satisfactoria. Y en este sentido, se ha de innovar (Fernández-Batanero *et al.*, 2024) e ir incorporando herramientas digitales en la educación para que puedan conformar un proceso que haga emerger un nuevo ambiente educativo, en el que se vienen redefiniendo los roles tradicionalmente asumidos por docentes y estudiantes, siendo la relación didáctica más cercana y comprometida en generar aprendizajes relevantes, funcionales y significativos (Sánchez-Rivas *et al.*, 2020).

## Financiación

## Referencias bibliográficas

# HERRAMIENTAS Y RECURSOS DIGITALES PARA EL OCIO EN EL ALUMNADO CON TEA

*Inmaculada García-Martínez*
Facultad de Ciencias de la Educación. Universidad de Granada (España)
*Antonio Luque de la Rosa*
Facultad de Ciencias de la Educación. Universidad de Almería (España)
*Óscar Gavín-Chocano*
Facultad de Humanidades y Ciencias de la Educación. Universidad de Jaén (España)

## Introducción

El Trastorno del Espectro Autista (TEA) es un conjunto de alteraciones del neurodesarrollo que afectan a la comunicación, la interacción social y los patrones de comportamiento.

El alumnado con TEA está, en base al artículo 73 de la Ley Orgánica 3/2020, de 29 de diciembre (LOMLOE), por la que se modifica la Ley Orgánica 2/2006, de 3 de mayo (LOE), dentro del grupo de alumnado que presenta necesidades educativas especiales. Estas necesidades se materializan en un conjunto de barreras que limitan su acceso, presencia, participación o aprendizaje debido a su condición de discapacidad. Este hecho es lo que determina que vayan a requerir determinados apoyos y atenciones educativas específicas, por un periodo de su escolarización o a lo largo de toda ella, para la consecución de los objetivos de aprendizaje adecuados a su desarrollo. Al hablar del TEA nos estamos refiriendo a una amplia gama de niños muy diversos con necesidades educativas muy distintas. Ello hace muy difícil establecer unas pautas específicas para diseñar intervenciones específicas con el alumnado con TEA.

Este hecho también preocupa si se considera su tendencia ascendente de prevalencia, pues según el propio Ministerio de Educación, Formación Profesional y Deportes (2022), el alumnado diagnosticado con TEA ha aumentado significativamente en la última década, representando el 28,1% del alumnado diagnosticado con Necesidades Específicas de Apoyo Educativo.

El alumnado con TEA presenta cierta singularidad que le hace único. El DSM-V (APA, 2014) incluye dentro del TEA al autismo, al síndrome de Asperger, síndrome de Rett y el trastorno generalizado del desarrollo no especificado. Según los criterios del DSM-IV, la diferencia entre Asperger y autismo está determinada por las habilidades en el lenguaje, donde el primero manifiesta un dominio superior respecto al segundo. El trastorno generalizado del desarrollo no especificado (TGD-NE) a diferencia del resto que se engloban dentro del TEA, no posee unos criterios claros que permitan un diagnóstico claro y preciso. En consecuencia, su diagnóstico se basa en que no se trata de otro trastorno por no cumplir con su sintomatología principal, compartiendo muchas de las manifestaciones y criterios propios asociados al TEA. Dentro de su perfil de necesidades, se observa cómo el ámbito social y emocional y el del lenguaje y la comunicación son los que están más afectados. Este hecho implica que sus relaciones con el entorno están comprometidas. Al respecto, muchos estudios se han centrado en analizar el perfil de este alumnado, señalando que a pesar de la gran variedad de perfiles que se aguardan dentro del TEA, todos ellos poseen un denominador común, su tendencia a reproducir conductas disruptivas, su rigidez mental y su deterioro social y emocional (Hirota y King, 2023; Lord *et al.*, 2020).

Al respecto, el desarrollo de una comunicación funcional desencadena en un fortalecimiento de las habilidades sociales, que propicia una mejor gestión emocional y el planteamiento de conductas adaptativas (Battaglia, 2017). Su intervención temprana garantiza la plena inclusión de este alumnado mediante estrategias que potencien la comunicación funcional y el desarrollo del lenguaje no solo para fortalecer las interacciones cotidianas, sino para reducir las conductas desafiantes que erosionan y fracturan las relaciones que establecen con su entorno (Lindgren *et al.*, 2020).

Examinando la literatura especializada sobre intervención educativa con alumnado con TEA, se ha encontrado diferencias en el enfoque y la metodología utilizada, destacando hasta hace unos años la terapia conductual aplicada (ABA), el entrenamiento en habilidades y las historias sociales. Estas técnicas han demostrado ser efectivas para enseñar rutinas, estructurar actividades y fomentar habilidades comunicativas. No obstante, estudios recientes apuntan hacia una tendencia a incluir herramientas digitales, ya que proporcionan nuevas oportunidades para personalizar y enriquecer las experiencias de aprendizaje.

En esta línea, diversas investigaciones han analizado el impacto que las TIC tienen en el desarrollo socioemocional y del lenguaje en el alumnado con TEA. Por ejemplo, Lane y Radesky (2019) en una revisión sistemática realizada sobre intervenciones con este alumnado encontraron que las herramientas tecnológicas aumentan el interés y la participación del alumnado al integrar elementos visuales y juegos interactivos. Asimismo, otros estudios han apuntado al importante papel que tienen las aplicaciones de comunicación aumentativa y alternativa (CAA) tanto para la expresión verbal, como para despertar su intención comunicativa y habilidades sociales en contextos educativos y recreativos.

Estas cualidades no solo afectan a su vida en el ámbito escolar, sino que trascienden al ámbito familiar y social (Begum y Mamin, 2019), desencadenando en una amplia variedad de estudios dirigidos a mejorar su inclusión plena.

En este contexto, las intervenciones no solo se centran en el uso de materiales analógicos o tradicionales, sino que, gracias al auge de la tecnología educativa, el uso de herramientas y recursos digitales adquiere una relevancia notable, especialmente en el ámbito del ocio y el tiempo libre (Folta *et al.*, 2022). Estas actividades representan una oportunidad para fomentar el desarrollo de habilidades sociales, comunicativas y cognitivas, además de proporcionar un espacio para la inclusión y el disfrute personal. Por tanto, este tipo de herramientas han demostrado ser una oportunidad para contribuir a los principios de individualización y personalización de la enseñanza, pues se adaptan a las necesidades específicas del alumnado con TEA.

Entre los recursos y herramientas digitales que se implementan para el diseño y desarrollo de las intervenciones educativas, sobresalen algunas relacionadas con la lectoescritura (Aguilar-Velázquez *et al.*, 2020), otras con su seguridad (Lázaro Cantabrana *et al.*, 2019) o con la mejora de la comunicación y las habilidades sociales (Díaz Escobar y Nistal Anta, 2020).

Específicamente al ocio, es importante precisar la diferencia entre este constructo y el tiempo libre. De esta manera, el tiempo libre se corresponde con el tiempo en el que una persona no tiene ningún tipo de obligación, ya sea laboral o académica. Por su parte, el ocio atiende a aquel tiempo libre, que la persona emplea para lo que desea (Puig Rovira y Trillas, 1996). Dentro de este enclave, el presente trabajo pretende explorar aquellas herramientas y recursos digitales que el alumnado con TEA puede utilizar en su tiempo de ocio, siempre desde una mirada educativa.

## Método

Este capítulo busca aportar una panorámica sobre las herramientas digitales y recursos tecnológicos que se utilizan para mejorar el ocio y tiempo libre del alumnado con TEA. Para ello, se ha realizado una revisión exhaustiva de la literatura en diferentes bases de datos académicas de prestigio: Scopus, Web of Science y Google Académico con la intención de sistematizar este proceso, cuyos pasos se describen a continuación.

- **Búsqueda bibliográfica**. Para definir la búsqueda se utilizaron las siguientes palabras clave: "TEA", "TIC", "ocio digital" y "aplicaciones para autismo". Además, para aumentar el rango de alcance de la búsqueda se utilizaron los operadores booleanos "AND" y "OR" en las diferentes búsquedas.
- **Selección de estudios relevantes**. Con vistas a acotar la búsqueda se plantearon dos criterios de inclusión principales: a) trabajos que incluyesen alguna aplicación o recurso digital específico para el alumnado con TEA o que pudiesen ser utilizados por ellos; y b) que las aplicaciones o herramientas digitales pudieran utilizarse con fines de ocio, directa o indirectamente. Por tanto, se desecharon aquellas aplicaciones y recursos que, aunque de utilidad en otros contextos educativos y/o terapéuticos, no eran aplicables al ocio y tiempo libre o que no pudieran ser utilizadas por alumnado con TEA.
- **Análisis de herramientas digitales**. Tras el proceso de revisión y filtrado, finalmente se identificaron aquellas aplicaciones y recursos específicos que cumplían tanto con los criterios establecidos como con el objetivo de la presente investigación y se procedió a su evaluación considerando para ello los siguientes parámetros: accesibilidad, propósito, coste y desarrollador.
- **Revisión crítica**. Los recursos seleccionados fueron sometidos a una revisión crítica, contrastándolos con la literatura científica disponible para garantizar su pertinencia y efectividad.

## Resultados

Tras el proceso de revisión, se encontraron nueve aplicaciones que se sintetizan a continuación en la tabla 1, considerando los siguientes indicadores: aplicación, accesibilidad, propósito, idioma, coste y desarrollador.

Tabla 1. Aplicaciones y recursos digitales para el ocio con alumnado con TEA

| *Aplicación* | *Accesibilidad* | *Propósito* | *Idioma* | *Coste* | *Desarrollador* |
|---|---|---|---|---|---|
| Pictello | Interface visual con imágenes y audio. No permite la personalización del contenido. | Herramienta para crear historias sociales personalizadas. | Español e inglés | Sí | Assistive Ware. |
| Social Express | Interfaz sencilla, aunque las transiciones entre actividades pueden resultar confusas. | Plataforma que ayuda a practicar interacciones sociales en entornos simulados. | Inglés | Sí | The Language Express, Inc. |
| Learn with Rufus | Interfaz sencilla y clara con imágenes grandes y botones de fácil acceso. | Aplicación para enseñar emociones y expresiones faciales. | Inglés | Sí | Rufus Robot, Inc. |
| Choiceworks | Interfaz visualmente sencilla, adaptable a las necesidades del usuario en cuanto a imágenes y texto. | Herramienta para estructurar rutinas diarias mediante pictogramas. | Español e inglés | Sí | Bee Visual, Inc. |
| Toontastic 3D | Interfaz altamente interactiva con gestos simples para manipular personajes y elementos. | Aplicación de animación que fomenta la creatividad y la narración. | Multilingüe | No | Google LLC. |
| Calm Counter | Interfaz sencilla y accesible con instrucciones clara y debidamente secuenciadas. | Recurso para gestionar emociones y regular el comportamiento. | Inglés | Sí | Touch Autism |
| Proloquo-2Go | Interfaz intuitiva con opciones de personalización y adaptabilidad a diferentes niveles de habilidad. | Comunicación aumentativa mediante pictogramas y voz. | Multilingüe | Sí | AssistiveWare, especialista en tecnologías para la accesibilidad. |
| Avaz | Interfaz clara y sencilla con la opción de personalizar el contenido según las necesidades del usuario. | Sistema visual interactivo para facilitar la comunicación. | Multilingüe | Sí | Avaz Inc., reconocido en aplicaciones para personas con TEA. |
| Autism Core Skills | Interfaz sencilla con instrucciones claras y actividades secuenciales. Diseño poco atractivo. | Desarrollo de habilidades sociales y comunicativas a través de juegos. | Inglés | Sí | Infiniteach, expertos en soluciones educativas. |

## Discusión y resultados

Las nueve aplicaciones halladas en la búsqueda ofrecen grandes beneficios para el alumnado con TEA, abriéndole un amplio espectro de oportunidades para explorar, aprender y desenvolverse de manera más autónoma.

En líneas generales, puede concluirse que están centradas en la mejora de la comunicación y el lenguaje, pues esta habilidad es esencial para garantizar la plena inclusión y participación del alumnado con TEA, en distintos escenarios y momentos.

Todas ellas están diseñadas o adaptadas para el alumnado con TEA, encontrándose diferentes niveles de dificultad. De esta manera, se observa como de las diez aplicaciones descritas, la aplicación AVAZ es la que destaca más por su accesibilidad y es la más recomendada para edades tempranas, siendo más "incompleta" para alumnos con TEA de mayor edad o aquellos con mayor dominio del lenguaje, a comparación de otras como Proloquo2Go. Con vistas a organizar el tiempo del alumnado con TEA y reducir la ansiedad que generan los cambios en su rutina, destacan las aplicaciones Choiceworks y Pictello, pues ayudan a estructurar sus actividades diarias de una forma clara y secuencial para alcanzar determinados propósitos y les hace tener la sensación de poseer control sobre lo que hacen, hecho que les reconforta. Pese al gran potencial de ambas, es destacable la variedad de opciones de accesibilidad y personalización que presenta Choicework. Esta aplicación está estructura en torno a tres aspectos fundamentales: rutinas diarias, gestión emocional y tareas pendientes; permitiéndoles así organizar y gestionar su tiempo libre y anticipar sus tareas, sin dejar de lado el componente emocional. Para ello, la aplicación le ofrece una sección diseñada para que el usuario pueda identificar sus emociones en un momento dado y en función de la emoción expresada, le sugiere ciertas estrategias para autorregularse.

También enfocado en la interacción social con vistas no solo a reforzar uno de los ámbitos de desarrollo más deteriorados de este alumnado, se encuentran recursos como Learn with Rufus y Social Express, que contribuyen a mejorar tanto la comprensión emocional, como las habilidades sociales de este alumnado en diferentes escenarios. A través de una variabilidad de funcionalidades en entornos simulados, los usuarios pueden trabajar el componente emocional a partir de la identificación de las diferentes expresiones faciales, tonos de voz y gestos que indican emociones o intenciones en otros, así como la comprensión de las emociones, tanto las suyas propias como las de los demás. Además, un aspecto destacable es que dichas herramientas ofrecen diferentes situaciones hipotéticas para que los usuarios aprendan, no solo a gestionar sus propias emociones y a autorregularse en determinados contextos, sino que también sean conscientes de cómo sus emociones influyen en las decisiones que van tomando tanto a nivel intrapersonal como interpersonal. Por último, destaca la herramienta Toontastic 3D, que aporta el componente lúdico, a través del fomento de la creatividad y el aprendizaje interactivo. Los usuarios a partir de un formato de historia predefinido, bien sea de aventura, comedia o misterio, pueden crear una historia personalizada en cuanto a personajes y animación. Dentro de la propia aplicación, pueden elegir entre una gran variedad de personajes como animales, personas o criaturas fantásticas, así como diferentes escenarios en 3D o si lo prefieren, diseñar y dibujar los suyos propios, aspecto que fomenta aún más si cabe la creatividad. Además, ofrece la opción de poder animar a los personajes utilizando diferentes gestos y expresiones faciales y grabando diálogos para que sus personajes cobren vida.

Por tanto, puede afirmarse que la incorporación de herramientas digitales en las intervenciones educativas con el alumnado con TEA no solo facilita la superación de barreras comunicativas y sociales, sino que también abre un abanico de oportunidades para potenciar sus habilidades. De manera general, estas aplicaciones pueden ofrecer un entorno seguro y estructurado donde los estudiantes pueden explorar, aprender y desarrollarse a su propio ritmo, corroborándose dos de los principios fundamentales de la educación inclusiva, la individualización y personalización de la enseñanza. La posibilidad de personalizar sus experiencias de aprendizaje es una garantía de que el alumnado va a recibir el apoyo que precisa en cada momento, lo que le permitirá que alcance su máximo potencial.

No obstante, a pesar de los múltiples beneficios que puede reportar el uso de este tipo de aplicaciones y recursos con el alumnado con TEA, es preciso señalar los inconvenientes que conllevan. Entre ellos, destaca que la mayoría de las herramientas y recursos seleccionados son de pago, lo que restringe su uso a todo el mundo por cuestiones económicas. Además, algunos de los recursos incluidos requieren dispositivos específicos, hito que limita su accesibilidad o incluso, en su versión gratuita, ciertas funcionalidades dentro de la aplicación requieren de un pago previo para poder disfrutar de un aprovechamiento total de las opciones avanzadas de las que dispone la propia aplicación. Por otro lado, la aplicación de estas herramientas, pese a estar todas ellas diseñadas para ser accesibles a todos los usuarios, en determinadas ocasiones requieren de formación previa y supervisión para poder utilizarlas de forma óptima y sacarles el mayor rendimiento posible dentro de las posibilidades que ofrecen.

Además de lo anterior, es importante puntualizar que el uso de estos dispositivos deben realizarse desde un prisma responsable y seguro pues, pese a los beneficios que pueden tener para conseguir el desarrollo integral del alumnado, un uso inadecuado o excesivo puede desencadenar en relaciones de dependencia, produciéndose efectos contrarios a lo esperado, en materia de inclusión plena, consolidación de las relaciones sociales e interacción física con los diferentes entornos y agentes con los que se relaciona. Para mitigar estas cuestiones, siempre que se implementen intervenciones de cualquier tipo con este alumnado, se recomienda que exista cierto equilibrio entre el uso de dispositivos electrónicos y el planteamiento de actividades no digitales que potencien la interacción personal.

Finalmente, estas herramientas destacan por su capacidad para adaptarse a las preferencias y necesidades individuales del alumnado, posicionándose como una alternativa idónea para mejorar el ocio y la inclusión del alumnado con TEA. Además, también fomentan la inclusión en actividades de ocio y tiempo libre, contribuyendo al desarrollo integral del alumnado con TEA. El mero hecho de integrar la tecnología, que resulta atractiva para el alumnado, y promover la creatividad en el proceso educativo y recreativo, contribuye a la creación de un espacio donde las diferencias se convierten en fortalezas, enriqueciendo tanto a los estudiantes como a sus comunidades escolares y familiares.

No obstante, su implementación debe ser planificada y monitorizada, de manera que forme parte de una estrategia más amplia que contemple tanto los aspectos educativos como recreativos.

## Financiación

## Referencias bibliográficas

# HERRAMIENTAS Y RECURSOS DIGITALES COMO APOYO A LAS FAMILIAS DEL ALUMNADO CON TEA

*Pedro Román-Graván*
Facultad de Ciencias de la Educación. Universidad de Sevilla (España)
*Eloy López-Meneses*
Universidad Pablo de Olavide (España)
*Carmen Siles Rojas*
Facultad de Ciencias de la Educación. Universidad de Sevilla (España)

## Introducción

El papel de las familias es fundamental en la atención y desarrollo de los estudiantes con Trastorno del Espectro Autista (TEA). Su participación no solo influye positivamente en el progreso del niño o de la niña, sino que también mejora la calidad de vida de todo el núcleo familiar.

Esta implicación de la familia tiene que ser en todas las fases del proceso, desde el diagnóstico y el tratamiento, y es esencial para el desarrollo óptimo del niño/a con TEA (Baña Castro, 2015). Esta colaboración permite una comprensión más profunda de las necesidades de la persona TEA y facilitará la implementación de estrategias efectivas tanto en el hogar como en entornos educativos.

Pero la familia no solo proporciona apoyo emocional y/o físico, sino que también influye en la toma de decisiones y calidad de vida del individuo con TEA. Baña Castro (2015) destaca que "la familia pasa a ser el principal y más permanente apoyo para el individuo, de su actuación van a depender muchas de las expectativas, posibilidades y bienestar de la persona" (p. 325). Un entorno familiar que promueve la independencia y la toma de decisiones contribuye significativamente al desarrollo personal del niño o niña TEA.

En ocasiones, el diagnóstico de este tipo de trastornos en un miembro de la familia puede generar una crisis que afecta la dinámica familiar. Giné (2001) señala que el nacimiento de un hijo con TEA provoca siempre, en mayor o menor medida, una crisis que se caracteriza por:

- un fuerte impacto psicológico y emocional para toda la familia,
- un proceso de adaptación y redefinición del funcionamiento familiar,
- cambios en la relación de pareja que tiene la guarda y custodia, y
- la necesidad de ayuda y de asesoramiento.

Por esta razón, es crucial que las familias reciban el apoyo necesario para adaptarse a esta nueva realidad y puedan recibir todo el sustento posible.

La relación entre la familia, los profesionales de la salud y la educación es vital para el desarrollo del niño/a con TEA. Una comunicación abierta y constante permitirá el diseño y puesta en marcha de estrategias efectivas en todos los entornos donde el niño/a interactúa. Además, la inclusión de la familia en programas de intervención temprana ha demostrado ser beneficiosa para el desarrollo de habilidades comunicativas y sociales en niños con TEA (Bejarano Martín, 2019).

Recientemente, las herramientas y recursos digitales han emergido como aliados fundamentales para empoderar y apoyar a las familias de personas con TEA. Estas tecnologías facilitan el acceso a información sobre el propio trastorno, mejoran la comunicación y proporcionan plataformas de apoyo, contribuyendo significativamente al bienestar familiar.

Las plataformas digitales ofrecen a las familias acceso inmediato a información actualizada sobre el TEA, qué estrategias de intervención pueden empezar a realizar y recursos educativos a los que pueden acceder. Por ejemplo, la Fundación Orange ha desarrollado aplicaciones específicas que permiten a los niños/as y adultos con TEA reproducir texto o pictogramas sin necesidad de leerlos, facilitando la comunicación y el aprendizaje (Fundación Orange, s.f.). Este acceso a recursos tan especializados permite a las familias comprender mejor el autismo y sus características, lo que es esencial para proporcionar un apoyo adecuado a sus hijos (Movimiento Azul, s.f.).

Las aplicaciones móviles y herramientas digitales han revolucionado la forma en que las familias se comunican con sus hijos con TEA. Por ejemplo, aplicaciones como Proloquo2Go (disponible para Apple y no para Android: https://apps.apple.com/es/app/proloquo2go/id308368164) y Choiceworks (disponible igualmente para Apple y no para Android: https://apps.apple.com/es/app/choiceworks/id486210964) ofrecen herramientas visuales que facilitan la comunicación y la organización de actividades diarias, adaptándose a las necesidades únicas de cada individuo.

Estas herramientas no solo promueven la inclusión, sino que también fomentan la conexión entre individuos que comparten experiencias similares. Las comunidades en línea proporcionan a las familias espacios para compartir experiencias, obtener asesoramiento y recibir apoyo emocional. Estas plataformas facilitan el acceso a recursos educativos, comunidades de apoyo y terapias innovadoras que permiten a las personas con autismo explorar su potencial en un entorno seguro y comprensivo (Movimiento Azul, s.f.). Además, la participación en grupos de apoyo en línea puede reducir el aislamiento y proporcionar estrategias prácticas para el manejo diario del TEA.

A continuación, relacionamos diferentes páginas web de asociaciones TEA de diferentes comunidades autónomas que pueden ser de interés:

- Federación Autismo Madrid (Comunidad de Madrid), entidad sin ánimo de lucro destinada a mejorar la calidad de vida de las personas con TEA y sus familias en la Comunidad de Madrid. https://www.autismomadrid.es
- Federación Autismo Andalucía (Andalucía), federación que agrupa a diversas asociaciones andaluzas dedicadas al apoyo de personas con TEA y sus familias. https://www.autismoandalucia.org
- Federación Autismo Castilla y León (Castilla y León), organización que coordina y representa a las asociaciones de autismo en Castilla y León, ofreciendo recursos y apoyo a la comunidad TEA. https://www.autismocastillayleon.org
- Federación Autismo Comunidad Valenciana (Comunidad Valenciana), federación que trabaja para mejorar la calidad de vida de las personas con TEA en la Comunidad Valenciana, agrupando a diversas asociaciones locales. https://www.autismocv.org
- Federación Autismo Extremadura (Extremadura), su objetivo principal es coordinar, impulsar y potenciar las actividades de sus miembros, promoviendo la creación y consolidación de entidades que defiendan los derechos de las personas con TEA en la comunidad autónoma de Extremadura. https://autismoextremadura.org

## Retos y necesidades de las familias de alumnado con TEA

Las familias de niños/as TEA desempeñan un papel fundamental en el desarrollo y bienestar de sus hijos, enfrentándose a numerosos y diarios desafíos. Estas dificultades abarcan aspectos fundamentales como la comunicación, el acceso a recursos y el manejo de conductas, además de necesidades específi-cas en los ámbitos educativo, social y emocional. Para abordar estas demandas de manera efectiva, resulta esencial establecer una intervención colaborativa entre las familias y los profesionales, y que esta colaboración permita ofrecer un apoyo integral y coherente.

En este apartado, se describen los principales retos que las familias enfrentan en su día a día, destacando cómo estos afectan a su calidad de vida y a la de sus hijos. Asimismo, se analizan las necesidades específicas que surgen en distintos ámbitos de la vida del alumnado con TEA, subrayando la importancia de una intervención adaptada y centrada en sus particularidades. Finalmente, se pone de relieve la relevancia de la colaboración activa entre las familias y los profesionales, como un eje fundamental para diseñar y aplicar estrategias que favorezcan el desarrollo integral y la inclusión social del alumnado con TEA.

### Principales desafíos en el día a día: comunicación, acceso a recursos, manejo de conductas, entre otros

Las familias de niños con TEA se enfrentan a múltiples desafíos en su vida diaria, especialmente en áreas como la comunicación, el acceso a recursos (no solo educativos sino de diversa índole), y en el manejo de conductas. Estos retos impactan significativamente la dinámica familiar y requieren estrategias específicas para abordarlos.

La comunicación es uno de los principales desafíos para las familias de niños con TEA. Los niños con este trastorno suelen presentar dificultades en el desarrollo del lenguaje y en la comprensión de las interacciones sociales, lo que puede generar frustración tanto en ellos como en sus familias. Según el Instituto Nacional de la Sordera y Otros Trastornos de la Comunicación (NIDCD), los niños con un trastorno del espectro autista generalmente están ensimismados y parecen vivir en un mundo privado en el que tienen una habilidad limitada de comunicarse y de interactuar bien con los demás (NIDCD, s.f.), esta limitación en la comunicación puede llevar a malentendidos y dificultades en la convivencia diaria.

El acceso a recursos educativos y terapéuticos adecuados es esencial para el desarrollo de los niños con TEA. Sin embargo, muchas familias tienen serias dificultades para obtener estos servicios, ya sea por limitaciones económicas, falta de información o escasez de profesionales especializados. Un informe de Save the Children destaca que el costo promedio mensual de criar a un niño en España ha aumentado a 758 euros, un 13% más que en 2022, lo que agrava la situación económica de las familias y dificulta el acceso a recursos necesarios (El País, 2024). Además, la falta de personal especializado en centros educativos, como Auxiliares Técnicos Educativos (ATE) (Varios Autores, 2016; Clavijo Gamero & Fernández González, 2022), complica la integración y el desarrollo adecuado de estos niños en el entorno escolar (Cadena SER, 2024).

Los niños con TEA pueden exhibir conductas desafiantes, como comportamientos repetitivos, resistencia a cambios en la rutina y dificultades para regular sus emociones. Estas conductas pueden generar estrés en el entorno familiar y requerir intervenciones específicas para su manejo. La Fundación Conectea señala que el autismo está interfiriendo en las relaciones interpersonales. Las familias de niños con autismo enfrentan varios tipos de desafíos. El desafío comienza temprano y dura toda la vida (Fundación Conectea, 2020). La implementación de estrategias de intervención conductual y el apoyo de profesionales especializados son fundamentales para abordar estos comportamientos y mejorar la calidad de vida de las familias.

Los desafíos mencionados no solo afectan la dinámica interna de la familia, sino que también tienen repercusiones en su vida social y emocional. El estrés constante, la necesidad de adaptar rutinas y la posible falta de comprensión por parte de la sociedad pueden llevar al aislamiento y a problemas de salud mental en los cuidadores. Es esencial que las familias reciban apoyo psicológico y social para enfrentar estos retos y mantener un equilibrio en su bienestar.

## Necesidades específicas en el ámbito educativo, social y emocional

Los estudiantes con TEA presentan necesidades específicas en los ámbitos educativo, social y emocional y éstas requieren intervenciones personalizadas para facilitar su desarrollo integral. En el entorno educativo, es fundamental que las familias entiendan que los docentes tengan que adaptar sus enseñanzas a las habilidades y necesidades individuales de cada estudiante con TEA. Esto implicará llevar a cabo estrategias que promuevan una educación inclusiva y de calidad (Fernández-Cerero y Román-Graván, 2024), avanzando en el desarrollo de la legislación educativa y flexibilizando la oferta educativa existente para responder a las particularidades de estos estudiantes. Además, es esencial promover la educación emocional dentro del aula y adaptar las actividades para facilitar la participación directa de los alumnos con TEA.

En el ámbito social, las familias deben asimilar que los alumnos con TEA suelen enfrentarse a desafíos en la interacción con sus pares y en la comprensión de las normas sociales. La implementación de programas de intervención en habilidades sociales y emocionales es crucial para mejorar estas competencias, facilitando su integración y participación en entornos sociales diversos. Además, es importante fomentar entornos que promuevan la inclusión y la comprensión de las diferencias individuales, contribuyendo a una convivencia más armoniosa y respetuosa.

En el ámbito emocional, es esencial que las familias entiendan, comprendan y asimilen las limitaciones y dificultades que los estudiantes con TEA pueden enfrentarse en su vida diaria, para poder desarrollar estrategias y habilidades que les ayuden a regular sus emociones y mejorar su bienestar emocional y el de las personas con TEA. La implementación de programas de intervención en inteligencia emocional que trabajen emociones como la alegría, tristeza, miedo, asco y enfado, utilizando métodos como pictogramas y juegos, puede ser beneficiosa para este propósito.

## Importancia de una intervención colaborativa entre familias y profesionales

La intervención colaborativa entre familias y profesionales es esencial para promover el desarrollo integral y la calidad de vida de las personas con TEA. Este enfoque conjunto permite una comprensión más profunda de las necesidades individuales y facilita la implementación de estrategias efectivas en diversos entornos.

La cooperación entre familias y profesionales en la intervención de personas con TEA ofrece múltiples ventajas. Según la AETAPI (Asociación Española de Profesionales del Autismo, 2021), los nuevos modelos de intervención ponen de relieve la importancia de la colaboración entre familias y profesionales para promover la calidad de vida de las personas con TEA. Esta colaboración asegura que las intervenciones sean coherentes y adaptadas a las necesidades específicas de cada individuo, tanto en el hogar como en entornos educativos y comunitarios.

Cuando la familia y los terapeutas trabajan juntos, se garantiza que las estrategias y herramientas utilizadas en las sesiones terapéuticas también se implementen en el hogar y otros entornos. Esto ayuda a la generalización de habilidades, es decir, que el niño/a pueda aplicarlas en diferentes contextos. De esta manera, las intervenciones son más efectivas cuando hay coordina-ción entre los profesionales y la familia. Los padres pueden reforzar en casa lo aprendido en terapia, favoreciendo el desarrollo del lenguaje, la comunicación, la autonomía y la regulación emocional. Por lo tanto, un enfoque colaborativo permite que las familias entiendan mejor las necesidades de su progenitor/a y cómo pueden ayudarlo/a. Esto reduce la frustración, el estrés y mejora la calidad de vida tanto del niño/a como de su entorno familiar (Universidad de Tucumán, 2021).

Una comunicación abierta y constante entre familias y profesionales es esencial para el éxito de las intervenciones. Además, proporcionar formación y recursos a las familias les permite comprender mejor el TEA y participar de manera más efectiva en el proceso de intervención. La Asociación Navarra de Autismo (APANAG) enfatiza la importancia de programas de intervención con familias en la primera infancia para organizar contextos que optimicen el desarrollo infantil (APANAG, s.f.).

## Herramientas y recursos digitales disponibles

### Aplicaciones móviles y plataformas digitales para familias

Las aplicaciones móviles y plataformas digitales se han convertido en herramientas esenciales no solo para los educadores sino también para las familias de personas con TEA, ofreciendo recursos que facilitan la comunicación, el aprendizaje y la inclusión social. Estas tecnologías proporcionan soluciones innovadoras y adaptables a las necesidades individuales de cada persona con TEA, permitiendo a las familias acceder a estrategias de intervención y apoyo de manera más efectiva.

Existen diversas aplicaciones diseñadas específicamente para apoyar a las familias y personas con TEA en diferentes áreas:

– **LetMeTalk** (http://letmetalk.info, https://www.autismohuelva.org/aplicacion-let-me-talk).
  Aplicación gratuita de comunicación aumentativa y alternativa (CAA) diseñada para ayudar a personas con dificultades en el habla o el lenguaje a comunicarse a través de pictogramas, imágenes y texto. Está especialmente pensada para niños y niñas con TEA) personas con parálisis cerebral, afasia, síndrome de Down o cualquier otra condición que afecte la capacidad de comunicación verbal. También puede ser útil en contextos educativos inclusivos, para facilitar la interacción y participa-ción del alumnado con necesidades comunicativas. La aplicación incluye más de 9 000 pictogramas del portal ARASAAC y permite al usuario crear frases combinando imágenes y palabras. Además, se pueden añadir fotografías personalizadas, adaptar el vocabulario y organizar el contenido en carpetas temáticas, lo que favorece su uso tanto en el hogar como en la escuela. No requiere conexión a internet ni registro, lo que la convierte en una herramienta accesible, intuitiva y versátil para docentes, familias, logopedas y terapeutas. En definitiva, se trata de un recurso muy valioso para fomentar la autonomía comunicativa y la inclusión de personas no verbales o con lenguaje emergente.

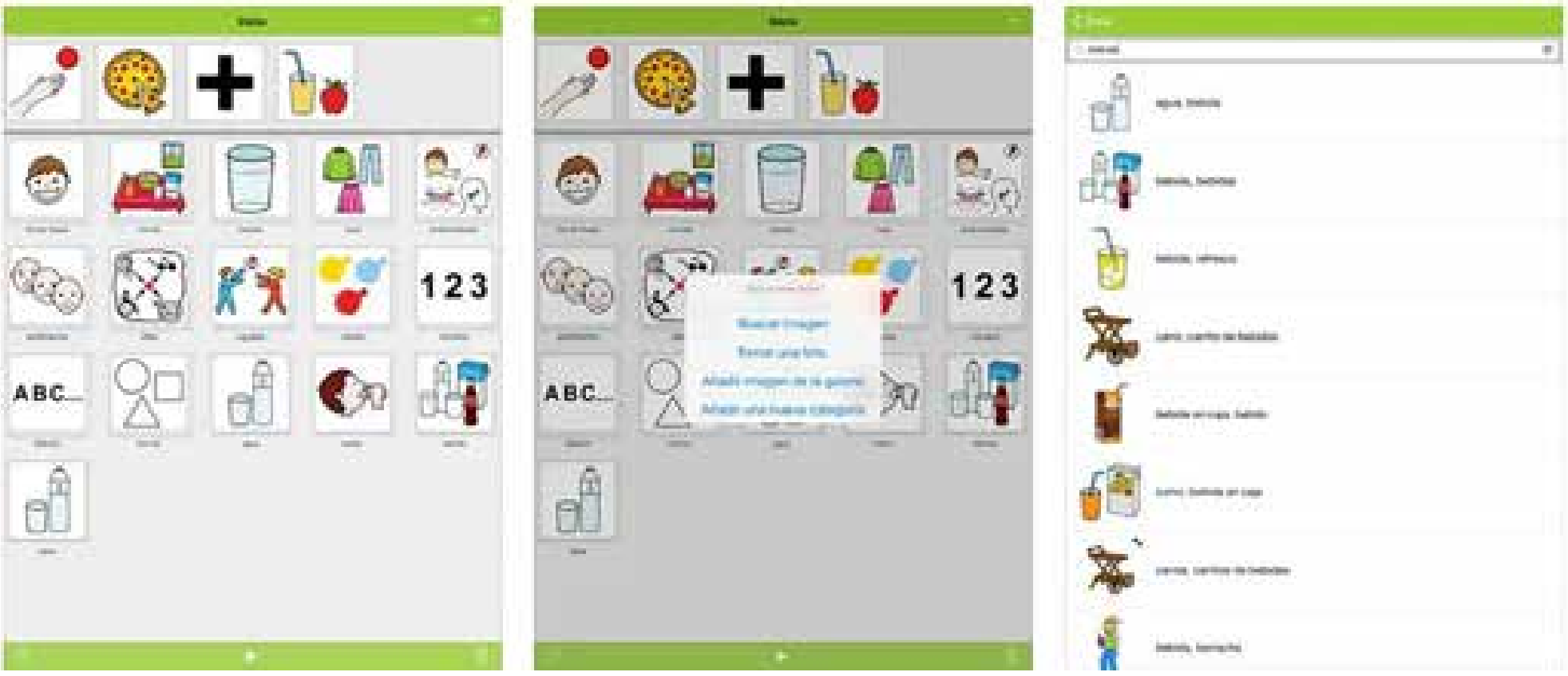

– **Proloquo2Go** (https://www.assistiveware.com/es/productos/proloquo-2go).
  Esta aplicación de comunicación aumentativa y alternativa está diseñada para personas que tienen dificultades para hablar. Ofrece una interfaz personalizable que permite a los usuarios comunicarse mediante símbolos y texto, facilitando la expresión de necesidades y emociones. Está disponible exclusivamente para dispositivos iOS, como iPhone y iPad, no existe una versión oficial para Android.

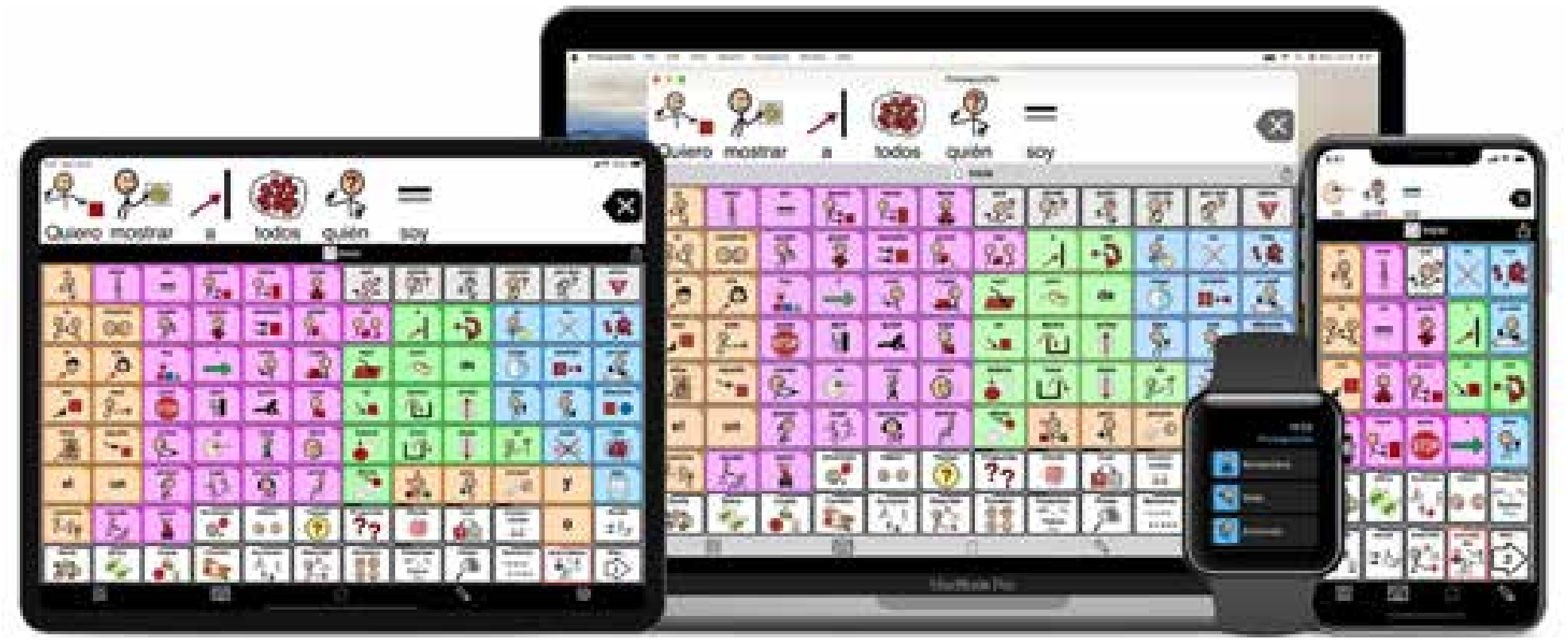

– **TEAyudo a Jugar** (https://fundacionorange.es/aplicaciones/teayudo-a-jugar).
Desarrollada por la Fundación Orange y la Universidad de Murcia, esta aplicación tiene como objetivo favorecer la inclusión social de los niños con autismo u otros trastornos del desarrollo. Propone momentos de juego de manera visual, facilitando la interacción y el desarrollo de habilidades sociales. La aplicación está disponible tanto para Android como para iOS.

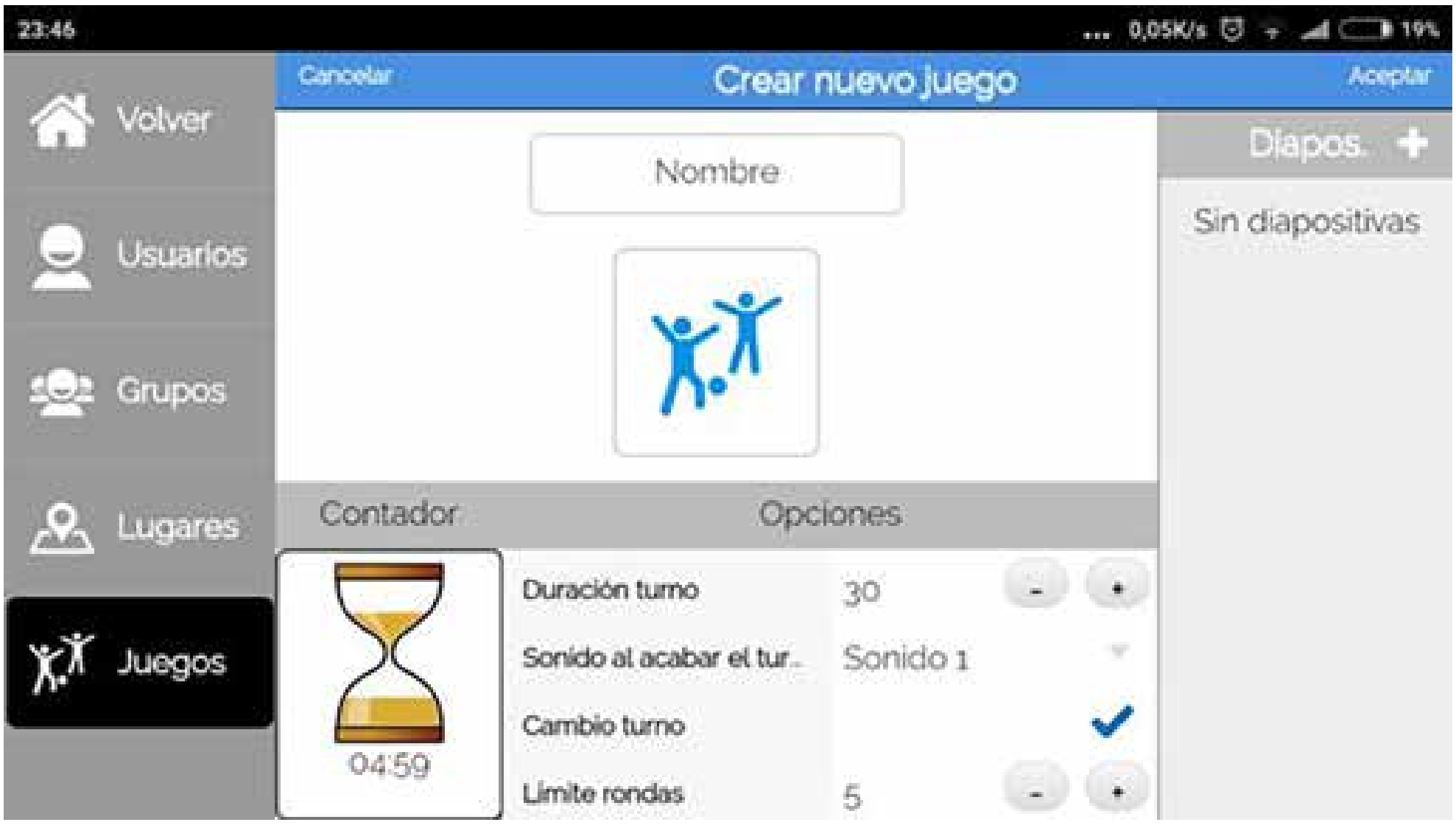

- **Choiceworks** (https://www.beevisual.com).
  Aplicación educativa que ayuda a niños y niñas con TEA a desarrollar rutinas, gestión emocional y habilidades de espera mediante tableros visuales. Permite personalizar horarios, estrategias de afrontamiento y temporizadores para fomentar la autonomía y autorregulación. Sus herramientas incluyen un tablero de rutinas, un tablero de emociones y un tablero de espera, facilitando la organización y el control de impulsos. Disponible en iOS, es una de las apps más recomendadas para apoyar el desarrollo de habilidades diarias en niños con necesidades especiales.

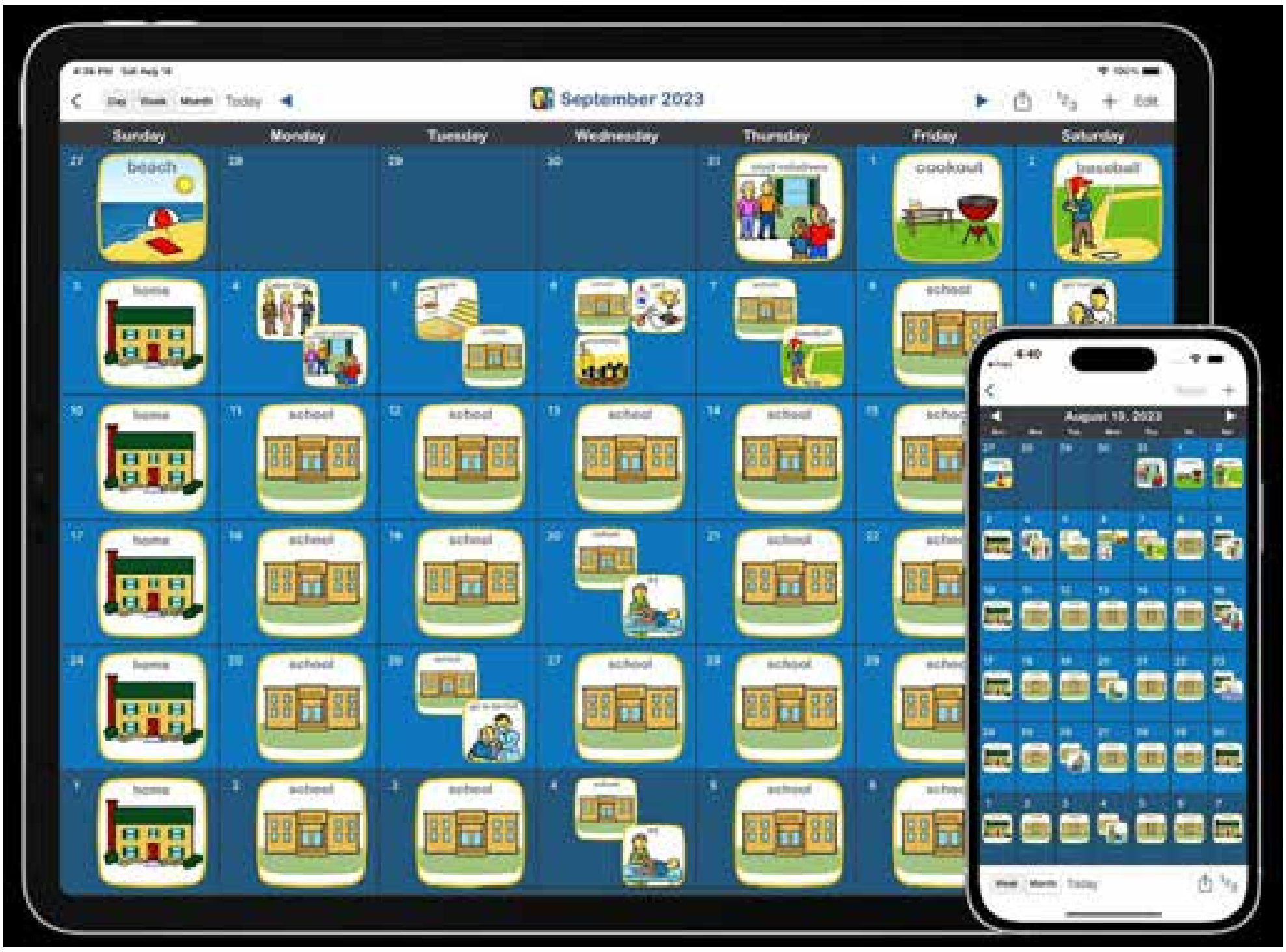

- **Día a Día** (https://fundacionorange.es/aplicaciones/dia-a-dia).
  Es un diario visual diseñado para personas con autismo o dificultades de comunicación. Permite al niño registrar y revisar sus actividades diarias de manera gráfica y estructurada, facilitando la anticipación de eventos y fomentando la comunicación sobre sus experiencias. Esta disponible tanto para Android como para iOS.

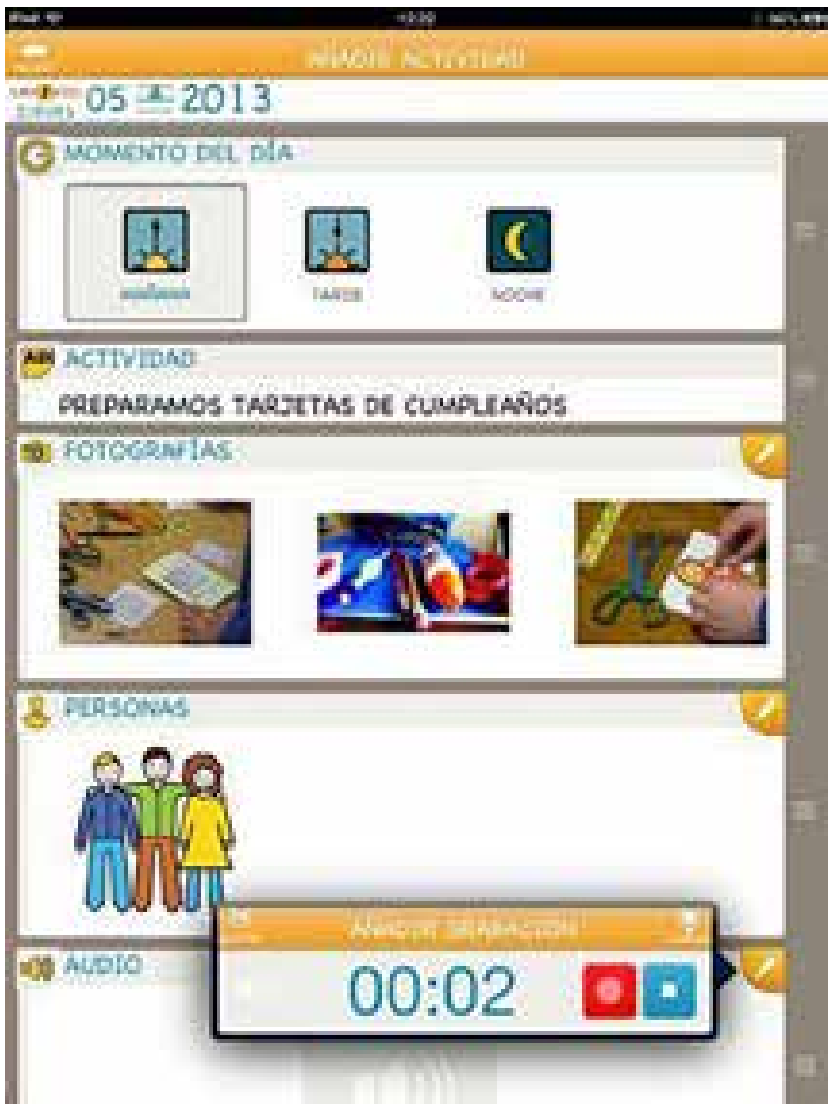

– **DictaPicto** (https://fundacionorange.es/aplicaciones/dictapicto-tea). Convierte el lenguaje oral en información visual en tiempo real, transformando frases en pictogramas. Esto mejora la comprensión del entorno para personas con TEA, facilitando la anticipación y secuenciación de actividades cotidianas. También está disponible para Android y para iOS.

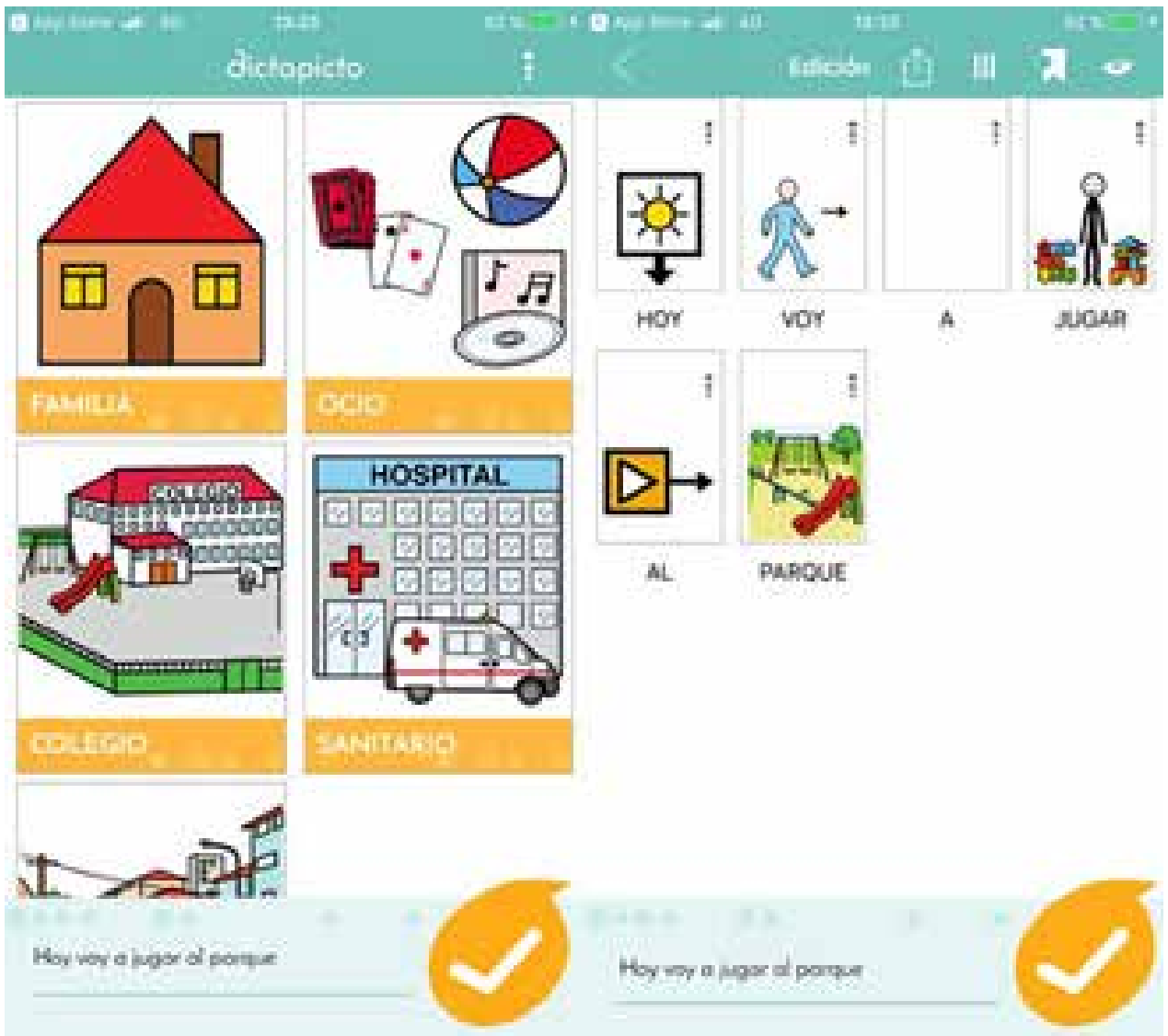

- **Pictogram Agenda** (https://www.pictogramagenda.es)
  Es una agenda visual electrónica que ayuda a las personas con TEA a comprender y anticipar las actividades diarias. Al reducir la ansiedad frente a lo nuevo o inesperado, facilita la adaptación a diferentes situaciones y promueve la autonomía. Está disponible para iOS y para dispositivos Android.

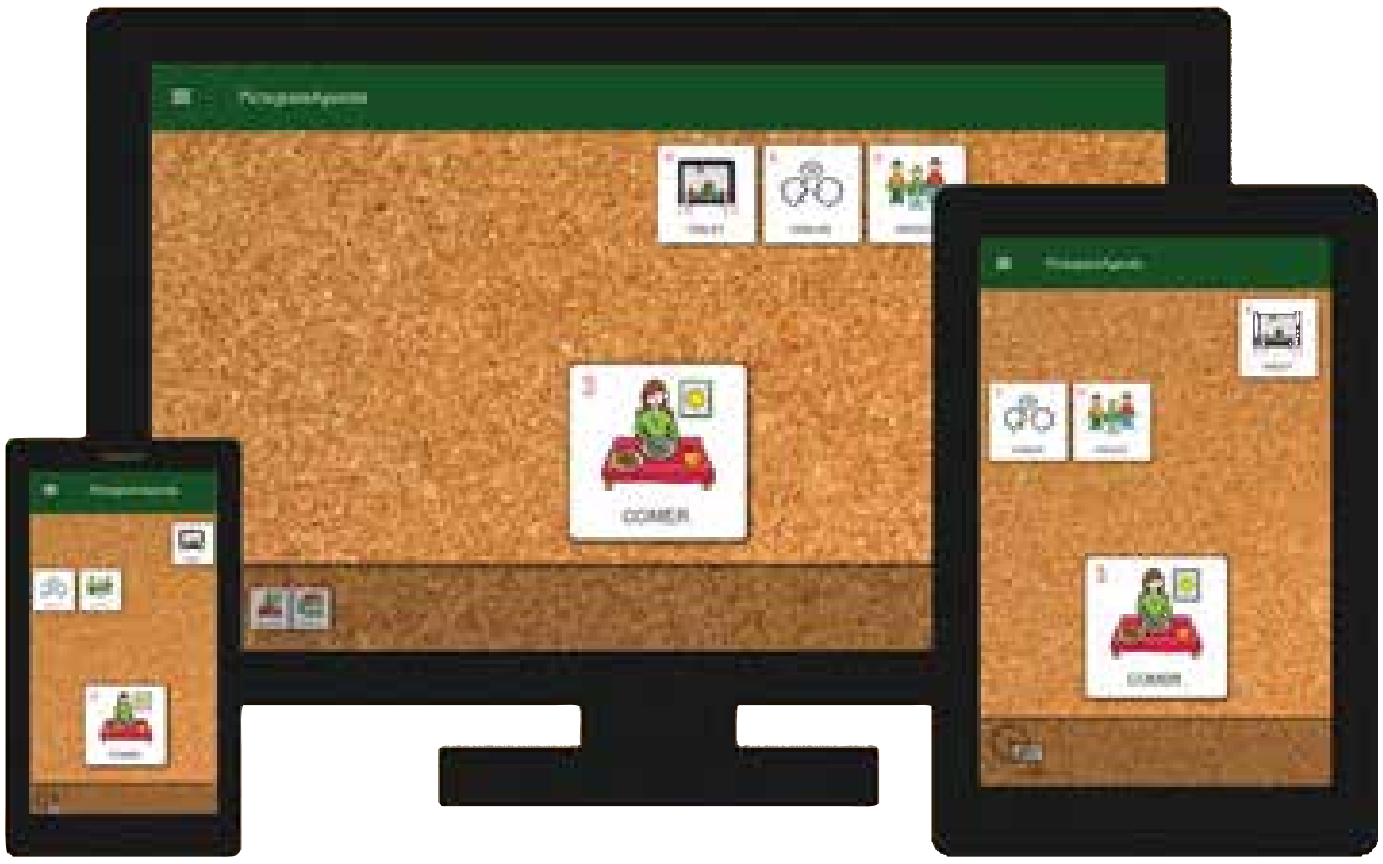

Además de las aplicaciones móviles, existen plataformas digitales que proporcionan recursos y comunidades de apoyo para las familias como son Auticmo (https://auticmo.com) y Lingokids (https://lingokids.com).

El uso de aplicaciones móviles ha demostrado ser beneficioso en el desarrollo de habilidades cognitivas y comunicativas en niños con TEA. Según un estudio, las aplicaciones educativas mejoran estas habilidades, proporcionando un entorno controlado y predecible que facilita el aprendizaje.

Es importante que las familias seleccionen aplicaciones que se adapten a las necesidades específicas de sus hijos y que promuevan un uso equilibrado de la tecnología. La supervisión y participación de los padres/madres en el uso de estas herramientas es esencial para maximizar sus beneficios y evitar posibles efectos negativos asociados al uso excesivo de dispositivos digitales.

## Grupos de apoyo y comunidades en línea

Los grupos de apoyo y las comunidades en línea desempeñan un papel crucial en el acompañamiento de las familias de personas con TEA. Estos espacios proporcionan soporte emocional, información valiosa y la oportunidad de conectar con otras familias que se enfrentan o han enfrentado a experiencias similares, facilitando el intercambio de información, estrategias y recursos efectivos (TrastornosMentales.org, 2023).

Participar en grupos de apoyo TEA permite a las familias compartir experiencias, obtener asesoramiento y recibir apoyo emocional. Estos grupos ofrecen un entorno donde las familias pueden expresar sus inquietudes y aprender de las vivencias de otras, lo que contribuye a reducir el aislamiento y a fortalecer la resiliencia.

Las plataformas digitales han ampliado el alcance de los grupos de apoyo, permitiendo la creación de comunidades en línea que trascienden las barreras geográficas. Estas comunidades ofrecen foros de discusión, recursos educativos y la posibilidad de establecer conexiones significativas con otras familias. Estas comunidades virtuales también facilitan el acceso a recursos educativos y programas de formación para familias. Organizaciones como la Fundación ConecTEA (https://www.fundacionconectea.org) ofrecen servicios de información y orientación en diferentes ámbitos de la vida, incluyendo el sanitario, educativo, social y laboral, proporcionando apoyo en la fase inicial del diagnóstico y más allá.

Algunos ejemplos de estas comunidades virtuales en la red social Facebook o grupos de Telegram son:

- Grupos de Facebook:
    * Grupo de apoyo, hijos con autismo (https://www.facebook.com/groups/2518989754791398): espacio dedicado a compartir experiencias, consejos y apoyo entre padres y familiares de niños con autismo.

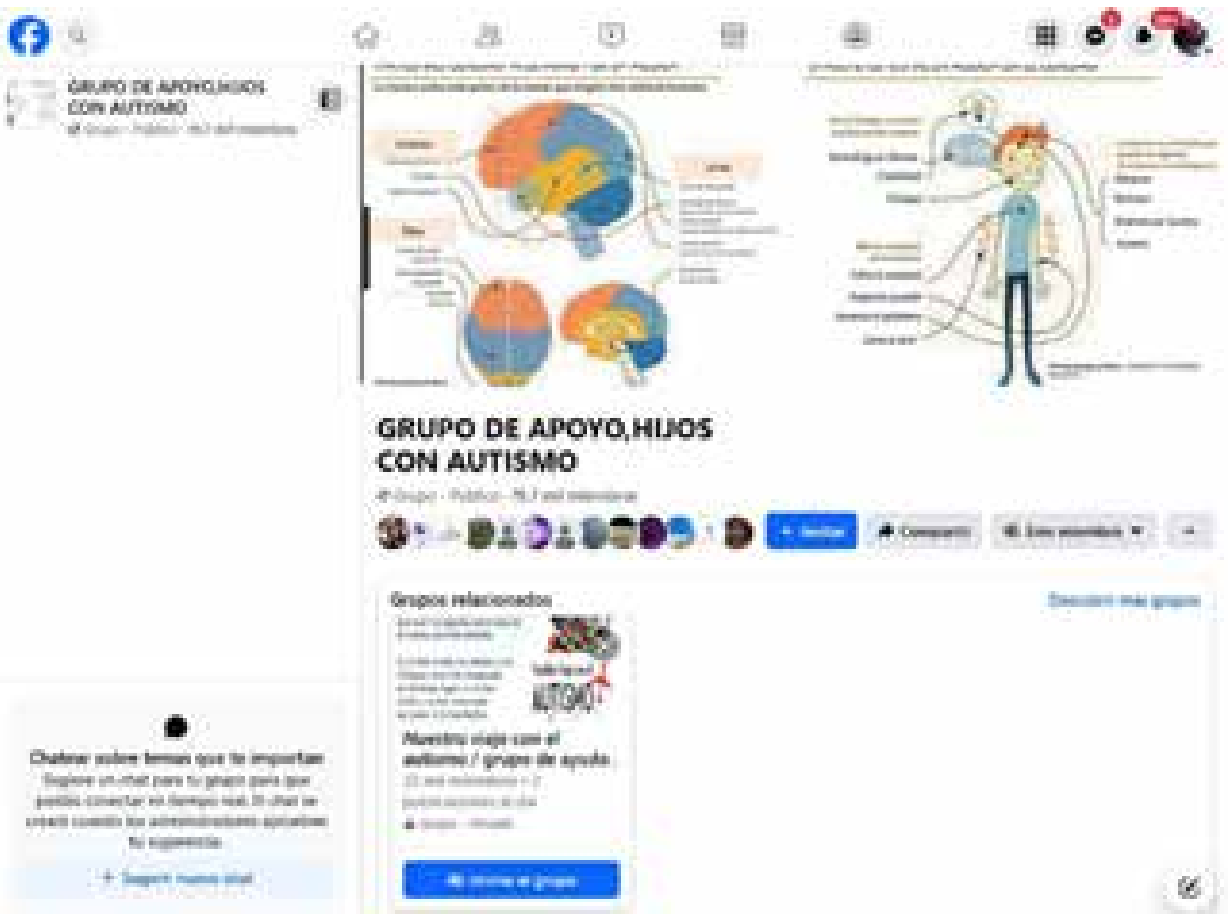

* Autismo Día a Día (https://www.facebook.com/autismodiaadiacorp): comunidad que ofrece talleres educativos y recursos para comprender y abordar el autismo en la vida cotidiana.

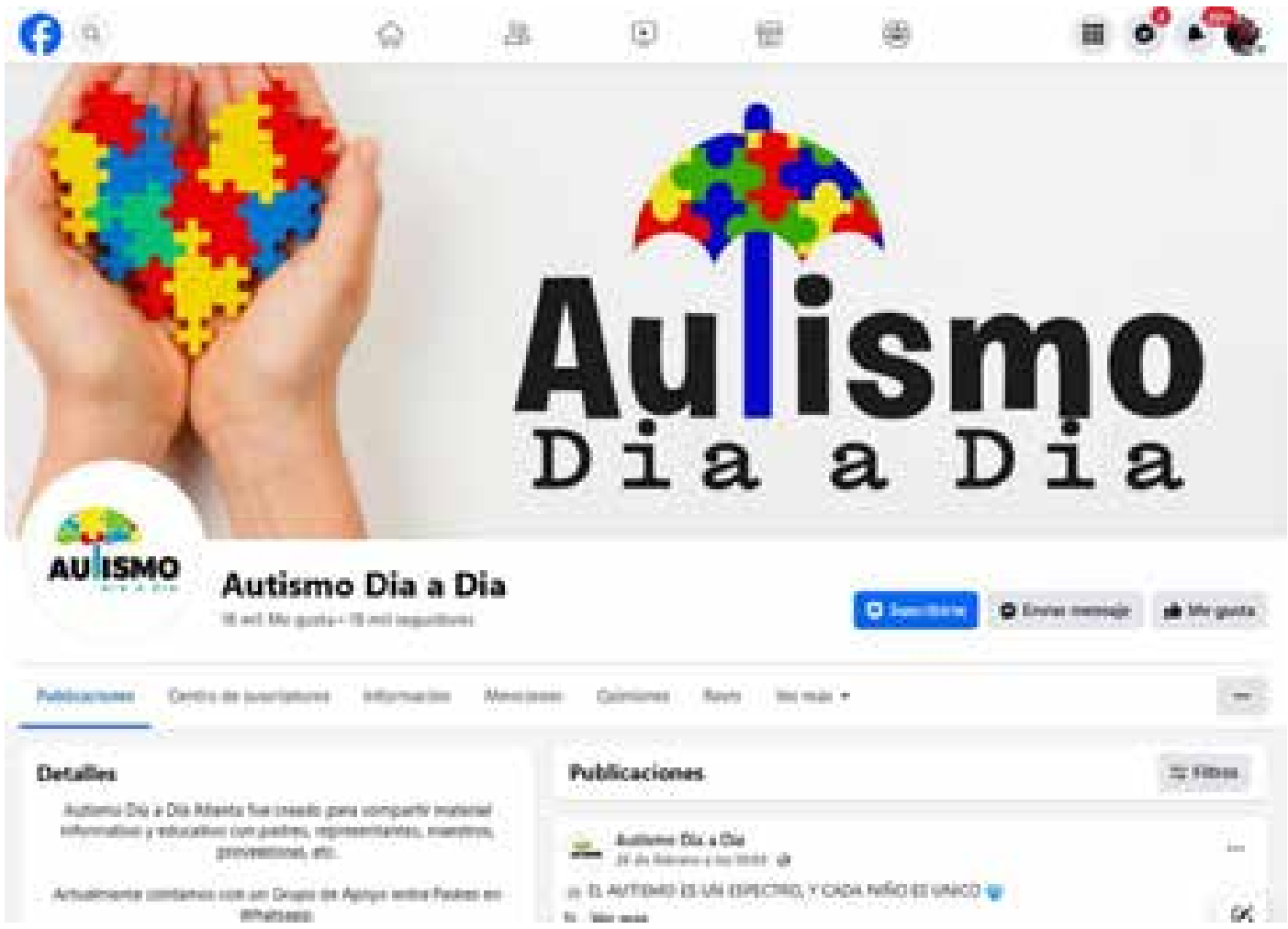

– Canales y Grupos de Telegram:
  * Autismo con Daniel Millán López (https://t.me/danielmillanlopez): Canal dirigido por el psicólogo Daniel Millán López, especializado en autismo, que ofrece información y recursos sobre TEA.

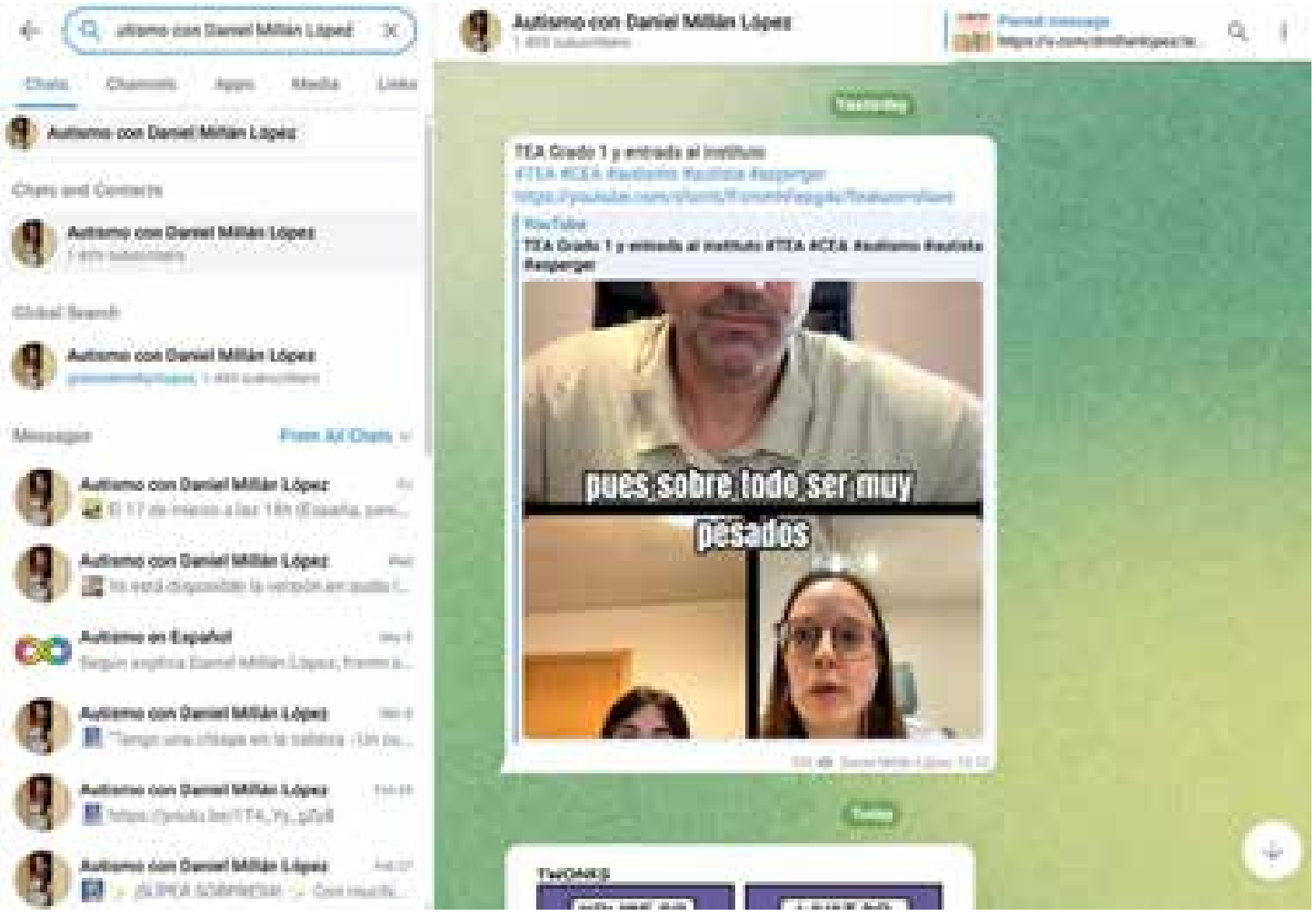

* Autismo en Español (https://t.me/autismo): Chat general sobre Trastorno del Espectro Autista (TEA) en español.

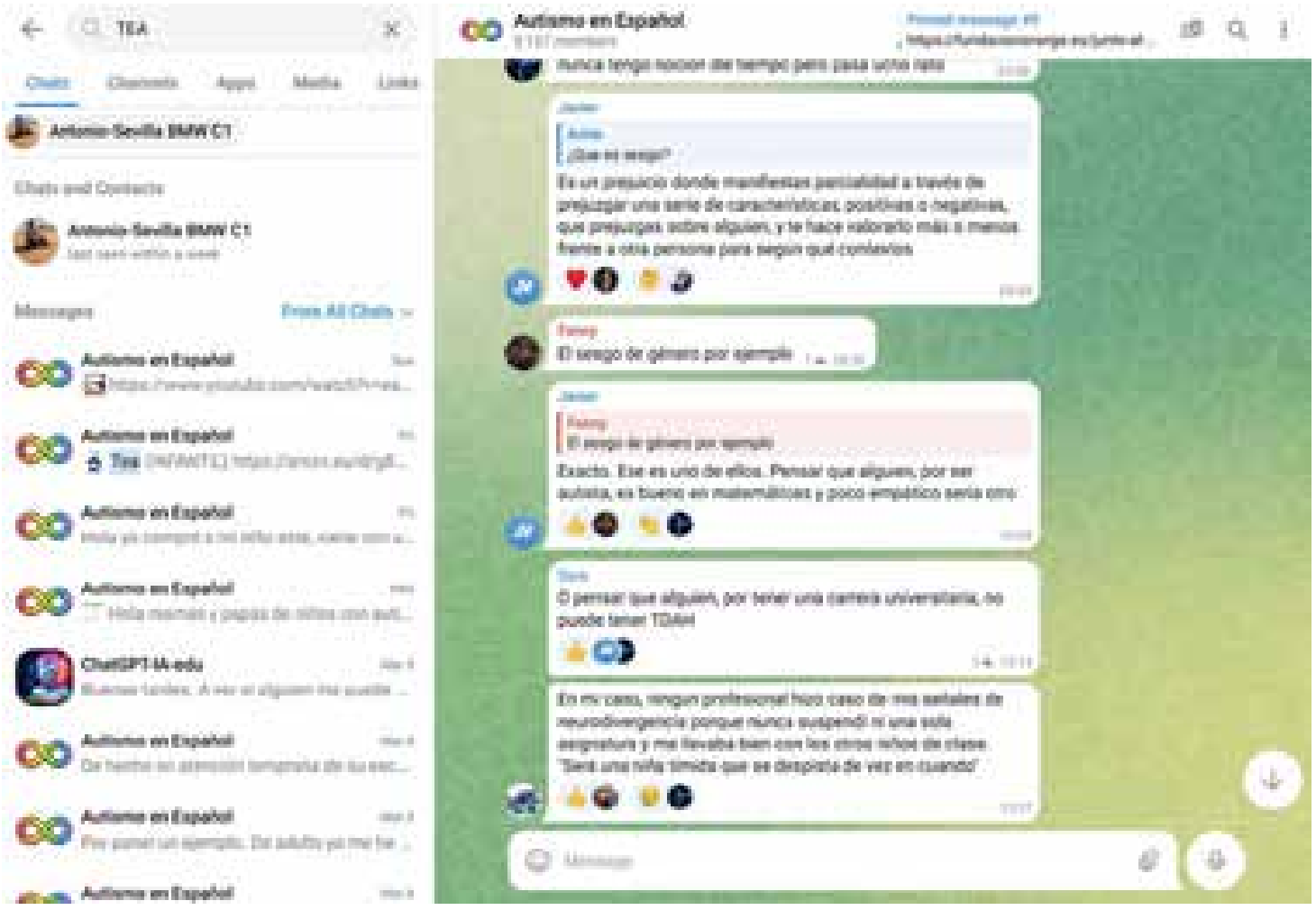

## Desafíos de los recursos digitales para las familias

El uso de recursos digitales para apoyar a personas con TEA ofrece muchos beneficios, pero también conlleva desafíos para las familias. A continuación, se presentan algunos de los principales obstáculos a los que se pueden enfrentar:

- En primer lugar, la selección de herramientas adecuadas, por ejemplo, existen muchas aplicaciones y plataformas para niños con TEA, pero no todas están diseñadas con base en la evidencia científica o adaptadas a las necesidades individuales del niño/a o del adolescente. Una familia puede probar varias aplicaciones de comunicación alternativa antes de encontrar la más efectiva para su hijo/a. Una posible solución sería buscar asesoramiento de terapeutas y educadores/as especializados para elegir herramientas digitales adecuadas.
- Un segundo aspecto sería buscar un equilibrio entre el uso de pantallas y otras actividades, ya que el uso excesivo de dispositivos digitales puede generar dependencia, afectando la socialización, la actividad física y la regulación emocional. Por ejemplo, un niño con TEA puede volverse obsesivo/a con una aplicación educativa y mostrar frustración cuando no se le permite usarla. Una posible solución sería establecer horarios de uso y combinar la tecnología con actividades sensoriales y de interacción presencial.

- Un tercer obstáculo a los que se pueden enfrentar sería la brecha digital de acceso y los costes que implican adquirir la tecnología, ya que muchas aplicaciones y dispositivos especializados pueden ser costosos, lo que limita el acceso a familias con menos recursos. Por ejemplo, aplicaciones como Proloquo2Go pueden ser altamente efectivas, pero su precio es elevado, lo que impide su adquisición por algunas familias. Una posible solución sería explorar opciones gratuitas o de bajo costo, como LetMeTalk o el uso de pictogramas físicos, y buscar programas de apoyo institucional.
- Un cuarto problema sería la falta de formación de las familias en el uso de recursos digitales, ya que algunos progenitores no están familiarizados con las aplicaciones o dispositivos y encuentran muchas dificultades para utilizarlos correctamente. Por ejemplo, una madre descarga una App de comunicación aumentativa, pero no sabe cómo programarla con el vocabulario adecuado para su hijo/a. Una posible solución sería que accediese a las guías, tutoriales o formaciones ofrecidas por los desarrolladores de aplicaciones o asociaciones especializadas en TEA, por ejemplo, a través de videos en streaming de plataformas como YouTube, Vimeo, etc.
- Un quinto obstáculo estaría relacionado con los riesgos asociados al acceso a internet, ya que los niños con TEA pueden ser más vulnerables a contenidos inapropiados o a interacciones no seguras en plataformas en línea. Por ejemplo, una persona con TEA que usa juegos en línea puede no comprender las normas de seguridad y compartir información personal sin saberlo. Una posible solución sería usar controles parentales, configurar perfiles supervisados y educar sobre seguridad digital de manera adaptada a su nivel de comprensión.
- Un sexto problema estaría relacionado con las dificultades en la generalización del aprendizaje digital al entorno real, es decir, algunos niños con TEA pueden aprender habilidades a través de aplicaciones, pero no trasladarlas a situaciones del mundo real. Por ejemplo, un niño que usa un comunicador aumentativo en casa puede no sentirse cómodo usándolo en la escuela o en espacios públicos. Una posible solución sería practicar el uso de la tecnología en diferentes entornos con apoyo progresivo de la familia y los profesionales.

## Estudios de caso y buenas prácticas

Aquí le proponemos dos estudios de caso basados en buenas prácticas del uso de tecnologías digitales por familias con hijos/as con TEA, junto con tres interrogantes para cada uno de ellos.

### Estudio de caso 1 (Facilitando la comunicación con Proloquo2Go)

María, una niña de 6 años con TEA y dificultades en el lenguaje verbal (TEA Nivel 2 o 3, según el DSM-5), ha comenzado a utilizar Proloquo2Go, una aplicación de Comunicación Aumentativa y Alternativa (CAA). Sus padres han trabajado con su terapeuta para personalizar la aplicación con imágenes y palabras relacionadas con su entorno diario. Después de varias semanas de uso constante, María ha logrado expresar sus necesidades básicas y emociones con mayor claridad, reduciendo sus crisis de frustración. Sin embargo, sus padres se preguntan cómo pueden integrar aún más la aplicación en diferentes contextos, como la escuela y las salidas familiares.

Interrogantes: ¿cómo puede la familia fomentar el uso de la aplicación en entornos fuera del hogar para mejorar la comunicación de María?; ¿qué estrategias pueden implementar los docentes para integrar Proloquo2Go en el aula y promover la interacción de María con sus compañeros?; y ¿cómo se puede evaluar si el uso de la aplicación está mejorando realmente la autonomía y la comunicación de María a largo plazo?

### Estudio de caso 2 (Desarrollo de habilidades sociales con Minecraft Education)

Sofía, una adolescente de 12 años con TEA, tiene dificultades para participar en juegos sociales con sus compañeros. Sus padres han decidido permitirle jugar en un entorno supervisado de Minecraft Education Edition, donde puede construir, colaborar y comunicarse con otros jugadores en un ambiente estructurado. Después de varios meses, Sofía ha mostrado mayor iniciativa para interactuar con sus compañeros en línea y ha transferido algunas de estas habilidades a situaciones del mundo real. Sin embargo, sus padres están preocupados por el tiempo de pantalla y la posibilidad de que su interés por el juego limite otras interacciones sociales presenciales.

Interrogantes: ¿cómo puede la familia equilibrar el tiempo de juego digital con otras actividades sociales y físicas?; ¿qué estrategias pueden implementarse para trasladar las habilidades sociales adquiridas en el juego a interaccio-nes fuera del entorno digital?; y ¿qué riesgos pueden surgir del uso excesivo de plataformas digitales y cómo pueden mitigarse en el caso de Sofía?

## Conclusiones

En este capítulo se ha analizado el papel fundamental que desempeñan las herramientas y recursos digitales en el apoyo a las familias de niños y niñas con TEA. Se ha destacado cómo las tecnologías pueden facilitar la comunicación, la organización de rutinas, el acceso a recursos educativos y el bienestar emocional de las familias, promoviendo la inclusión y la calidad de vida de estas personas.

Las aplicaciones móviles y plataformas digitales han demostrado ser aliadas clave en la intervención y el desarrollo del alumnado con TEA. Herramientas como Proloquo2Go, Choiceworks o Pictogram Agenda ofrecen soluciones prácticas para mejorar su comunicación y autonomía. Asimismo, los grupos de apoyo y comunidades en línea proporcionan espacios de acompañamiento emocional e intercambio de experiencias para las familias, reduciendo el aislamiento y fomentando la colaboración con otros cuidadores y profesionales.

Sin embargo, a lo largo del capítulo también se han señalado desafíos importantes en el uso de estos recursos digitales, tales como la selección de herramientas adecuadas, la brecha digital, el acceso económico a dispositivos y aplicaciones, y la necesidad de formación en su uso. La generalización del aprendizaje digital al mundo real sigue siendo un reto clave para garantizar que los beneficios de estas tecnologías se reflejen en la vida cotidiana de personas con TEA.

En este sentido, es fundamental seguir avanzando en el desarrollo de herramientas digitales más inclusivas, accesibles y adaptadas a la diversidad de necesidades dentro del espectro autista. Se recomienda fortalecer la formación de las familias en el uso de estas tecnologías, impulsar políticas públicas que faciliten el acceso a dispositivos y aplicaciones especializadas, y fomentar la investigación sobre la efectividad de los recursos digitales en la intervención con alumnado TEA.

Finalmente, la implementación exitosa de tecnologías en la vida de las personas con TEA requiere de un enfoque colaborativo entre familias, profesionales de la educación y la salud, y desarrolladores tecnológicos. Solo a través de esta sinergia será posible garantizar que las herramientas digitales se conviertan en verdaderos facilitadores del desarrollo, la inclusión y la autonomía del alumnado con TEA.

## Referencias bibliográficas

# INTELIGENCIA ARTIFICIAL COMO APOYO A LA EDUCACIÓN DEL ALUMNADO CON TEA

*José Fernández-Cerero*
Facultad de Ciencias de la Educación. Universidad de Sevilla (España)
*Marta Montenegro-Rueda*
Facultad de Ciencias de la Educación y del Deporte (Melilla). Universidad de Granada (España)

## Introducción

Actualmente, el número de alumnado con Necesidades Educativas Especiales (NEE) en el sistema educativo ha ido incrementando significativamente durante los últimos años. Esta tendencia refleja una mayor concienciación sobre la importancia de la inclusión educativa y subraya la necesidad de estrategias y herramientas innovadoras para satisfacer sus necesidades particulares. En este sentido, se ha evidenciado una gran cantidad de estudiantes diagnosticados con Trastorno del Espectro Autista (TEA). El TEA suele ser caracterizado por una serie de características principales en la comunicación social, el comportamiento repetido y el procesamiento sensorial. Atendiendo a la comunicación social, las personas diagnosticadas con TEA suelen presentar dificultades en la interpretación y la expresión de emociones y en la comprensión del lenguaje no verbal (Hodges *et al.*, 2020). En cuanto a la segunda característica, estos individuos suelen enfocarse intensamente en temas específicos, mostrando patrones de actividad limitada y repetitiva (Napolitano *et al.*, 2022). Esto puede incluir el uso repetitivo de objetos, movimientos estereotipados y una necesidad marcada de mantener rutinas o resistir cambios en su entorno. Finalmente, abordando el procesamiento sensorial, se destaca principalmente la hiperreactividad o hiporreactividad a diversos estímulos sensoriales, reaccionando de manera extrema a sonidos, luces o texturas o, por el contrario, la expresión de poca respuesta a estos estímulos (Miller & Wallace, 2019).

La inclusión de estudiantes con TEA en aulas ordinarias es un objetivo prioritario en los sistemas educativos modernos, ya que permite su participa-ción en actividades académicas y sociales junto a sus compañeros. Sin embargo, la efectividad de esta inclusión depende de factores como la formación del profesorado, el diseño de programas específicos y el acceso a recursos adecuados (Kurth, 2015). Además, se ha demostrado que la colaboración entre docentes, familias y especialistas mejora los resultados académicos y sociales de estos estudiantes (Anderson *et al.*, 2018). En este sentido, la Inteligencia Artificial (IA) se presenta como una herramienta innovadora para apoyar el aprendizaje y el desarrollo de habilidades en estudiantes con TEA. La IA permite la creación de intervenciones personalizadas y adaptativas, optimizando la enseñanza mediante tecnologías como asistentes virtuales, robots sociales y plataformas de aprendizaje adaptativo. Estas herramientas pueden abordar necesidades específicas del alumnado con TEA, tales como la mejora de habilidades sociales, la regulación emocional y la reducción de estímulos sensoriales que generan incomodidad (Xing, 2024). El uso y aplicación de la IA ha demostrado tener beneficios significativos en la mejora de la comunicación entre los docentes y los estudiantes a través de estrategias personalizadas en modelos predictivos (Lampos *et al.*, 2021). Además, el uso de plataformas educativas impulsadas por inteligencia artificial y la realidad virtual inmersiva ha demostrado progresos relevantes en el desarrollo de habilidades emocionales y sociales, contribuyendo de manera significativa a la integración social y académica de los estudiantes con TEA (Atturu & Naraganti, 2024, Herrero & Lorenzo, 2019).

En este capítulo, se explora cómo la IA puede actuar como un recurso clave para apoyar tanto a los docentes como a los estudiantes con TEA, ofreciendo soluciones innovadoras y personalizadas. Se presentarán herramientas específicas que han demostrado su eficacia en el ámbito educativo, ilustrando su potencial para mejorar las habilidades sociales, el aprendizaje adaptativo y la regulación emocional. Asimismo, se reflexionará sobre los desafíos que conlleva la implementación de estas tecnologías, considerando aspectos éticos, de equidad y de accesibilidad para garantizar su integración efectiva en un entorno educativo inclusivo.

## Impacto de la Inteligencia Artificial como apoyo al alumnado con TEA

La incorporación de la IA en la educación inclusiva ha abierto nuevas oportunidades para transformar el aprendizaje y fomentar la inclusión educativa, especialmente para los estudiantes con TEA. La UNESCO define la educación inclusiva como un enfoque basado en la equidad, promoviendo la par-ticipación de todos los estudiantes, independientemente de sus capacidades, necesidades o contextos (UNESCO, 2020). Este modelo subraya la importancia de adaptar el entorno educativo para eliminar barreras, respetar la diversidad y garantizar oportunidades equitativas de aprendizaje de todos los estudiantes.

El alumnado con TEA enfrenta retos en el entorno educativo, entre los que destacan las dificultades en la comunicación verbal y no verbal, los problemas para interpretar señales sociales y establecer relaciones interpersonales, así como necesidades sensoriales específicas que puedan afectar su concentración y bienestar. También enfrentan desafíos en la adaptación al entorno, en la autorregulación emocional y en la gestión de situaciones nuevas o imprevistas (Papoutsi *et al.*, 2018; Al-Saadi & Al-Tani, 2022). Sin embargo, estos retos se vuelven más significativos durante la transición de la escolarización de primaria a secundaria (de la Torre González, 2020). Para muchos estudiantes, este cambio representa un desafío que puede generar inquietud o ansiedad. Por lo tanto, resulta fundamental implementar medidas específicas que reduzcan una adaptación más fluida y reduzcan el impacto negativo de este proceso.

Los avances tecnológicos alcanzados en los últimos años han traído múltiples beneficios a la sociedad (Maroto-Gómez *et al.*, 2023). Diversos estudios han destacado la relevancia de las tecnologías emergentes en la educación inclusiva, subrayando cómo herramientas innovadoras pueden mejorar la interacción y el aprendizaje adaptativo para estudiantes con TEA (Fernández Cerero *et al.*, 2024). En este contexto, el uso de la IA ha experimentado un crecimiento significativo debido a su capacidad para transformar el aprendizaje. Esta tecnología ofrece herramientas personalizadas que eliminan barreras, fomentan la inclusión y se adaptan a la diversidad del alumnado (Chu *et al.*, 2022; Verdú *et al.*, 2017). Gracias a la IA, es posible proporcionar retroalimentación rápida y personalizada, lo que permite a los estudiantes mejorar su rendimiento académico (Dever *et al.*, 2020; Castellani *et al.*, 2024). Entre las innovaciones destacadas, los sistemas de tutoría inteligente y las plataformas de aprendizaje adaptativo han demostrado ser eficaces al atender las necesidades individuales de los estudiantes, especialmente aquellos con necesidades específicas (Van Seters *et al.*, 2012; Castellani *et al.*, 2024). Estas herramientas no solo ajustan materiales y entornos de aprendizaje según las características personales de los usuarios, sino que también integran estrategias como la gamificación, que fomenta la participación activa y mejora el rendimiento académico, incluso en alumnado con TEA (Stambuk-Castellano *et al.*, 2022).

Asimismo, los entornos de aprendizaje inteligentes, diseñados con técnicas de IA, no solo ofrecen acceso a recursos digitales desde cualquier lugar, sino que proporcionan orientación, pistas y apoyo en el momento adecuado, adaptándose tanto a los contenidos como al progreso del estudiante (Hwang, 2014). Estos sistemas permiten el desarrollo de tareas complejas que trascienden los límites de los métodos tradicionales y contextuales (Zapata-Ros, 2018), promoviendo así un aprendizaje más inclusivo y efectivo. Asimismo, la IA está revolucionando la educación al facilitar la creación de entornos personalizados y accesibles, donde se potencian las capacidades de los estudiantes y se responde a sus necesidades individuales, mejorando la calidad y equidad del aprendizaje (Montiel-Ruiz & López Ruiz, 2023).

## Herramientas y aplicaciones de IA como apoyo al alumnado con TEA

Las herramientas y aplicaciones basadas en IA destacan por su capacidad para abordar necesidades específicas relacionadas con habilidades sociales, emocionales, comunicativas y académicas, ofreciendo soluciones innovadoras y accesibles. Por tanto, a continuación, se presenta una selección de herramientas que el profesorado de ESO puede utilizar para apoyar al alumnado con TEA.

Estas herramientas ilustran cómo la tecnología basada en IA puede ser una aliada clave en la personalización de la enseñanza, la inclusión y el apoyo al desarrollo integral del alumnado con TEA. Al integrar estas soluciones en el aula, el profesorado puede potenciar la participación y el progreso de estos estudiantes, al tiempo que facilita la labor educativa.

Tabla 1. Herramientas y aplicaciones de IA como apoyo al alumnado con TEA.

| *Nombre* | *Tipo de Herramienta* | *Descripción* | *Área de apoyo* |
|---|---|---|---|
| Autism AI Screening System | Aplicación | Herramienta basada en IA para la detección e identificación de rasgos de TEA. Combina algoritmos de aprendizaje profundo y cuestionarios interactivos para evaluar patrones de comportamiento, comunicación y habilidades sociales de forma precisa y eficiente. | Optimización de la detección temprana de TEA, accesibilidad para profesionales y familias (Shahamiri & Thabtah, 2020). |

Tabla 1. Herramientas y aplicaciones de IA como apoyo al alumnado con TEA.

| *Nombre* | *Tipo de Herramienta* | *Descripción* | *Área de apoyo* |
|---|---|---|---|
| Emotion AI Chatbot | Programa informático | Chatbots que simulan conversaciones naturales, diseñados para ayudar a estudiantes con TEA a desarrollar habilidades de reconocimiento y regulación emocional mediante interacciones interactivas. | Mejora en el reconocimiento emocional y las relaciones interpersonales (Lozano-Martínez *et al.*, 2011). |
| Proloquo-2Go | Aplicación | Herramienta de Comunicación Aumentativa que utiliza IA para generar comunicación personalizada a través de símbolos y texto, especialmente útil para estudiantes con dificultades de lenguaje. | Mejora de la comunicación expresiva y receptiva (Collete *et al.*, 2019). |
| Cognoa | Plataforma | Plataforma que combina IA con evaluaciones clínicas para identificar y monitorizar las necesidades de estudiantes con TEA, proporcionando recursos para padres y educadores. | Detección temprana y seguimiento del desarrollo (Megerian *et al.* 2022). |
| Zeno Robot | Robot interactivo | Robot interactivo que utiliza IA para ayudar a niños con TEA a desarrollar habilidades sociales, emocionales y comunicativas mediante juegos y actividades guiadas. | Desarrollo de habilidades sociales y comunicativas (Salvador *et al.*, 2015). |
| Speech Blubs | Aplicación | Herramienta diseñada para mejorar las habilidades de habla y lenguaje en niños con TEA, utilizando actividades interactivas basadas en IA. | Apoyo en el desarrollo del habla y lenguaje (Kyriakaki *et al.*, 2023). |
| Read&Write by Texthelp | Software de accesibilidad | Herramienta basada en IA que ofrece lectura en voz alta, predicción de palabras y ayudas para la comprensión, adaptada a estudiantes con necesidades específicas como TEA. | Apoyo en la lectura, escritura y comprensión (Ozdowska *et al.*, 2021). |
| EmoTeach | Aplicación | Herramienta que emplea IA para enseñar a los estudiantes con TEA a reconocer y responder a emociones en rostros y situaciones sociales mediante juegos y actividades. | Mejora del reconocimiento y regulación emocional (Barba *et al.*, 2022). |
| AutiSpark | Aplicación | Diseñada específicamente para niños con TEA, incluye actividades basadas en IA para enseñar conceptos básicos, habilidades motoras, sociales y académicas de manera estructurada y accesible. | Refuerzo de habilidades básicas y motoras (Zakia *et al.*, 2024). |

## Desafíos y limitaciones del uso de la IA con el alumnado con TEA

La integración de la IA en la educación de alumnado con TEA presenta retos específicos que requieren atención en los ámbitos pedagógico, tecnológico y ético. La preparación docente para trabajar con herramientas de IA no solo debe incluir el aprendizaje técnico, sino también el desarrollo de competencias pedagógicas inclusivas que garanticen una educación equitativa y adaptada a las necesidades de este alumnado.

### Capacitación docente y competencias inclusivas

La necesidad de formar a los docentes es un aspecto fundamental para la transformación de los entornos educativos en espacios que promuevan la igualdad y el respeto por la diversidad. La implementación de programas de capacitación continua en competencias inclusivas no sólo mejora la calidad de la educación, sino que también fomenta un clima escolar más acogedor y respetuoso, donde cada estudiante tiene la oportunidad de desarrollarse plenamente y participar activamente en su proceso de aprendizaje. Para que la IA sea eficaz en entornos educativos que incluyen a estudiantes con TEA, la formación docente debe enfocarse en competencias digitales inclusivas. Esto implica enseñar a los docentes a identificar las necesidades individuales del alumnado con TEA y a aplicar estrategias que maximicen los beneficios de estas tecnologías, respetando principios de accesibilidad y equidad (Erickson *et al.*, 2022). La formación específica debe incluir:

- Diseño de programas personalizados: Capacitar a los docentes en el uso de herramientas de IA adaptadas a los estilos de aprendizaje y características específicas del alumnado con TEA (Smith *et al.*, 2021).
- Colaboración interdisciplinaria: Promover la cooperación entre expertos en IA y el profesorado para desarrollar soluciones efectivas y éticas (Johnson *et al.*, 2020).

### Equidad en el acceso y barreras tecnológicas

El acceso equitativo a tecnologías basadas en IA sigue siendo un desafío, particularmente para centros educativos en zonas rurales o con recursos limitados. En el caso de alumnado con TEA, estas barreras económicas y la falta de recursos tecnológicos pueden limitar más aún su inclusión en la educación. De este modo, se propone el desarrollo de nuevas herramientas y plataformas accesibles de código abierto para el aprendizaje de estudiantes con TEA (Wilson *et al.*, 2023), así como fomentar el desarrollo de políticas educativas, que promuevan la financiación económica para la adquisición de tecnologías educativas inclusivas (Davis *et al.*, 2023). Para que estas políticas sean efectivas, resulta fundamental complementarlas con programas de formación docente que promuevan el uso ético de la tecnología educativa. Esto incluye prestar especial atención a la privacidad y seguridad de los datos de los estudiantes, así como a la implementación de estrategias que ayuden a identificar y reducir posibles sesgos en las herramientas digitales (Farooqi *et al.*, 2024). También es clave incentivar la colaboración entre docentes, especialistas en inteligencia artificial y familias, ya que este trabajo conjunto permite desarrollar soluciones personalizadas y éticas que respondan mejor a las necesidades del alumnado con TEA. Por último, el establecimiento de redes de apoyo entre docentes y el uso de simulaciones prácticas contribuyen significativamente a mejorar la preparación del profesorado para aplicar estas tecnologías en el aula. En conjunto, estas acciones son esenciales para avanzar hacia un modelo educativo más inclusivo y accesible para todos los estudiantes.

### Privacidad y seguridad de datos

El uso de herramientas de IA con alumnado con TEA requiere recopilar datos personales para personalizar el aprendizaje, lo que genera riesgos relacionados con la privacidad y la seguridad de la información sensible. Es fundamental establecer protocolos de protección de datos y cumplir con normativas internacionales como el Reglamento General de Protección de Datos (GDPR) (Chen *et al.*, 2021). De este modo es recomendable la implementación de tecnologías con cifrado robusto para proteger la privacidad de los estudiantes, así como la transparencia en el uso y procesamiento de datos educativos (Lee *et al.*, 2022). Siguiendo esta línea, es necesario desarrollar determinados marcos regulatorios que permitan regular el uso de la IA en el ámbito educativo, donde se incluya aspectos claros relacionados con la seguridad y privacidad (Che, 2024)

## Conclusiones

La IA se presenta como una herramienta transformadora para la educación inclusiva, especialmente en el apoyo al alumnado con TEA. Su capacidad para personalizar el aprendizaje y adaptarse a las necesidades específicas permite abordar desafíos relacionados con la comunicación, la regulación emo-cional y las necesidades sensoriales. Herramientas como plataformas web, robots interactivos y aplicaciones móviles han demostrado su eficacia en el desarrollo de habilidades sociales, emocionales y académicas, fomentando la integración de este alumnado en entornos educativos. Sin embargo, su implementación plantea desafíos importantes, como la necesidad de capacitar al profesorado en competencias digitales inclusivas, principalmente en el uso de la IA, garantizar la equidad en el acceso a estas tecnologías y establecer mecanismos robustos de protección de datos. Superar estas barreras será crucial para aprovechar el potencial de la IA en la construcción de entornos educativos más inclusivos y equitativos. En este sentido, también se hace evidente la necesidad de formar a los docentes para que puedan implementar mediante estrategias efectivas el uso de las TIC en el alumnado con TEA.

## Financiación

## Referencias bibliográficas

# HERRAMIENTAS Y RECURSOS DIGITALES PARA EL DESARROLLO DEL PENSAMIENTO LÓGICO-MATEMÁTICO DEL ALUMNADO CON TEA

*Rocío Piñero-Virué*
Facultad de Ciencias de la Educación. Universidad de Sevilla (España)

## Introducción

Expondremos una breve información a modo de guía para poder reconocer las principales características del sujeto con trastorno del espectro autista (TEA), así como, la construcción de su proceso de aprendizaje. Posteriormente, nos vamos a centrar en el marco legal de la Ley Orgánica 3/2020, de 29 de diciembre, por la que se modifica la Ley Orgánica 2/2006, de 3 de mayo, de Educación (LOMLOE, 2020) donde se explicitan las Competencias claves, potenciando una visión de la Competencia matemática y en ciencia, tecnología e ingeniería, señalando como factor necesario en el aprendizaje el desarrollo del pensamiento lógico-matemático en el niño. Por lo que, vamos a ofrecer unas pautas y orientaciones para trabajar esta temática con educandos TEA en el marco de una escuela inclusiva; destacando las estrategias y recursos que favorecen el trabajo en equipo. De este modo, mostraremos que una escuela para todos requiere de una continuada labor del ámbito escolar al familiar y viceversa, destacando la comunicación entre todos los miembros de la comunidad educativa como eje central, además de la disposición de los servicios necesarios para atender a la diversidad; y en este sentido, proponemos el apoyo de herramientas y recursos digitales tecnológicos más un enfoque multinivel como respuesta al aula heterogénea. Este proceso se ha de convenir bajo un clima de respeto y armonía, potenciando las fortalezas del sujeto TEA, así como las de todos los alumnos, apoyados con recursos humanos y materiales que cubran las necesidades del alumnado; trabajamos en un ambiente de colaboración y cooperación sustentados en la comunicación como base para concebir y desarrollar una plena equidad e igualdad de oportunidades. El alumnado TEA requiere de una serie de adaptaciones y atenciones que han de ser focalizadas y obradas partiendo de una adecuada evaluación para que el tratamiento a seguir sea individualizado, puesto que se llega a generalizar cuando se habla de los trastornos y dificultades, pero cada uno es diferente. A través de la siguiente Figura 1 podremos observar los principales descriptores a desarrollar.

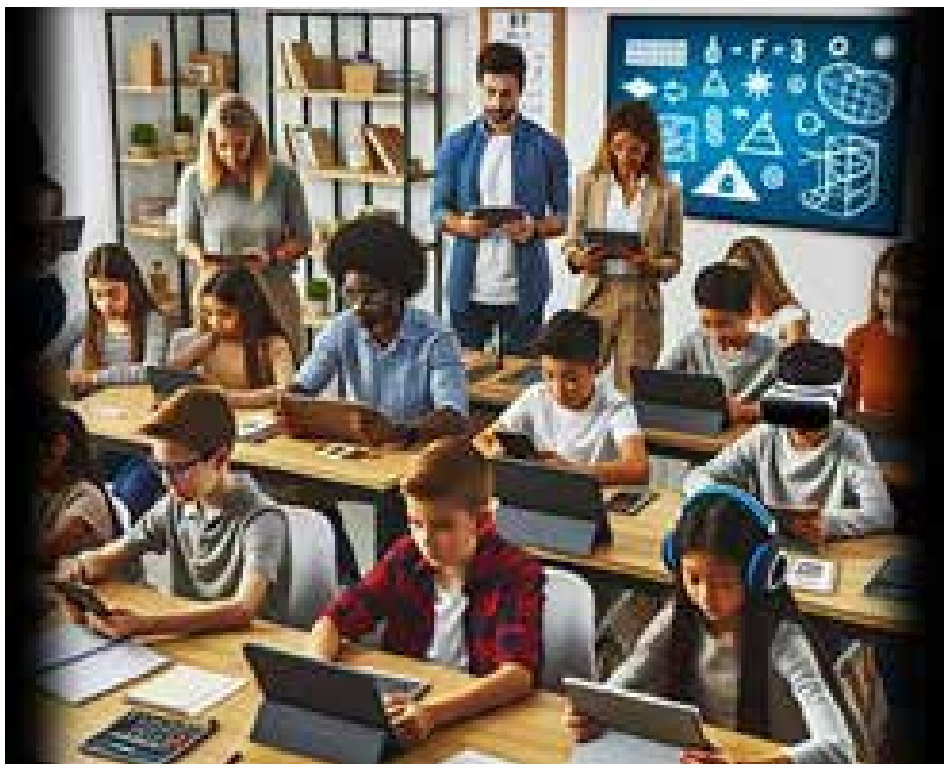

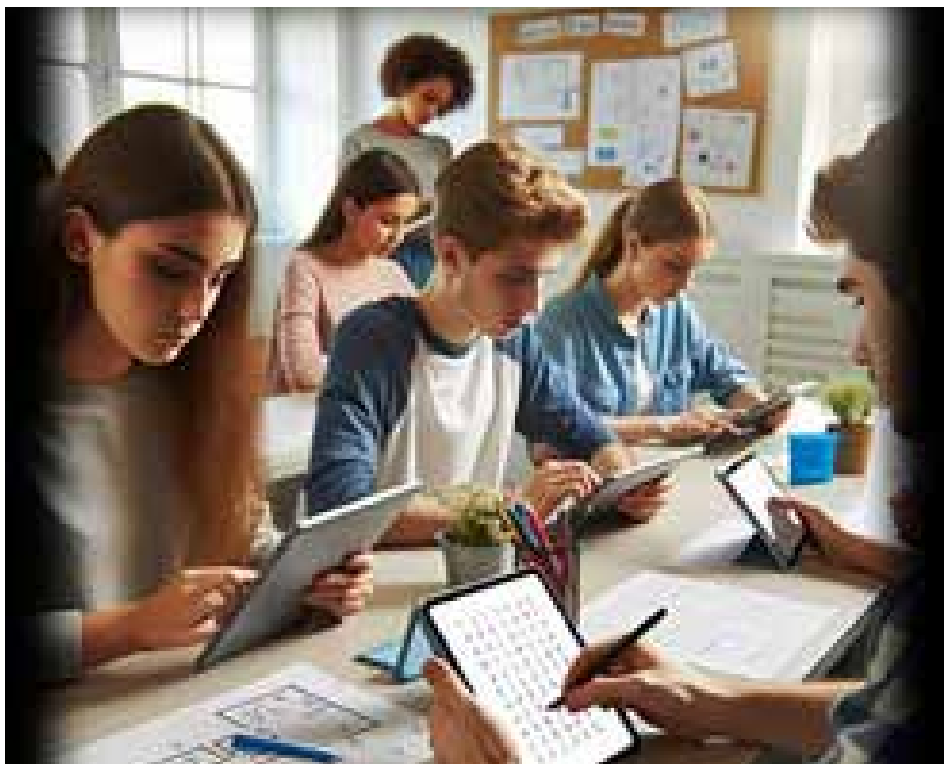

Figura 1. Desarrollo del planteamiento.

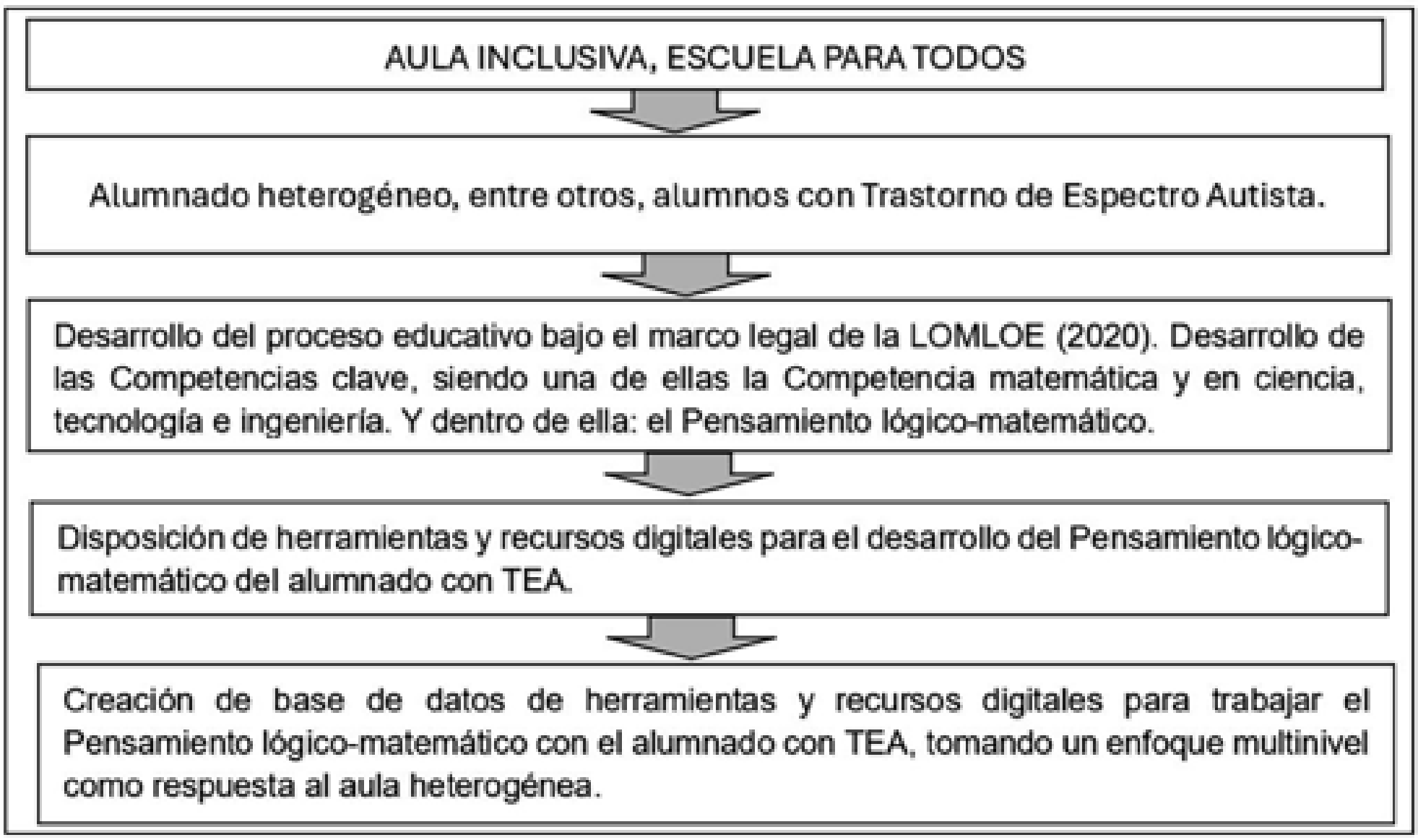

Nota: Fuente de elaboración propia.

## El proceso de aprendizaje de un alumno TEA

Según la Asociación Americana de Psiquiatría [APA] (2014), el trastorno del espectro autista (o "del autismo", TEA o ASD, por sus siglas en inglés) es un trastorno del neurodesarrollo cuyas principales características se encuentran relacionadas con las limitaciones en la comunicación e interacción social. Para conocer las características de un sujeto con TEA, hemos acudido a diferentes fuentes documentales como Cuesta (2009), Taylor (2015), y Fernández-Batanero (2020), entre otros, donde se especifican las características clínicas que pueden presentar los discentes con TEA en el comportamiento, la comunicación, el lenguaje, la motricidad, los sentidos y las emociones: Referente al nivel intelectual, sus habilidades pueden variar desde la discapacidad intelectual a capacidades superiores; en el nivel comunicativo, sí presentan alteraciones verbales y no verbales, por lo que puede que no empleen el lenguaje y, en cambio otros, que sí posean una lingüística fluida, aunque no son capaces de mantener una comunicación recíproca funcional, por tanto, les afecta a la interacción social al no saber cómo iniciar o mantener esa comunicación o no tienen en cuenta las reacciones de la otra persona reflejando un escaso interés por los demás, escasas o nulas relaciones de amistad, ya que no pueden entender las emociones de los demás, al igual que el exiguo o ningún contacto visual con los demás; en el nivel conductual, presentan intereses limitados, a los que se unen las conductas de comportamientos repetitivos como balancearse, girar o aletear con las manos, además de las dificultades para afrontar cambios o improvisarlos, ya que siguen patrones rígidos de comportamiento y no aceptan esos cambios. Todo ello, les dificulta desenvolverse de manera adecuada con sus iguales o mayores en el contexto educativo, o en el social. También pueden presentar dificultades en diferentes áreas cognitivas, como: la atención, la memoria, la flexibilidad cognitiva y/o la planificación; pueden tener dificultad para comprender instrucciones verbales complejas, puesto que el hecho de procesar la información de manera global o establecer conexiones entre ideas o conceptos les puede dificultar. Sin embargo, juega a su favor, el tener la gran fortaleza cognitiva en retener detalles o la memorización de grandes cantidades de información sobre un determinado tema.

Por ello, una evaluación exhaustiva de estas características nos va a ayudar a trabajar con el alumno en el ámbito educativo, en este caso, para que pueda construir el proceso de aprendizaje a su medida, sirviéndonos de técnicas, estrategias, recursos y herramientas que potencien de manera favorable esta labor. La enseñanza individualizada enfocada desde la perspectiva inclusiva nos lleva a promover con todos los educandos en un mismo espacio fomen-tando la fortaleza de algunos y superando las dificultades de otros. Así, nos planteamos una enseñanza sin barreras incentivando el trabajo en equipo como base de la escuela para todos. Para Echeita (2017):

> La educación inclusiva se trata de una meta que quiere ayudar a transformar los sistemas educativos para que todo el alumnado, sin restricciones, limitaciones ni eufemismos respecto a ese TODOS, tenga oportunidades equiparables y de calidad para el desarrollo pleno de su personalidad (p. 18).

Y en esta escuela, regida por un marco legal donde se imparte docencia en diferentes conocimientos, tanto conceptuales como procedimentales y actitudinales, se desarrollan entre otros, el pensamiento lógico-matemático.

## La Competencia matemática y en ciencia, tecnología e ingeniería y la necesidad de desarrollar el pensamiento lógico-matemático

En el marco legal de la LOMLOE (2020), las Competencias claves que se recogen para la etapa de la enseñanza básica, expuestas en la Figura 2, así como el referente a la Competencia matemática y en ciencia, tecnología e ingeniería, en la Figura 3:

Figura 2. Competencias clave (LOMLOE, 2020).

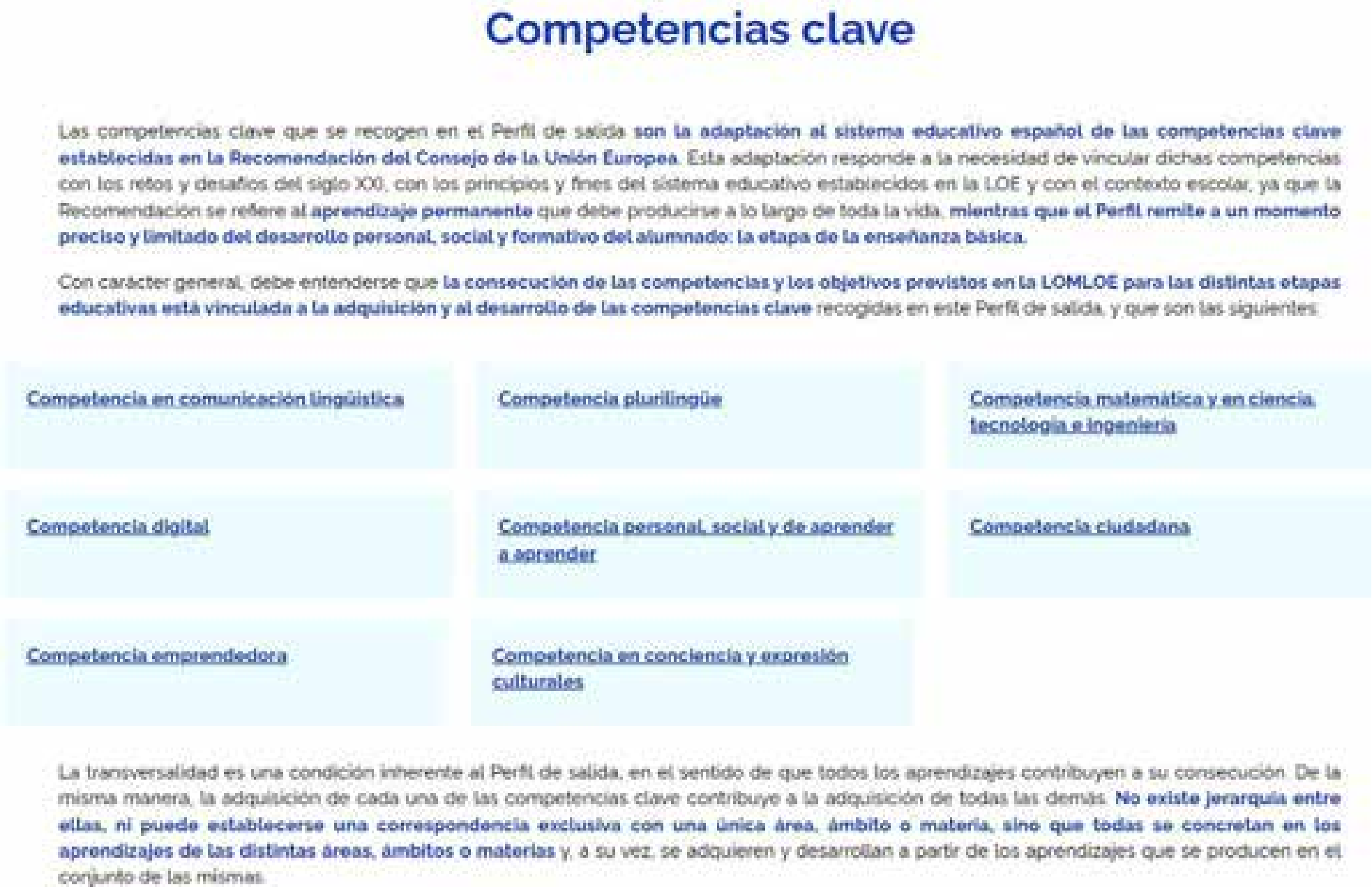

### Competencias clave

Las competencias clave que se recogen en el Perfil de salida **son la adaptación al sistema educativo español de las competencias clave establecidas en la Recomendación del Consejo de la Unión Europea**. Esta adaptación responde a la necesidad de vincular dichas competencias con los retos y desafíos del siglo XXI, con los principios y fines del sistema educativo establecidos en la LOE y con el contexto escolar, ya que la Recomendación se refiere al **aprendizaje permanente** que debe producirse a lo largo de toda la vida, **mientras que el Perfil remite a un momento preciso y limitado del desarrollo personal, social y formativo del alumnado: la etapa de la enseñanza básica.**

Con carácter general, debe entenderse que **la consecución de las competencias y los objetivos previstos en la LOMLOE para las distintas etapas educativas está vinculada a la adquisición y al desarrollo de las competencias clave** recogidas en este Perfil de salida, y que son las siguientes:

- Competencia en comunicación lingüística
- Competencia plurilingüe
- Competencia matemática y en ciencia, tecnología e ingeniería
- Competencia digital
- Competencia personal, social y de aprender a aprender
- Competencia ciudadana
- Competencia emprendedora
- Competencia en conciencia y expresión culturales

La transversalidad es una condición inherente al Perfil de salida, en el sentido de que todos los aprendizajes contribuyen a su consecución. De la misma manera, la adquisición de cada una de las competencias clave contribuye a la adquisición de todas las demás. **No existe jerarquía entre ellas, ni puede establecerse una correspondencia exclusiva con una única área, ámbito o materia, sino que todas se concretan en los aprendizajes de las distintas áreas, ámbitos o materias** y, a su vez, se adquieren y desarrollan a partir de los aprendizajes que se producen en el conjunto de las mismas.

Nota: Elaborada a partir de la web: https://educagob.educacionfpydeportes.gob.es/curriculo/curriculo-lomloe/menu-curriculos-basicos/ed-primaria/competencias-clave.html

Figura 3. Competencia matemática y en ciencia, tecnología e ingeniería (LOMLOE, 2020).

**Competencia matemática y en ciencia, tecnología e ingeniería**

La competencia matemática y competencia en ciencia, tecnología e ingeniería (competencia STEM por sus siglas en inglés) entraña la comprensión del mundo utilizando los métodos científicos, el pensamiento y representación matemáticos, la tecnología y los métodos de la ingeniería para transformar el entorno de forma comprometida, responsable y sostenible.

La competencia matemática permite desarrollar y aplicar la perspectiva y el razonamiento matemáticos con el fin de resolver diversos problemas en diferentes contextos.

La competencia en ciencia conlleva la comprensión y explicación del entorno natural y social, utilizando un conjunto de conocimientos y metodologías, incluidas la observación y la experimentación, con el fin de plantear preguntas y extraer conclusiones basadas en pruebas para poder interpretar y transformar el mundo natural y el contexto social.

La competencia en tecnología e ingeniería comprende la aplicación de los conocimientos y metodologías propios de las ciencias para transformar nuestra sociedad de acuerdo con las necesidades o deseos de las personas en un marco de seguridad, responsabilidad y sostenibilidad.

Nota: Elaborada a partir de la web: https://educagob.educacionfpydeportes.gob.es/curriculo/curriculo-lomloe/menu-curriculos-basicos/ed-primaria/competencias-clave.html

Figura 4. Descriptores operativos de la Competencia matemática y en ciencia, tecnología e ingeniería (LOMLOE, 2020).

Descriptores operativos

| AL COMPLETAR LA EDUCACIÓN PRIMARIA, EL ALUMNO O LA ALUMNA… | AL COMPLETAR LA ENSEÑANZA BÁSICA, EL ALUMNO O LA ALUMNA… |
|---|---|
| STEM1. Utiliza, de manera guiada, algunos métodos inductivos y deductivos propios del razonamiento matemático en situaciones conocidas, y selecciona y emplea algunas estrategias para resolver problemas reflexionando sobre las soluciones obtenidas. | STEM1. Utiliza métodos inductivos y deductivos propios del razonamiento matemático en situaciones conocidas, y selecciona y emplea diferentes estrategias para resolver problemas analizando críticamente las soluciones y reformulando el procedimiento, si fuera necesario. |
| STEM2. Utiliza el pensamiento científico para entender y explicar algunos de los fenómenos que ocurren a su alrededor, confiando en el conocimiento como motor de desarrollo, utilizando herramientas e instrumentos adecuados, planteándose preguntas y realizando experimentos sencillos de forma guiada. | STEM2. Utiliza el pensamiento científico para entender y explicar los fenómenos que ocurren a su alrededor, confiando en el conocimiento como motor de desarrollo, planteándose preguntas y comprobando hipótesis mediante la experimentación y la indagación, utilizando herramientas e instrumentos adecuados, apreciando la importancia de la precisión y la veracidad y mostrando una actitud crítica acerca del alcance y las limitaciones de la ciencia. |
| STEM3. Realiza, de forma guiada, proyectos, diseñando, fabricando y evaluando diferentes prototipos o modelos, adaptándose ante la incertidumbre, para generar en equipo un producto creativo con un objetivo concreto, procurando la participación de todo el grupo y resolviendo pacíficamente los conflictos que puedan surgir. | STEM3. Plantea y desarrolla proyectos diseñando, fabricando y evaluando diferentes prototipos o modelos para generar o utilizar productos que den solución a una necesidad o problema de forma creativa y en equipo, procurando la participación de todo el grupo, resolviendo pacíficamente los conflictos que puedan surgir, adaptándose ante la incertidumbre y valorando la importancia de la sostenibilidad. |
| STEM4. Interpreta y transmite los elementos más relevantes de algunos métodos y resultados científicos, matemáticos y tecnológicos de forma clara y veraz, utilizando la terminología científica apropiada, en diferentes formatos (dibujos, diagramas, gráficos, símbolos…) y aprovechando de forma crítica, ética y responsable la cultura digital para compartir y construir nuevos conocimientos. | STEM4. Interpreta y transmite los elementos más relevantes de procesos, razonamientos, demostraciones, métodos y resultados científicos, matemáticos y tecnológicos de forma clara y precisa y en diferentes formatos (gráficos, tablas, diagramas, fórmulas, esquemas, símbolos…), aprovechando de forma crítica la cultura digital e incluyendo el lenguaje matemático-formal con ética y responsabilidad, para compartir y construir nuevos conocimientos. |
| STEM5. Participa en acciones fundamentadas científicamente para promover la salud y preservar el medio ambiente y los seres vivos, aplicando principios de ética y seguridad y practicando el consumo responsable. | STEM5. Emprende acciones fundamentadas científicamente para promover la salud física, mental y social, y preservar el medio ambiente y los seres vivos; y aplica principios de ética y seguridad en la realización de proyectos para transformar su entorno próximo de forma sostenible, valorando su impacto global y practicando el consumo responsable. |

Nota: Elaborada a partir de la web: https://educagob.educacionfpydeportes.gob.es/curriculo/curriculo-lomloe/menu-curriculos-basicos/ed-primaria/competencias-clave.html

Centrados en el pensamiento lógico-matemático podemos resumir que se trata de la capacidad de razonar de manera estructurada y sistemática para resolver problemas y comprender conceptos matemáticos; implicando habilidades como: razonamiento lógico, siendo la capacidad de seguir una secuencia de pasos lógicos para llegar a una conclusión; resolución de problemas, intentando identificar, analizar y encontrar soluciones a problemas matemáticos; abstracción como la habilidad de generalizar conceptos y reconocer patrones; y el análisis y síntesis, referente a poder descomponer problemas complejos en partes más simples y combinar diferentes elementos para formar un todo coherente. Por tanto, estas habilidades son fundamentales en el área de las matemáticas, de la ciencia, la ingeniería y la informática. De ahí, que se conjuguen todas estas áreas en la competencia expuesta (competencia matemática y en ciencia, tecnología e ingeniería) y se trabajen nuevos métodos como el STEAM (Ciencia, Tecnología, Ingeniería, Arte y Matemáticas).

Por tanto, el desarrollo del pensamiento lógico-matemático en los niños es fundamental para el desarrollo de su capacidad de resolver problemas y comprender el mundo que les rodea. Comenzando a edades tempranas se puede ir fomentando a través de diversas estrategias y actividades. Según Piaget (1936), cuando expone las etapas del desarrollo de la inteligencia, podemos hablar de cuatro: en la Etapa Sensorimotora (0-2 años), los niños exploran el mundo a través de sus sentidos y acciones; comienzan a entender conceptos básicos como la permanencia de los objetos y las relaciones espaciales. En la Etapa Preoperacional (2-7 años), los niños desarrollan habilidades de conteo y empiezan a entender conceptos de cantidad y número; utilizan el juego simbólico para representar y manipular ideas matemáticas. En la Etapa de Operaciones Concretas (7-11 años), los niños comienzan a pensar lógicamente sobre objetos y eventos concretos; por lo que pueden realizar operaciones mentales como la clasificación, seriación y conservación. Y en la Etapa de Operaciones Formales (11 años en adelante), los adolescentes desarrollan la capacidad de pensar de manera abstracta y lógica; pudiendo resolver problemas complejos y entender conceptos matemáticos avanzados.

Para desarrollar este campo del pensamiento lógico-matemático en el niño, fundamentado en el aprendizaje por descubrimiento y en el aprendizaje colaborativo y cooperativo, hemos de trabajar estrategias incluyendo juegos y actividades lúdicas como el conteo, clasificación, y resolución de problemas, utilizando juegos de mesa y rompecabezas, de esta forma, el aprendizaje será más atractivo; enlazando las experiencias cotidianas, involucrando a los alumnos en actividades diarias que requieran el uso de habilidades matemáticas, como cocinar (medir ingredientes) o hacer compras (contar dinero); usando materiales manipulativos, ofreciéndoles objetos que puedan manipular para entender conceptos matemáticos, como bloques, cuentas y figuras geométricas, para posteriormente, puedan pasar a operar con objetos abstractos; e incluso, fomentando la lectura de textos que incluyan conceptos matemáticos para discutir las historias y reforzar el aprendizaje, además de conectar diferentes conceptos. Del mismo modo, se hace necesario: fomentar la resolución de problemas, ayudando a los niños a formular hipótesis y establecer predicciones, lo que contribuye a su capacidad de razonamiento; deducir reglas; realizar experimentos; relacionar conceptos mediante mapas mentales; o fomentar la noción de objeto permanente; entre otras.

De esta forma, el desarrollo lógico-matemático es un proceso que implica adquirir habilidades para razonar, identificar patrones, resolver problemas, y comprender el mundo a través de la lógica y las matemáticas. Al igual que en el aprendizaje del resto de los saberes y competencias de las diferentes áreas, ha de existir una coordinación entre familia y escuela debido a la importancia de cohesionar el entorno familiar y escolar, puesto que el contexto donde se desenvuelven los niños juega un papel crucial en el desarrollo de su pensamiento lógico-matemático. Es importante que tanto en casa como en la escuela se les brinde un ambiente rico en estímulos y oportunidades para explorar y aprender, continuando el desarrollo de la actividad educativa en casa, para que puedan ayudar a los niños a construir una base sólida en el pensamiento lógico-matemático. En la LOMLOE (2020), se especifica que, los procesos matemáticos como el razonamiento, la resolución de problemas, las conexiones y la prueba son considerados "ejes" de la competencia matemática.

## El desarrollo del pensamiento lógico-matemático en el niño TEA

Situados en una filosofía de escuela para todos, y matizando que cada individuo es diferente y necesita de una enseñanza a su medida, hemos de considerar, que cada uno va a trabajar mejor con una estrategia u otra, así como con un recurso u otro. Por ello, hemos de ir observando el proceso de cada sujeto para adaptarnos lo máximo posible. En el caso del alumnado con TEA ocurre lo mismo, el trastorno es diferente en cada niño, y aunque esté diagnosticado como TEA, el trabajo con cada uno de ellos será acorde a sus características y a su evolución. Para Calderón-Bajaña *et al.* (2024) en el caso de los estudiantes con TEA, se presentan dificultades que impiden la correcta adquisición de los conocimientos necesarios para el desarrollo de las operaciones lógico-matemáticas, al presentar una muy baja generalización de los aprendizajes, selectividad atencional, resistencia a actividades nuevas o cambiantes y una falta de consistencia en las reacciones a la estimulación, por lo que se les ha de ofrecer otras técnicas de enseñanza para que puedan verse beneficiados.

Por ello, además de las estrategias y recursos que expusimos en el apartado anterior y que nos puede servir de gran ayuda para llevar a cabo con todo el alumnado, también vamos a tratar de manera generalizada una serie de estrategias acordes al TEA para poder cultivar de la manera más propicia posible el desarrollo lógico-matemático de los discentes con TEA (adaptando el proceso a los estilos de aprendizaje), apoyándonos en el uso de materiales visuales, actividades lúdicas y prácticas, y la conexión con situaciones de la vida real. Por tanto, podrá ser de gran ayuda: descomponer las tareas en otras más pequeñas; proporcionar un ambiente estable y predecible; evitar cambios inesperados; favorecer que los educandos tomen decisiones; utilizar materiales visuales atractivos y estructurados; fomentar la autonomía en la ejecución de las actividades; procurar momentos en que el estudiante está motivado; darles más tiempo para procesar la información; ejecutar la actividad en momentos secuenciados, repitiendo el proceso y las normas para su desarrollo; tener presente las preferencias e intereses del estudiante; ofrecer posibilidades de ensayos repetidos; y llevar a cabo siempre que sea posible un aprendizaje sin error.

Acorde a estas estrategias, las actividades lúdicas como: realizar puzles y rompecabezas; clasificar objetos de acuerdo a su tamaño, forma o color; reconocer figuras geométricas; realizar experimentos; relacionar conceptos mediante mapas mentales; Conexión con situaciones de la vida real; conectar las matemáticas con situaciones del mundo real; formular, plantear, transformar y resolver problemas a partir de situaciones de la vida cotidiana; usar medios tecnológicos interactivos; y/o usar programas y ayudas visuales, favorecerán la adquisición de dicho aprendizaje. Hemos de tener presente, que pueden tener problemas con las habilidades del lenguaje (tanto de expresión como de comprensión); pueden tener problemas en las habilidades motoras; pueden presentar dificultades para recordar secuencias; organizar la secuencia con una adecuada estructura y tiempo; pueden mostrar desinterés o aburrimiento, por lo que es importante hacer las actividades interactivas y divertidas, por lo que podemos servirnos de las tecnologías como medios que por sus características se adaptan a las necesidades de cada sujeto. Y nos centramos aquí, en las TIC como medios que facilitan el proceso educativo. Según Franceschette & Zapata-Cardona (2025) en muchos casos, reformular los problemas matemáticos o proporcionar indicaciones sobre el proceso de resolución puede ser crucial para que las personas con TEA logren resolverlos. Así, se hace fundamental conocer las prácticas que ya emprende el profesorado para garantizar la inclusión escolar.

## Base de datos de herramientas y recursos digitales para trabajar el pensamiento lógico-matemático con el alumnado con TEA

Para desarrollar el pensamiento lógico matemático con alumnos con TEA, podemos utilizar diferentes herramientas digitales puesto que cada vez con mayor asiduidad las TIC van siendo investigadas y estudiadas para que nos sirva como fuente de apoyo para atender a la diversidad. Siguiendo a García-Peñalvo (2018) las tecnologías educativas y/o del aprendizaje es un campo de investigación muy activo y en constante evolución, con una orientación interdisciplinar, contribuyendo en la mejora del aprendizaje. Y en este caso exponemos la siguiente Tabla 1 para conocer algunas de ellas:

Tabla 1. Herramientas y recursos digitales para potenciar el pensamiento lógico-matemático con el alumnado con TEA.

| *Herramientas y recursos* | *Descripción* | *Dirección web* |
|---|---|---|
| Khan Academy | Página web con cursos de matemáticas gratuitos, organizados por niveles y temas. | https://es.khanacademy.org/ |
| Kids Free | Aplicación que contiene juegos para el desarrollo de la memoria, la atención, la clasificación, la discriminación visual y auditiva, el razonamiento lógico, la secuenciación, la estructuración espacial y las coordinaciones viso-manuales. Tareas divertidas, interactivas y que proporcionan retroalimentación. | https://www.commonsensemedia.org/lists/free-apps-for-kids |
| Conecta 4 | Es un juego de estrategias personales que emplea el usuario para dar resolución al problema haciendo uso de la percepción, atención, memoria, concentración y anticipación. Gana el jugador que consiga conectar 4 fichas consecutivas del mismo color. | https://play.google.com/store/apps/details?id=com.lochmann.viergewinntmultiplayer&hl=es |
| CPA | Comunicador personal adaptable. Es un sistema de comunicación para personas con problemas graves de comunicación (autismo, trastornos neurológicos, discapacidades motoras, afasias). Reproduce voz y facilita la sintaxis mediante ordenación de imágenes. | http://www.comunicadorcpa.com/ |
| Proyect@ Emociones 2 | El Proyect@ Emociones Software es una aplicación que apoya al desarrollo de la empatía en los niños dentro del espectro del autismo. | https://sid-inico.usal.es/recursos_internet/proyect-emociones-2/ |

Tabla 1. Herramientas y recursos digitales para potenciar el pensamiento lógico-matemático con el alumnado con TEA.

| *Herramientas y recursos* | *Descripción* | *Dirección web* |
|---|---|---|
| Números encadenados | Aplicación que estimula el desarrollo de la atención, la discriminación visual y la velocidad de respuesta motriz mediante retos lúdicos. Di- senada para todas las edades, consiste en encadenar unas piezas con otras, las piezas se pueden encadenar si tienen la misma forma o el mismo color con dificultad creciente y antes de que se termine el tiempo. | https://www.glc.us.es/~jalonso/exercitium/cuadriseguidos-y-numeros-encadenados/ |
| Jade Autism | Contiene actividades para mejorar la atención y el razonamiento matemático. | https://www.jadend.tech/en |
| Picto TEA | Facilita la comunicación con el entorno mediante la comunicación por pictogramas digitales en lugar de las tarjetas físicas. | https://www.fundacionqualis.org.ar/2018/03/07/pictotea/ |
| PictoSonidos | Ayuda a aprender vocabulario. Es una aplicación donde a través de los pictogramas, con sonidos y locuciones asociadas, se ayuda a la comprensión de conceptos y a incrementar el vocabulario de personas con trastornos de comunicación oral, que aprenden más fácilmente a través de imágenes. | https://www.pictosonidos.com/ |
| Soyvisual | Ayuda a construir oraciones. | https://www.soyvisual.org/ |
| MathWorld | Página web con recursos gratuitos sobre matemáticas. | https://mathworld.wolfram.com/ |
| WolframAlpha | Buscador de respuestas a problemas matemáticos. | https://www.wolframalpha.com/ |
| Torre de Hanói | Es un juego lógico-matemático utilizado para estimular el pensamiento lógico y el razonamiento espacial, así como el desarrollo de habilidades y estrategias para la solución de problemas. Es un recurso didáctico para la aproximación al concepto de logaritmo. | https://www.geogebra.org/m/NqyWJVra |
| Math TV | Videos explicativos sobre diversos temas de matemáticas. | https://www.mathtv.com/ |
| Unicoos | Video lecciones que en algunos casos incluyen materiales complementarios. | https://www.unicoos.com/ |

Tabla 1. Herramientas y recursos digitales para potenciar el pensamiento lógico-matemático con el alumnado con TEA.

| *Herramientas y recursos* | *Descripción* | *Dirección web* |
|---|---|---|
| Logic Land | Aplicación para desarrollar el razonamiento lógico. Juego para el desarrollo que ofrece tareas de secuencia de figuras, adivina como luce una figura, define cuantos elementos, encuentra la forma entre otras, enfocadas al desarrollo del razonamiento lógico, habilidad matemática, inteligencia espacial, así como atención y memoria de niños. | https://play.google.com/store/apps/details?id=com.hedgehogacademy.shapesfree&pli=1 |
| Hojas de cálculo | Como Excel y OpenOffice Calc. | https://docs.google.com/spreadsheets/d/1o1o7K1Z6Jz2BiXH7F9nJFBXpcKlgGdndA621W1Ar7J8/edit?hl=es&gid=0#gid=0 |
| Herramientas de matemáticas interactivas | Aula Planeta. | https://www.aulaplaneta.com/2015/09/08/recursos-tic/25-herramientas-para-ensenar-matematicas-con-las-tic |
| Enigmas y juegos de lógica | Como los de Puzzleclopedia, BuscoAcertijos o Acertijos Matemáticos. | https://www.psicoactiva.com/puzzleclopedia/ |
| Math Cilenia | Minijuegos para practicar operaciones básicas, ideal para alumnos de primaria. | https://math.cilenia.com/es |
| Geogebra | Software matemático multiplataforma que permite crear simulaciones interactivas para comprender conceptos de álgebra y geometría de manera visual. | https://www.geogebra.org/classic?lang=es |
| Ábaco Online | Herramienta para representar números y aprender a sumar de manera gráfica. | https://www.online-calculator.com/es/online-abacus/ |
| Math Papa: | Calculadora de álgebra que resuelve ecuaciones paso a paso, ayudando a los alumnos a entender el proceso. | https://www.mathpapa.com/ |
| Tarjetas Numéricas y de Conteo: | Actividades interactivas que incluyen tarjetas numéricas, escalas numéricas y juegos de asociación número-cantidad. | https://es.pinterest.com/rafy2020rv/tarjetas-de-conteo-imagen-y-n%C3%BAmero/ |
| Rueda Numérica y Memotest: | Juegos para trabajar la asociación de números y cantidades, así como el seguimiento de patrones con formas geométricas. | https://wordwall.net/es-ar/community/memotest-n%C3%BAmeros |

Nota: Elaboración propia.

Podemos argumentar los múltiples beneficios de estos recursos y herramientas digitales, puesto que son interactivas (atractivas y motivadoras y, permiten la interacción directa con los conceptos matemáticos); visualización (facilitan y potencian la comprensión de conceptos abstractos a través de representaciones visuales y manipulativas); adaptabilidad (la tecnología permite ajustar el nivel de dificultad y personalizar las actividades dependiendo de las necesidades individuales). Según Guerrero-Jirón *et al.* (2020) las TIC asumen un rol fundamental en la mejora continua de la clase, lo que exige la fusión entre las TIC de aprendizaje y el conocimiento.

## Un enfoque multinivel como respuesta al aula heterogénea

El enfoque multinivel es una estrategia educativa diseñada para atender la diversidad en el aula, permitiendo que todos los estudiantes, independientemente de sus habilidades y conocimientos previos, puedan aprender juntos, facilitando, a través de las adaptaciones del currículum, el progreso de cada alumno según sus necesidades y capacidades. De acuerdo con Hernández-Caravaca (2018), mediante la puesta en práctica de la enseñanza multinivel se favorece el respeto a las diferencias del grupo y se fomenta la igualdad de oportunidades puesto que se plantean distintas situaciones de aprendizaje que dan respuesta a los diferentes niveles de competencia curricular de los estudiantes. En este sentido, el diseño de actividades multinivel constituye una forma de atender la diversidad en el aula, facilitando el nivel de competencia curricular a cada estudiante. Por ello, podemos exponer como ventajas del enfoque multinivel: la promoción del aprendizaje inclusivo (participando todo los estudiantes en las mismas actividades, pero con diferentes niveles de complejidad); diversidad y equidad (reconociendo y valorando que cada discente es diferente y se ha de manifestar las mismas oportunidades de aprendizaje y éxito a todos, trabajando cada uno a su ritmo ajustado a sus necesidades); adaptación del currículo (modificando y adaptando el currículo a cada sujeto enfocado en la enseñanza individualizada); ambiente de apoyo (donde todos dan y reciben apoyo: mejorando la autoestima, reduciendo la competitividad, reconociendo los logros y capacidades, trabajando en equipo, asignando roles y responsabilidades). En definitiva, el aula multinivel se convierte en un espacio acorde a las características de cada individuo, donde se diseñan proyectos interdisciplinares basados en la colaboración y el aprendizaje constructivo.

## Conclusión

En el marco de una educación inclusiva donde nos situamos, es fundamental que los centros educativos y los docentes adopten un enfoque de enseñanza personalizada para atender a la diversidad dentro de un mismo espacio. Siguiendo a Vélez & Rivadeneira (2024), para que el estudiante con TEA logre desarrollar un proceso acorde a sus características y de manera correcta, es necesario trabajar con recursos que sean llamativos para ellos, para que los estudiantes puedan participar en clase y conseguir sus metas. Planteamos en este caso, las herramientas y recursos digitales para el desarrollo del pensamiento lógico-matemático del alumnado con TEA, trabajados en un aula multinivel para que el currículum se adapte a sus necesidades y todos los alumnos alcancen sus logros en función de sus ritmos de aprendizaje.

## Financiación

## Referencias bibliográficas

## Webgrafía

# PERFIL BIOGRÁFICO DE LOS AUTORES

## Víctor Abella-García

Profesor Contratado Doctor en el Departamento de Ciencias de la Educación de la Universidad de Burgos. Doctor en Ciencias de la Educación por la Universidad de Valladolid, obtuvo la calificación de sobresaliente cum laude con una tesis centrada en la formación docente en entornos digitales. Previamente se tituló como Maestro de Educación Primaria y como Licenciado en Psicopedagogía, consolidando una trayectoria centrada en la mejora de la enseñanza, la inclusión y el desarrollo profesional docente.

Su investigación se orienta a la aplicación crítica de las tecnologías educativas, la competencia digital docente, la alfabetización digital, el uso de recursos digitales con fines inclusivos y el impacto de las metodologías activas en la motivación y el rendimiento del alumnado. Ha participado en más de una decena de proyectos competitivos nacionales e internacionales y en contratos de transferencia con administraciones y centros educativos, reforzando su compromiso con una universidad conectada al entorno.

Cuenta con numerosas publicaciones en revistas indexadas, capítulos de libro y participaciones en congresos. En el ámbito docente, imparte asignaturas en grados y másteres de educación, tutoriza trabajos académicos y participa como formador en programas de desarrollo profesional. Es miembro activo de RUTE y la SEP.

e-mail: vabella@ubu.es, https://orcid.org/0000-0001-9406-9313

## Pedro-José Arrifano-Tadeu

Profesor de Educación Matemática en el Instituto Politécnico da Guarda (IPG, Portugal) y coordinador del Centro de Estudios en Educación e Innovación (CI&DEI) — sede de Guarda desde diciembre de 2018. Posee formación sólida: grado en Educación Matemática (Universidad de Coimbra, Portugal), máster en Educación Matemática (UBI – Covilhã) y doctorado en Didáctica de la Matemática (UTAD). Además, realizó un postdoctorado en “Problem-Solving Skills in Kindergarten” en la Sakarya University (Turquía).

Su trayectoria institucional incluye haber sido decano de la Escuela de Educación, Comunicación y Deporte del IPG (2015–2019) y miembro de comisiones ejecutivas de centros de investigación.

En el ámbito científico, desarrolla líneas de investigación relacionadas con educación matemática, TIC y educación digital, innovación educativa, inclusión, multiculturalismo, educación infantil y primaria, pensamiento computacional, metodologías activas, robótica educativa y discapacidad.

Ha participado en numerosos proyectos europeos (Erasmus+, EIT-HEI, etc.), publicado en revistas internacionales y dirigidos trabajos sobre TIC, inclusión y formación docente. Entre sus publicaciones recientes destaca el estudio sobre “dificultades tecnológicas en la enseñanza de matemáticas en secundaria en Italia” (2025).

Su perfil combina experiencia docente, liderazgo institucional, investigación internacional y compromiso con la innovación educativa y la inclusión.

e-mail: ptadeu@ipg.pt, https://orcid.org/0000-0002-0698-400X

## Vanesa Ausín-Villaverde

Profesora Titular de Universidad en Didáctica y Organización Escolar en la Universidad de Burgos, donde desarrolla su labor docente e investigadora desde hace más de una década. Licenciada en Psicopedagogía, Diplomada en Magisterio y Doctora en Ciencias de la Educación, ha orientado su trayectoria a la mejora de la calidad educativa, la equidad en el aula, el desarrollo de competencias docentes y la integración de tecnologías emergentes con fines inclusivos.

Ha participado en numerosos proyectos competitivos y contratos de transferencia centrados en competencias digitales docentes, uso de herramientas tecnológicas en contextos escolares y universitarios, y evaluación de prácticas educativas equitativas. Su producción científica incluye artículos en revistas JCR y Scopus, capítulos de libro y manuales docentes sobre formación inicial y continua del profesorado, TIC, realidad aumentada e inclusión educativa.

En el ámbito docente imparte asignaturas en los grados de Educación Infantil, Primaria y Pedagogía, así como en el Máster de Profesorado, y ha dirigido TFG, TFM y tesis doctorales. Participa en actividades de formación permanente del profesorado y mantiene vínculos con centros educativos. Miembro activo de grupos de investigación y redes como RUTE, presenta con regularidad trabajos en congresos nacionales e internacionales sobre inclusión, competencia digital y metodologías activas.

e-mail: vausin@ubu.es, https://orcid.org/0000-0002-8943-6251

## Vanesa Delgado-Benito

Profesora Titular de Universidad en el área de Didáctica y Organización Escolar de la Universidad de Burgos. Es Diplomada en Magisterio (Educación Infantil), Licenciada en Pedagogía y Doctora en Ciencias de la Educación, con calificación Cum Laude, Mención de Doctorado Europeo y Premio Extraordinario de Doctorado. Ha obtenido diversas becas de investigación y destaca su estancia predoctoral en la Università degli Studi di Roma Tre.

En el ámbito docente ha sido profesora asociada, ayudante y titular, y cuenta con evaluación DOCENTIA con la máxima calificación ("actividad docente muy destacada"). Posee el Certificado de Formación Inicial en Docencia Universitaria y ha impartido cerca de 50 cursos de formación del profesorado sobre recursos digitales, herramientas 2.0 y evaluación formativa. Desde 2015 forma a doctorandos en competencias digitales y uso de la red para la investigación.

En investigación, pertenece a un grupo activo desde 2010 y ha participado en más de veinte proyectos, incluidos estudios interdisciplinarios sobre trastornos del movimiento y el proyecto europeo FORDYSVAR sobre aprendizaje inclusivo mediante realidad virtual. Ha publicado artículos en revistas indexadas, más de treinta capítulos de libro y participado en numerosos congresos. Es evaluadora de revistas científicas y miembro de SEP y RUTE.

e-mail: vdelgado@ubu.es, https://orcid.org/0000-0001-8168-7120

## María-Dolores Díaz-Noguera

Profesora Titular de Universidad en el Departamento de Didáctica y Organización Educativa de la Universidad de Sevilla. Doctora en Ciencias de la Educación, desarrolla desde 1995 una sólida trayectoria científica centrada en innovación docente, educación inclusiva y transformación digital desde perspectivas críticas, éticas y feministas. Ha participado en numerosos proyectos de I+D+i financiados por el Ministerio de Ciencia e Innovación y por convocatorias autonómicas, destacando CODITEA, PID2023-149083OA-I00 y MindGuard, vinculados a la inclusión, la accesibilidad y el uso de tecnologías emergentes para la mejora educativa y social.

Su actividad investigadora incluye el diseño y validación de escalas psicopedagógicas, análisis de actitudes docentes hacia la inteligencia artificial, estudios sobre gamificación, motivación y bienestar educativo, así como investigaciones sobre transformación digital universitaria. Ha publicado en revistas indexadas de impacto como IJERPH, Education Sciences, Sustainability o IJEM, y es autora de numerosos capítulos en editoriales como Dykinson, Octaedro y Wolters Kluwer.

Además, participa activamente como ponente, evaluadora y miembro de equipos editoriales en congresos y revistas especializadas. Su trayectoria destaca por el rigor científico, el compromiso inclusivo y su papel como referente en pedagogía crítica, equidad digital y transformación educativa en el ámbito iberoamericano.

e-mail: noguera@us.es, https://orcid.org/0000-0002-0624-4079

## José-María Fernández-Batanero

Catedrático de Universidad de Didáctica y Organización Educativa de la Universidad de Sevilla. Su actividad investigadora se centra en tres líneas de trabajo: formación del profesorado, atención a la diversidad y tecnología educativa como apoyo a las personas con necesidades educativas especiales.

Forma parte del Grupo de Investigación "GID" (HUM-390) y del Grupo de Tecnología Educativa de la Universidad de Sevilla.

Director de la Red Latinoamericana de TIC aplicadas a personas con discapacidad (ReLaTICyD) y Miembro del Consejo Consultivo del Centro de Estudos em Educaçao e Inovaçao (CI&DEI, Portugal). Miembro nato de la Cátedra Institucional de Educación en Tecnologías Emergentes, Gamificación e Inteligencia Artificial (EduEmer) de la Universidad Pablo de Olavide (Sevilla).

Evaluador de la Agencia Nacional de Evaluación y Prospectiva (ANEP) y asesor de la Comisión Nacional de Investigación Científica y Tecnológica – CONICYT, del Gobierno de Chile. Investigador en el prestigioso ranking World's Top 2% Scientists List 2024 Stanford University.

Ha impartido docencia en los tres niveles del sistema educativo, siendo en la actualidad funcionario docente en Primaria y Universidad (Primaria excedencia voluntaria). Está en posesión de 6 diplomas a la Excelencia Docente Universitaria y de la Insignia de Oro de la Ciudad de Sevilla en reconocimiento a la Excelencia Docente.

e-mail: batanero@us.es, https://orcid.org/0000-0003-4097-5382

## José Fernández-Cerero

Doctorando con contrato predoctoral FPU en el área de Didáctica y Organización Escolar de la Universidad de Sevilla, adscrito al grupo de investigación: Didáctica: Análisis Tecnológico y Cualitativo de los Procesos de Enseñanza-Aprendizaje (HUM-390).

Sus líneas de trabajo se centran en la formación del profesorado, la integración de tecnologías de la información y la comunicación (TIC) y la educación inclusiva, especialmente en contextos de diversidad.

Como investigador y docente, participa activamente en proyectos de innovación educativa, orientados a cómo las TIC pueden mejorar el aprendizaje y la inclusión en entornos escolares.

Su vocación científica se basa en considerar la educación como motor de equidad, desarrollo social y bienestar, con una clara apuesta por la transformación educativa mediante la tecnología.

José Fernández Cerero combina su formación académica y experiencia docente con un enfoque inclusivo, crítico y proactivo, orientado a favorecer la diversidad y la calidad educativa mediante prácticas innovadoras. Si deseas, puedo hacer también una versión en inglés o más sintética.

e-mail: jfcerero@us.es, https://orcid.org/0000-0002-2745-6986

## Inmaculada García-Martínez

Profesora Titular en el Departamento de Pedagogía de la Universidad de Granada. Doctora en Ciencias de la Educación con Premio Extraordinario de Doctorado, es Licenciada en Psicopedagogía y Diplomada en Magisterio en Educación Especial, lo que fundamenta su trayectoria centrada en la inclusión, la atención a la diversidad y la innovación educativa.

Ha participado en múltiples proyectos autonómicos, nacionales e internacionales sobre competencias docentes inclusivas y digitales, educación del alumnado con discapacidad y diseño universal para el aprendizaje. Destaca su implicación en investigaciones vinculadas al TEA, las tecnologías accesibles y la ética profesional docente.

Cuenta con numerosas publicaciones en revistas indexadas en JCR y Scopus, así como capítulos de libro en editoriales de prestigio como Octaedro, Narcea, Graó y Dykinson. Ha realizado estancias de investigación en universidades de referencia, como Macquarie (Australia) y Bolonia (Italia), fortaleciendo su proyección internacional y su participación en redes científicas.

En el ámbito docente, imparte clases en los grados de Educación Infantil y Primaria y en el Máster de Educación Inclusiva, abordando innovación metodológica, diversidad y justicia educativa. También ha dirigido TFG, TFM y tesis doctorales. Su trabajo se caracteriza por un compromiso firme con la equidad, la transformación educativa y la responsabilidad social.

e-mail: igmartinez@ugr.es, https://orcid.org/0000-0003-2620-5779

## Óscar Gavín-Chocano

Profesor en el área de Didáctica y Organización Escolar del Universidad de Jaén (UJAen).

Su investigación se centra en la competencia digital docente, la educación inclusiva, la inteligencia emocional, la resiliencia y el bienestar de estudiantes y profesionales de la educación, así como en los procesos de enseñanza-aprendizaje en contextos diversos.

Ha publicado numerosos artículos en revistas indexadas, con un perfil de más de 50 publicaciones, que abordan temas como auto-concepto, motivación, estrés académico, discapacidad intelectual, educación inclusiva y metodologías digitales.

Además, su trayectoria incluye participación en proyectos de I+D+i relacionados con la inclusión, la formación docente y la calidad educativa; ha abordado la salud emocional, el uso de TIC en educación y la mejora del bienestar de colectivos vulnerables. Su enfoque combina el rigor científico con un compromiso ético hacia la equidad, la inclusión y la transformación social mediante la educación.

e-mail: ogavin@ujaen.es, https://orcid.org/0000-0002-1975-5003

## Elena Hernández-de-la-Torre

Profesora Titular en el área de Didáctica y Organización Escolar de la Universidad de Sevilla. Pertenece al grupo de investigación I.D.E.A. (Innovación, Desarrollo, Evaluación y Asesoramiento) del instituto IEDU. Cuenta con una producción científica extensa: 90 publicaciones entre artículos, capítulos, ponencias y libros, además de dirigir cuatro tesis doctorales.

Sus líneas de trabajo incluyen la formación docente, redes escolares, inclusión educativa, diversidad, tutorización y mentoría del profesorado, así como el uso de tecnologías educativas y la atención a la diversidad.

Sus investigaciones han abordado temáticas como mentoría de docentes principiantes, redes educativas virtuales, educación inclusiva para alumnado con altas capacidades o con diversidad funcional, formación para el uso de TIC en contextos de discapacidad, y liderazgo educativo.

También ha participado en al menos 12 proyectos competitivos nacionales y autonómicos sobre innovación, escuela inclusiva, evaluación institucional y formación del profesorado.

Su trayectoria combina docencia universitaria, investigación aplicada, asesoramiento y transferencia de resultados hacia centros escolares, contribuyendo a una educación más equitativa, colaborativa e inclusiva.

e-mail: eht@us.es, https://orcid.org/0000-0001-6390-1955

## Carlos Hervás-Gómez

Profesor Titular de Universidad en el Departamento de Didáctica y Organización Educativa de la Universidad de Sevilla, donde desarrolla su labor docente e investigadora desde 2001. Licenciado en Filosofía y Ciencias de la Educación y Doctor en Educación, centra su trayectoria en tecnología educativa, formación del profesorado y evaluación docente. Su investigación se orienta especialmente al uso de tecnologías emergentes y de la inteligencia artificial en contextos educativos, así como al desarrollo de competencias digitales inclusivas y a la atención a la diversidad.

Ha participado en numerosos proyectos I+D financiados, incluidos los proyectos nacionales CODITEA (PID2022-138346OB-I00), MindGuard (PII/2024/0018) y PID2023-149083OA-I00, todos ellos centrados en inclusión, autismo y tecnologías asistenciales. Su producción científica incluye artículos en revistas indexadas como EJER, Contemporary Educational Technology, EKS, IJEM o IJERPH, además de más de diez libros y monografías, entre ellos The Education Revolution through Artificial Intelligence (Octaedro, 2024), y obras coeditadas como Conexiones digitales y Sembrando el futuro.

En el ámbito docente, forma a futuros profesionales en competencias digitales, evaluación y diseño didáctico con TIC. Participa activamente en congresos y comités científicos y mantiene un enfoque pedagógico basado en la innovación, la accesibilidad y una digitalización educativa ética y humanista.

e-mail: hervas@us.es, https://orcid.org/0000-0002-0904-9041

## Eloy López-Meneses

Profesor Titular de Universidad en el Departamento de Educación y Psicología Social de la Universidad Pablo de Olavide. Doctor en Filosofía y Ciencias de la Educación por la Universidad de Sevilla, obtuvo el Premio Extraordinario de Doctorado y el Segundo Premio Nacional de Fin de Carrera.

Con más de dos décadas de experiencia docente universitaria, ha sido distinguido tres veces con la calificación de excelencia en el programa DOCENTIA y ha impartido más de 3 000 horas de formación especializada en tecnología educativa.

Su actividad investigadora se centra en tecnologías emergentes aplicadas a la educación, incluyendo realidad aumentada y virtual, inteligencia artificial, MOOC, u-learning y formación docente digital. Dirige el grupo de investigación EduInnovagogía (HUM-971) y ha participado en más de 25 proyectos competitivos, nueve de ellos nacionales, algunos como investigador principal. Lidera la Red Internacional Latinnova, integrada por equipos de 16 países.

Posee una amplia producción científica: 185 artículos (10 JCR Q1), más de 78 libros y 150 capítulos, además de 48 marcas educativas registradas. Ha dirigido 19 tesis doctorales y actualmente supervisa otras diez. Es editor de la revista IJERI y director de la colección "Innovation in Social Sciences" en Dykinson. Su trayectoria destaca por su impacto en la innovación educativa y la transformación digital.

e-mail: elopmen@upo.es, https://orcid.org/0000-0003-0741-5367

## Antonio Luque-de-la-Rosa

Profesor Titular de Universidad en el Departamento de Educación de la Universidad de Almería, donde desarrolla desde hace más de veinte años una sólida labor docente e investigadora.

Doctor en Innovación Educativa, su trayectoria se orienta a la inclusión, la atención a la diversidad, la evaluación docente y el uso de tecnologías aplicadas al aprendizaje. Ha participado en múltiples proyectos competitivos nacionales e internacionales, entre ellos el proyecto estatal CODITEA (PID2022-138346OB-I00), centrado en la capacitación docente para el uso de competencias digitales inclusivas con alumnado con TEA, así como diversas iniciativas de cooperación educativa en México, Guatemala, Ecuador y Honduras dirigidas a la igualdad de género, la atención a estudiantes con discapacidad y la mejora institucional.

Su producción científica es amplia y consolidada, con publicaciones en revistas de impacto como Sustainability, Education Sciences, Children, International Journal of Environmental Research and Public Health, REDIE o Frontiers in Psychology. Aborda temas como la realidad aumentada, el bienestar emocional, la educación artística, la innovación universitaria, el desempeño docente y la inclusión educativa. También es autor y coautor de libros en editoriales como Octaedro y Dykinson. Su labor destaca por la combinación de rigor académico, compromiso social e innovación pedagógica.

e-mail: aluque@ual.es, https://orcid.org/0000-0001-7981-029X

## Marta Montenegro Rueda

Profesora Ayudante Doctora en la Facultad de Ciencias de la Educación y del Deporte (Melilla), Universidad de Granada. Graduada en Educación Primaria y con tres másteres en Psicopedagogía, Lengua Española y Educación Especial, obtuvo el Doctorado en Educación por la Universidad de Sevilla con Mención Internacional y la calificación de Sobresaliente Cum Laude.

Su trayectoria investigadora comenzó en 2018 como personal I+D+i en la Universidad de Sevilla, donde posteriormente fue Personal Investigador en Formación y Profesora Sustituta Interina antes de incorporarse a la UGR.

Su investigación se centra en inclusión educativa y Tecnologías de la Información y la Comunicación, con especial atención al alumnado con necesidades educativas especiales y a la competencia digital docente. Pertenece al Grupo HUM-390 y ha realizado estancias en el Instituto

Politécnico de Guarda y en la Universidad de Granada. Ha participado en proyectos competitivos como PID2022-138346OB-I00 y PID2019-108230RB-I00.

Cuenta con más de 35 artículos en revistas indexadas, 80 capítulos de libro y más de 40 comunicaciones en congresos. Es editora invitada y revisora en revistas científicas y ha recibido premios al mejor artículo científico (2023 y 2024) y el Premio UNIA-Digital a la Investigación (2025). Su labor docente integra innovación, TIC y educación inclusiva.

e-mail: mmontenegro@ugr.es, https://orcid.org/0000-0003-4733-289X

## Gloria-Luisa Morales-Pérez

Doctora en Ciencias de la Educación y Licenciada en Pedagogía. Su trayectoria se centra en la innovación educativa, la integración de tecnologías digitales en la enseñanza, la atención a la diversidad y la formación del profesorado.

Ha liderado investigaciones sobre competencias tecnológicas en estudiantes adultos y ha diseñado y validado cuestionarios sobre uso de redes sociales y TIC como herramientas pedagógicas.

Su producción académica incluye estudios sobre robótica educativa, realidad aumentada, e-learning, educación inclusiva y formación digital en distintos contextos (incluyendo alumnos mayores y formación profesional).

Ha participado en proyectos de formación docente vinculados a la educación inclusiva, al uso de TIC para alumnado con discapacidad y a modelos de enseñanza-aprendizaje flexibles.

Su experiencia docente se desarrolla principalmente como profesora titular en la Escuela Universitaria de Osuna, donde actualmente trabaja, impartiendo asignaturas relacionadas con tecnologías educativas, formación del profesorado y aprendizaje a lo largo de la vida.

Además, colabora en el desarrollo, diseño y producción de contenidos educativos, materiales multimedia e-learning y formación para la empleabilidad, aportando una visión inclusiva y tecnológicamente actualizada.

e-mail: glorialuisamp@euosuna.org, https://orcid.org/0000-0002-4699-4093

## María José Navarro-Montaño

Profesora Titular de Universidad en el Departamento de Didáctica y Organización Educativa de la Universidad de Sevilla, adscrita a la Facultad de Ciencias de la Educación, donde imparte asignaturas relacionadas con la atención a la diversidad y las estrategias docentes inclusivas.

Forma parte del grupo de investigación “Didáctica: Análisis Tecnológico y Cualitativo de los Procesos de Enseñanza-Aprendizaje”, centrado en innovación educativa y uso de tecnologías para la mejora de la práctica docente.

Su trayectoria investigadora se ha orientado a la educación inclusiva, la formación del profesorado, los recursos didácticos para la diversidad, la planificación escolar con criterios de equidad y la integración de las TIC en contextos educativos. Ha participado en numerosos proyectos I+D dedicados a la diversidad, las metodologías activas y la formación digital docente, y cuenta con una producción científica amplia, que incluye libros, capítulos y artículos centrados en dirección escolar, inclusión y prácticas educativas innovadoras.

Su actividad combina docencia universitaria, publicación científica, participación en proyectos competitivos y transferencia de conocimiento orientada a mejorar la respuesta educativa a la diversidad y a promover entornos formativos más equitativos.

e-mail: maripe@us.es, https://orcid.org/0000-0003-2462-8348

## Rocío Piñero-Virué

Doctora en Educación y Profesora Sustituta en el Departamento de Didáctica y Organización Educativa de la Universidad de Sevilla. Su formación incluye la Diplomatura en Magisterio (Educación Infantil), la Licenciatura en Pedagogía y diversas microcredenciales especializadas en atención a la diversidad y respuesta educativa inclusiva.

Su trayectoria docente, iniciada en 2002-03, abarca titulaciones de Grado y Máster en Educación Infantil, Primaria, Pedagogía, Psicopedagogía, Actividad Física y Deportiva, así como el MAES. Ha dirigido TFG y TFM y mantiene una intensa labor tutorial coordinada con el SACU, atendiendo a alumnado con necesidades específicas de apoyo educativo desde un enfoque inclusivo y personalizado.

En investigación, forma parte del Grupo HUM-390 “Análisis Tecnológico y Cualitativo de los Procesos de Enseñanza-Aprendizaje”, centrado en diversidad, inclusión, tecnologías educativas y metodologías emergentes.

Su tesis doctoral dio lugar a la patente educativa “Las Aventur@s de Horacio el Ratón”, orientada a mejorar la comprensión lectora en alumnado con dificultades. Ha participado en proyectos autonómicos, nacionales e internacionales, publicado en revistas indexadas como Píxel-Bit o Social Sciences, y contribuido con capítulos y coordinaciones en editoriales de prestigio. Está acreditada por DEVA y ANECA para Ayudante Doctora, Contratada Doctora y Profesora de Universidad Privada.

e-mail: rpv@us.es, https://orcid.org/0000-0002-0120-0931

## Miguel-María Reyes-Rebollo

Profesor Titular de la Universidad de Sevilla, adscrito a la Facultad de Ciencias de la Educación y al Departamento de Didáctica y Organización Educativa. Cuenta con dos sexenios de investigación y una sólida formación multidisciplinar: es Licenciado en Filosofía y Ciencias de la Educación por la UNED, Doctor en Psicopedagogía y Doctor en Ciencias en el área de Gestión y Salud Ambiental por la Universidad de Huelva. Además, posee un Máster en Sistemas Integrados de Gestión Ambiental, así como títulos de especialización en Medio Ambiente, Comunicación Social y Dirección Técnica de Brigadas Helitransportadas para la extinción de incendios forestales.

Miembro del Grupo de Investigación HUM-390, ha participado en numerosos proyectos autonómicos, nacionales e internacionales centrados en innovación docente, inclusión, educación ambiental y tecnología educativa. Su trayectoria investigadora aborda la Educación Especial, la Atención a la Diversidad, la formación del profesorado y el uso de TIC en contextos educativos. Ha intervenido en congresos especializados y publicado artículos, capítulos y libros vinculados a estas líneas de trabajo.

Ha dirigido diversas tesis doctorales y ejerce docencia en materias relacionadas con diversidad, educación especial y didáctica. Su filosofía profesional se orienta a formar docentes capaces de construir una escuela inclusiva y comprometida con la atención a la diversidad.

e-mail: mmreyes@us.es, https://orcid.org/0000-0003-1496-521X

## Sonia Rodríguez-Cano

Doctora en Ciencias de la Educación por la Universidad de Burgos, donde también obtuvo las titulaciones de Licenciada en Pedagogía y Diplomada en Magisterio. Actualmente es CTDOC en la Facultad de Educación y cuenta con una sólida trayectoria en innovación pedagógica, tecnologías aplicadas al aprendizaje e inclusión educativa. Durante nueve años fue profesora asociada en Didáctica y Organización Escolar y compaginó esta labor con docencia en Educación Infantil. En gestión académica ha sido coordinadora de título y, desde 2023, Secretaria Académica de la Facultad de Educación.

Es fundadora y presidenta de la asociación ASIRE, desde la que ha impulsado cinco congresos educativos, dos de ellos como presidenta del Comité Científico. Posee amplia experiencia en proyectos europeos, destacando la coordinación de FORDYSVAR (Erasmus+), centrado en realidad aumentada y virtual para alumnado con dislexia, además de participar en IDEAL, CARS, TUT4IND y MINDivers4All. Colabora también en RISKREAL (H2020) y PROCEDIM.

Forma parte del grupo EDINTEC y de los Grupos de Innovación Docente 3I y TICEF. Ha participado en 17 proyectos, 25 eventos científicos y cuenta con numerosas publicaciones en revistas indexadas. Su labor incluye creación de recursos educativos, evaluación científica y transferencia a la práctica docente.

e-mail: srcano@ubu.es, https://orcid.org/0000-0002-4242-6865

## Pedro Román-Graván

Profesor de EGB, Licenciado en Ciencias de la Educación (Pedagogía), Experto Universitario en Evaluación Educativa y Doctor en Ciencias de la Educación. Es Profesor Titular de Universidad en el Departamento de Didáctica y Organización Educativa de la Universidad de Sevilla. Su línea de investigación principal se relaciona con los procesos educativos mediados por la Tecnología Educativa y de formación en contextos de diversidad, evaluación, robótica educativa, realidad aumentada, impresión 3D y drones.

Forma parte del Grupo de Investigación "GID" (HUM-390) y del Grupo de Tecnología Educativa de la Universidad de Sevilla. Evaluador de la Agencia Nacional de Evaluación y Prospectiva (ANEP).

Ha impartido docencia en los tres niveles del sistema educativo: Educación Primaria, Educación Secundaria Obligatoria y Universidad, siendo en la actualidad funcionario docente en Universidad de Sevilla.

Posee cuatro sexenios de investigación reconocidos por la Comisión Nacional Evaluadora de la Actividad Investigadora (CNEAI) ANECA: tres de Investigación y uno de Transferencia.

Ocupa la posición 12/51 entre los investigadores de Educación más citados en el Área de Didáctica y Organización Escolar de su Universidad. Desde 2012 hasta 2023 ha sido miembro de Comisión Asesora Aula de la Experiencia de Universidad de Sevilla, donde imparte la asignatura de Nuevas Tecnologías.

e-mail: proman@us.es, https://orcid.org/0000-0002-1646-9247

## Carmen Siles-Rojas

Profesora Titular de Universidad en el Departamento de Didáctica y Organización Educativa de la Universidad de Sevilla. Doctora en Pedagogía, acumula más de veinticinco años de experiencia docente e investigadora, con una trayectoria previa en distintos niveles del sistema educativo.

Ha impartido docencia en diplomatura, licenciatura, grado, máster y doctorado, especialmente en materias vinculadas a educación inclusiva, formación del profesorado, tecnologías aplicadas a la enseñanza y diseño de entornos de aprendizaje accesibles.

Su investigación se centra en innovación educativa, atención a la diversidad, metodologías activas mediadas por tecnología y capacitación docente para el uso inclusivo de recursos di-gitales. Ha participado en numerosos proyectos competitivos, entre ellos EDU2016-75232-P, PID2019-108230RB-100 y PID2022-138346OB-I00 (CODITEA), así como en iniciativas autonómicas orientadas a la mejora educativa. Forma parte del grupo HUM-390, donde ha contribuido al desarrollo de investigaciones sobre robótica educativa para alumnado con TEA y sobre competencias docentes digitales.

Cuenta con una amplia producción científica: más de cincuenta capítulos de libro, ocho libros y alrededor de veinte artículos en revistas especializadas. Ha participado en más de treinta congresos y es evaluadora en revistas indexadas. Su trayectoria destaca por el compromiso con la inclusión, la innovación didáctica y la mejora de la enseñanza universitaria.

e-mail: csiles@us.es, https://orcid.org/0000-0002-6854-4990

*Diciembre, 2025*

La Junta de Parlamentarios de Cantabria y la Preautonomía (1977-1983)
*Víctor Gautier Fernández*

Liberalismo, masonería y clandestinidad. La formación de sociabilidades políticas en los inicios de la España Contemporánea (1814-1825)
*Sergio Ceballos Coz*

Auge y declive de una saga de jándalos en Sevilla: la familia De la Cuadra (1795-1920)
*José Andrés Otero Campos*

Tras las huellas del silencio. Cuba y la España franquista, 1940-1958
*Katia Figueredo Cabrera*

Ciudades y corona. Fiscalidad, representación y gobierno en la monarquía hispánica en la Edad Moderna
*Ramón Lanza García y Roberto López Vela (eds.)*

El lenguaje de la secularización en América Latina. Contribuciones para un léxico
*Elisa Cárdenas Ayala y Francisco A. Ortega (coords.)*

El desafío del desarrollo. Trayectorias de los grandes economistas latinoamericanos del s. XX
*Juan Odisio y Marcelo Rougier (eds.)*

El león durmiente. Democracia, republicanismo y federalismo en España, 1812-1936
*Manuel Suárez Cortina*

De la harina al acero. José María Quijano y las forjas de Buelna (Cantabria, 1873-1914)
*Sara del Hoyo Maza*

El historiador como docente
*Carlos Larrinaga y Antonio J. Pinto (eds.)*

Los pósitos de pescadores. Una inusitada aventura reformista (1917-1943)
*Alberto Ansola Fernández*

Escapando del eurocentrismo. Una historia no-europea del mundo en la Edad Moderna
*Rafael Barquín*

Discurso sobre la democracia
*Carlos Nieto Blanco*

En la sociedad contemporánea, marcada por una evolución tecnológica acelerada, la educación se enfrenta al reto —y a la oportunidad— de incorporar nuevas herramientas digitales que transformen el aprendizaje, haciéndolo más inclusivo, flexible y significativo. Este libro se inscribe en esa línea de compromiso con la innovación educativa y la equidad, poniendo el foco en un colectivo que requiere una atención especializada y sensible: el alumnado con Trastorno del Espectro Autista (TEA).

Lejos de abordar la tecnología desde una perspectiva meramente instrumental, esta obra propone un enfoque profundamente humano, donde las tecnologías emergentes —como la realidad aumentada y virtual, los entornos inmersivos, los sistemas de comunicación aumentativa, las aplicaciones móviles o la inteligencia artificial— se convierten en aliadas para favorecer la autonomía, la autorregulación emocional, el desarrollo de habilidades sociales y la inclusión real del alumnado con TEA en los entornos educativos.

Cada capítulo ofrece una aportación rigurosa, fruto de la experiencia, la investigación y la reflexión crítica de profesionales del ámbito de la educación, la psicología, la tecnología y la atención a la diversidad. Desde estudios de caso hasta propuestas metodológicas, pasando por herramientas prácticas y marcos teóricos actualizados, los autores y autoras comparten conocimientos que nacen del trabajo colaborativo y del contacto directo con la realidad de las aulas.